\documentclass[10pt,aps,prb,twocolumn,showpacs,superscriptaddress]{revtex4-2}
 
\usepackage{amsfonts}
\usepackage{braket} 
\usepackage{mathrsfs}
\usepackage{amsmath}

\usepackage[backref=none,
bookmarksnumbered=true,
bookmarks=true,
bookmarksopen=true,
colorlinks=true,
citecolor=blue,
linkcolor=blue,
anchorcolor=green,
urlcolor=blue,unicode=false]{hyperref}
\usepackage{cleveref}

\pdfoutput=1
\usepackage{bm}
\usepackage{amssymb}
\usepackage{graphicx}
\usepackage{color}
\usepackage[dvipsnames]{xcolor}
\usepackage{simplewick}
\usepackage[version=3]{mhchem}
\usepackage{wasysym}
\usepackage{physics}
\usepackage{siunitx}
\usepackage{wrapfig}
\usepackage{empheq}
\usepackage[extra]{tipa}
\crefname{figure}{Fig.}{Fig.}
\Crefname{figure}{Fig.}{Fig.}
\crefname{equation}{Eq.}{Eq.}
\Crefname{equation}{Eq.}{Eq.}
\crefname{section}{Sec.}{Sec.}
\Crefname{section}{Sec.}{Sec.}
\crefname{appendix}{App.}{App.}
\Crefname{appendix}{App.}{App.}

\DeclareMathOperator{\sign}{sign}

\newcommand{\thetaJC}{\theta_{\text{JC}}}

\newcommand{\phiJC}{\phi_{\text{JC}}}
\newcommand{\phiR}{\phi_{\text{R}}}
\newcommand{\thetaR}{\theta_{\text{R}}}
\newcommand{\thetaJCR}{\theta_{\text{JC+R}}}
\newcommand{\OmegaR}{\Omega_{\text{R}}}
\newcommand{\OmegaJC}{\Omega_{\text{JC}}}
\newcommand{\OmegaRvec}{\vb{\Omega}_{\text{R}}}
\newcommand{\OmegaJCvec}{\hat{\vb{\Omega}}_{\text{JC}}}
\newcommand{\OmegaJCRvec}{\hat{\vb{\Omega}}_{\text{JC+R}}}
\newcommand{\Neff}{N_{\text{eff}}}

\begin{document}

\title{Bang-bang protocol for nondispersive qubit readout}

\newcommand{\colt}{Department of Physics and Institute for Quantum Information and Matter, California Institute of Technology, Pasadena, California 91125, USA}
\newcommand{\colta}{AWS Center for Quantum Computing, Pasadena, California 91125, USA}

\author{Nina del Ser}\email{ninadelser@gmail.com}
\affiliation{\colt}
\author{Yinan Chen}
\affiliation{\colt}
\author{Jacob Steiner}
\affiliation{\colt}
\author{Gil Refael}
\affiliation{\colt}
\affiliation{\colta}

\begin{abstract}
Fast, precise, and quantum-non-demolition (QND) readout of superconducting qubits is a fundamental component of high-fidelity quantum sensing and computation. Conventional approaches typically operate in the dispersive regime, where the qubit-resonator coupling $g$ is weak compared to the detuning $\Delta$. While exhibiting good QND properties, the readout rate is limited to $\sim g^2\sqrt{N}/\Delta\ll g$, where $N$ is the number of photons in the resonator. QND readout in the nondispersive regime, where the readout rate reaches its full potential $\sim g$, relies on parameter sweeps that may encounter resonances, leading to measurement-induced state transitions (MIST). In this work, we study a nondispersive readout protocol that replaces these sweeps by sudden quenches of the coupling constant, using a resonator that is preloaded with photons. We call this protocol bang-bang readout, and show that it realizes single-shot projective measurements. The fidelity and QNDness of the qubit post-measurement are remarkably high, with an error that decreases like $1/N$. We show that the protocol can also be implemented without preloading the resonator by instead strongly driving the qubit, e.g., with a classical flux drive. 
\end{abstract}

\maketitle

\section{Introduction}

Quantum measurement plays a central role in the development of quantum technologies. In particular, the task of reading out a qubit state in a single shot --- quickly, precisely, and non-destructively --- remains a significant bottleneck in the practical realization of quantum devices \cite{jakob2025,mude2025}. Qubit readout schemes rely on entangling the qubit with a classically distinguishable pointer apparatus. This approach dates back to Stern-Gerlach \cite{gerlach1989}, and underlies, e.g., the readout of quantum dots \cite{petersson2010,gachter2022,elzerman2004}, nitrogen-vacancy centers in diamond \cite{irber2021,hopper2018}, and trapped ions \cite{myerson2008}.

In superconducting qubits, readout is typically implemented by coupling the qubit to a harmonic readout resonator in the dispersive regime \cite{clerk2010introduction,krantz2019,blais2021circuit}, where the resonator frequency is strongly detuned from the qubit frequency. Here, the qubit-resonator interaction effectively shifts the resonator frequency away from its bare value by the dispersive shift $\pm\chi =\pm g^2/\Delta$, where $g$ and $\Delta$ are the qubit-resonator coupling and detuning, respectively. The resonator frequency will be either red- or blue-shifted, depending on the state of the qubit; this enables the build-up of entanglement between the qubit and resonator subsystems. By measuring either the frequency or the phase of the microwave radiation emitted by the resonator, it is then possible to infer the state of the qubit \cite{wallraff2004strong,blais2004cavity,duty2005}, thereby implementing a projective measurement  \cite{gambetta2006qubit,gambetta2007protocols,gambetta2008quantum}.

\begin{figure}
  \centering
  \includegraphics[width=\linewidth]{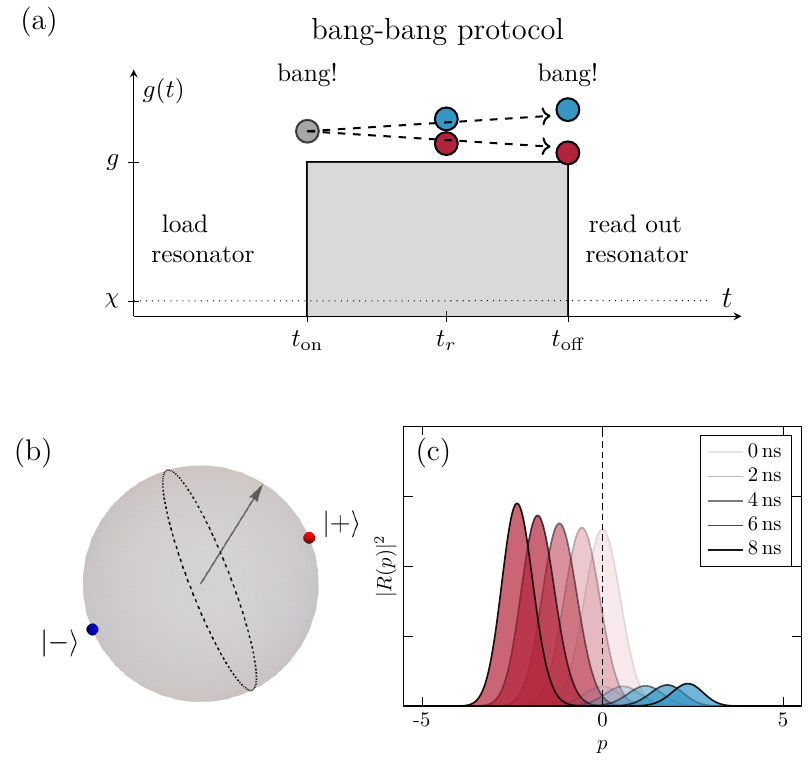}
  \caption{(a) Schematic of the bang-bang nondispersive readout protocol. (b) Bloch sphere representation of the $\ket{\pm}$ sweet-spot basis states of a qubit in the nondispersive JC model, \cref{eq:JCmodel}. $\ket{\pm}$ are related to the $S^z$-eigenstates $\ket{\uparrow,\downarrow}$ by \cref{eq:RabiModelEigenbasis}, with $\thetaR\to\thetaJC,\ \phiR\to\phiJC$. (c) Probability distribution of the resonator quantum state projected onto the $p$-coordinate. 
  At the readout time $t_r\approx$\SI{7}{\nano\second}, the distributions corresponding to $\ket{+}$ (red) and $\ket{-}$ (blue) become distinguishable.
  Parameters used in panels (b)--(c): $g/2\pi=$\SI{100}{\mega\hertz}, $\Delta/2\pi=$\SI{23}{\mega\hertz}, and $N=|\alpha_0|^2=9$.}
\label{fig:qubitSweetspotAngles}
\end{figure} 

This so-called dispersive readout \cite{wallraff2004strong,blais2004cavity,boissonneault2009dispersive,jeffrey2014fast,sete2015quantum,walter2017rapid} has a number of favorable properties: First, it is simple to implement, requiring little hardware overhead compared to approaches such as longitudinal readout \cite{didier2015fast,touzard2019,kerman2013,harpt2025} or Josephson bifurcation amplifier-based readout \cite{siddiqi2004rfdriven,siddiqi2006dispersive,lupascu2006highcontrast,boulant2007quantum,mallet2009singleshot,vijay2009invited}. Second, in the dispersive regime, the qubit Hamiltonian effectively commutes with the coupling to the readout resonator, such that, if initialized in a computational state, the qubit remains in this state post-measurement. In other words, repeated measurements do not alter the outcome. A readout satisfying this property is commonly referred to as quantum-non-demolition, or QND \cite{braginsky1980quantum,caves1980measurement}. QND readout is desirable as the post-measurement qubit state is known and can therefore be reused or actively reset without relying on uncontrolled dissipative relaxation. This is particularly important in mid-circuit measurements, where the measured system remains part of the subsequent quantum computation. Examples include repeated syndrome extraction and measurement-based feedback, used in quantum error correction \cite{kelly2015state,andersen2020repeated,rudinger2022characterizing}.

The advantages of dispersive readout come at the cost of reducing the effective coupling, and thereby the rate at which pointer states separate, from the bare value $g$ to $\abs{\chi}\sqrt{N} \ll g$. 
Modern high-power readout therefore increasingly approaches, and can in fact exceed, the critical photon number $N \gtrsim N_c = \Delta^2/4g^2$, beyond which the perturbative dispersive description breaks down and the full dressed qubit-resonator dynamics becomes important \cite{spring2024fast,swiadek2024enhancing,xiong2025}. In the strongly nondispersive regime, the pointer-state separation rate approaches the bare coupling scale $g$ \cite{sete2013catchdisperserelease}. While one might naïvely expect such readout to no longer be QND---as the qubit Hamiltonian does not commute with the qubit-resonator coupling \cite{boissonneault2009dispersive}---QNDness can largely be maintained if the resonator is adiabatically loaded and unloaded \cite{sete2013catchdisperserelease,govia2015unitaryfeedbackimproved,khezri2016measuring}. Alternatively, the resonator is preloaded and the qubit is subsequently brought close to resonance adiabatically \cite{sete2013catchdisperserelease}. These protocols effectively map the unentangled qubit-resonator basis in the large detuning limit to coherent-state-like superpositions within a single dressed eigenladder (commonly referred to as dressed coherent states), and back. In the dressed coherent state, the pointers separate at rate $g$, enabling fast readout. If the loading and unloading do not induce transitions between the eigenladders, the measurement remains QND.

At high powers, however, loading and unloading the resonator sweeps the coupled system through a range of photon occupations. In realistic multilevel qubits, this trajectory can encounter multiphoton resonances between dressed eigenladders, producing measurement-induced state transitions (MIST) \cite{sank2016measurementinduced,khezri2023measurementinduced,dumas2024measurementinduced,nesterov2024measurementinduced,connolly2025full}. Similarly, the loading sweep results in resonant coupling to parasitic two-level systems \cite{dai2026characterization}. While several strategies exist to mitigate these problems \cite{kurilovich2025highfrequency,li2026mitigating,mori2026suppression}, state transitions during measurement remain a major limiting factor of superconducting qubit technology \cite{fechant2025offset,hazra2025benchmarking}. This is the motivation for considering protocols that avoid sweeping the photon occupation of the coupled system altogether.  
 
One strategy to suppress MIST pioneered in recent experiments is to traverse such crossings diabatically 
\cite{wang2025probing,hoyau2026measurementinduced,lin2026mitigation}. Motivated by these developments, we consider a complementary protocol in which the resonator is populated while it is still decoupled from the qubit, after which the qubit-resonator interaction is suddenly switched on. We ask a more basic question: can such a maximally nonadiabatic coupling protocol nevertheless realize a fast and QND measurement? 
Specifically, we consider a protocol where the qubit and resonator are initially decoupled, while the resonator is preloaded to a large coherent state $\ket{\alpha_0}$, with $N = \abs{\alpha_0}^2$. The coupling is then suddenly switched on to a value $g \gg \Delta/(2\sqrt{N})$, allowing the pointer states to separate. They become distinguishable at the readout time $t_r = 4g^{-1}$. Then, at time $t_\textrm{off} \gtrsim t_r = 4g^{-1}$, the coupling is suddenly switched off, and the resonator state is subsequently measured projectively. We call this protocol ``bang-bang readout"; see also \cref{fig:qubitSweetspotAngles}(a) for a schematic of the procedure. At first sight, such a maximally nonadiabatic protocol might itself be expected to strongly disturb the qubit. Building on the large-field Jaynes-Cummings (JC) dynamics of Gea-Banacloche \cite{gea1991,gea1992,gea1992losspurity}, we show analytically that the opposite is true within the JC model: at large photon numbers, the quench-induced non-QNDness vanishes as $1/N$, while pointer separation occurs at the bare coupling rate $g$. 

The key insight is that, when initialized correctly in one of two basis vectors---denoted by $\ket{\pm}$ and referred to below as the sweet-spot basis vectors---the qubit state evolves during the readout but remains in a known and approximately pure state \cite{gea1991}. $\ket{\pm}$ are in general \emph{not} the same as $\ket{\uparrow,\downarrow}$, see \cref{fig:qubitSweetspotAngles}(b). This means that the computational basis of the qubit needs to be mapped to $\ket{\pm}$ prior to the readout experiment. In the subsequent dynamics, the quasi-pure states are generally time-dependent, $\ket{\pm(t)}$. For a general initial qubit state $\ket{\psi_0}=c_+ \ket{+} + c_- \ket{-}$, the full qubit-resonator system then evolves as $\ket{\Psi(t)}=c_+ \ket{+(t)} \ket{\alpha_+(t)} + c_- \ket{-(t)} \ket{\alpha_-(t)} + \order{N^{-1/2}}$. Once the resonator states become distinguishable, $\braket{\alpha_+(t)}{\alpha_-(t)} \simeq 0$, the qubit-resonator coupling is turned off, and the resonator state is measured, effectively projecting the qubit onto the time-dependent measurement basis $\ket{\pm(t)}$; see \cref{fig:qubitSweetspotAngles}(c). After the conclusion of the readout experiment, the basis vectors $\ket{\pm(t)}$ are mapped back to the computational basis vectors \cite{govia2015unitaryfeedbackimproved}. The states $\ket{\pm(t)}$ are in general not orthogonal, so this correction step is conditioned on the measurement outcome. 
We note that Ref.~\cite{gard2024fast} also proposed readout based on pulsed transverse qubit-resonator coupling. There, QNDness is obtained by optimizing the interaction pulse, such that the qubit undergoes an integer number of exchange-like oscillations and returns to its initial computational state. In contrast, our protocol does not rely on such a fine-tuned recurrence: approximate QNDness emerges parametrically at large $N$, provided the computational basis is mapped to the sweet-spot basis prior to readout. 

To arrive at these findings, we develop an analytical theory of the JC model in the regime of large photon numbers $N\gtrsim 1$. Our approach is based on a saddle-point approximation, and provides a systematic asymptotic expansion of any desired observable in powers of $1/\sqrt{N}$. Here, we retain only terms up to and including $\mathcal{O}(1/N^2)$, which is sufficient to obtain the leading-order errors in the quantities of interest for readout. This goes beyond the works of Gea-Banacloche and others, who investigated the leading-order behavior \cite{gea1991,gea1992losspurity,gea1992,GeaBanacloche1992,phoenix1991,Auffeves2003}. In particular, we use this approach to analytically find the purity of the reduced qubit density matrix, which identifies the sweet-spot basis, as well as to fully characterize our measurement protocol in terms of the post-measurement density matrices and the resulting fidelity and QNDness. This approach makes no assumptions about the relative sizes of $g,\Delta$ and $N$, allowing a smooth transition between the dispersive and nondispersive regimes and revealing several interesting results, such as a finite-$N$ asymmetry in the post-measurement purity, fidelity and QNDness between the two sweet-spot qubit states $\ket{\pm}$.

The rest of the paper is organized as follows. Section~\ref{sec:model} introduces the JC model. In Sec.~\ref{sec:JCmodelDynamics}, 
we develop an analytical framework to calculate the dynamics of all observables relevant for readout in the JC model. It uses a saddle-point method combined with self-consistent perturbation theory in $1/\sqrt{N}$ that may be of interest to some readers as a technical novelty. We use this approach to show that initializing the qubit in the sweet-spot basis leads to a plateau of the qubit purity at readout time to within $\mathcal{O}(1/N)$ of the ideal value unity. In Sec.~\ref{sec:SecondDrive}, we show that adding a second, classical drive on the qubit can further improve qubit purity at $\mathcal{O}(1/N)$. We note that improvements of the readout rate in the dispersive regime due to classical driving were previously shown in Refs.~\cite{touzard2019,ikonen2019}. Section \ref{sec:Measurements} contains the main results regarding the performance of our measurement protocol. To establish these, we use projective measurements on the resonator directly after the qubit-resonator coupling is turned off as a proxy for more realistic homodyne or heterodyne measurement schemes. This allows us to calculate the fidelity and QNDness of the proposed nondispersive bang-bang readout protocol. We find that, even deep in the nondispersive regime, these quantities reach the ideal value 1 within an error $\mathcal{O}(1/N)$ at readout time. Finally, we conclude in Sec.~\ref{sec:conclusion}. 
We postpone a detailed study of our protocol in a realistic system, including a multi-level qubit device, counter-rotating terms, and finite damping, to a future work \cite{nonDispersiveReadoutFluxonium2026}.

\section{Jaynes-Cummings model}
\label{sec:model}

A minimal description of a superconducting qubit readout experiment is provided by the Jaynes-Cummings (JC) model,  
\begin{equation}\label{eq:JCmodel}
   \hat{H}_{\text{JC}}=-\frac{\Omega}{2}\hat{S}^z+\omega\hat{a}^{\dagger}\hat{a}+g(\hat{a}\hat{S}^-+\hat{a}^{\dagger}\hat{S}^+),
\end{equation}
where the spin operators $\hat{S}^z$, $\hat{S}^{\pm} = (\hat{S}^{\mp})^\dagger$ describe the qubit, and $\hat{a}^\dagger,\hat{a}$ are bosonic creation and annihilation operators for the resonator photons. $\Omega$, $\omega > 0$ are the energies of the qubit and resonator, respectively, and $g$ is the qubit-resonator coupling constant, in units where $\hbar=1$. The qubit eigenstates are $\ket{\sigma}$, with $\hat{S}^z \ket{\sigma} = \sigma \ket{\sigma}$, with $\sigma = \uparrow,\downarrow$. 
Note that in our convention, $\hat{S}^{-}$ increases the qubit energy, while $\hat{S}^{+}$ decreases it. The JC model results from truncating physical artificial atom-resonator models to the two lowest energy levels forming the qubit and applying the rotating-wave approximation. The rotating-wave approximation is justified if $\Omega + \omega \gg g\sqrt{N}$. Alternatively, one can directly engineer the JC Hamiltonian \cite{FeiYan2018,rower2024}. The truncation to the lowest two qubit levels is less controlled, as it disregards the possibility of measurement-induced leakage to noncomputational states. Large anharmonicities as in, e.g., Fluxonium qubits \cite{manucharyan2009,nguyen2019,nguyen2022,bao2022} may alleviate this issue. 

For completeness and as a basis for comparison, we also introduce the effective Hamiltonian in the dispersive regime. At large detuning $\Delta=\omega-\Omega \gg 2g\sqrt{N}$, \cref{eq:JCmodel} may be approximated as the diagonal Hamiltonian
\begin{equation}\label{eq:dispersiveHamiltonian}
    \hat{H}_{\text{disp.}}= -\frac{1}{2}\left(\Omega-\chi\right)\hat{S}^z+\hat{a}^{\dagger}\hat{a}\left(\omega+\chi\hat{S}^z\right).
\end{equation}
In contrast to those of the JC model, the eigenstates of \cref{eq:dispersiveHamiltonian} are simple tensor products of Fock states and eigenstates of the $\hat{S}^z$ operator, $\ket{n,\sigma}=\ket{n}\ket{\sigma}$, with $n\in\mathbb{N}$ and $\sigma=\uparrow,\downarrow$. 

\section{Dynamics in the Jaynes-Cummings model}
\label{sec:JCmodelDynamics}

To understand qubit readout in the nondispersive regime, we need to solve for the dynamics of the qubit and resonator in the general Jaynes-Cummings model, \cref{eq:JCmodel}, going \emph{beyond} the dispersive approximation, \cref{eq:dispersiveHamiltonian}. Let us consider the simplest setup to treat mathematically which still yields all the salient features. We separately initialize the qubit in a pure state $\ket{\psi_0}$, and the resonator in a coherent state $\ket{\alpha_0}=\hat{D}(\alpha_0)\ket{0}$, where $\alpha_0$ is a coherent amplitude and $\hat{D}(\alpha)=e^{\alpha \hat{a}^\dagger-\alpha^*\hat{a}}$ is the bosonic displacement operator. The coherent cavity state naturally arises, for instance, when we coherently drive an empty resonator $\ket{0}$ with $\hat{H}_d(t)=\omega\alpha_0 e^{-i\omega t}\hat{a}^{\dagger}+\text{h.c.}$. The full qubit-resonator system in our experiment is thus initially described by the wavefunction
\begin{equation}\label{eq:qubitCavityInitialState}
    \ket{\Psi_0}=\ket{\psi_0}\ket{\alpha_0}.
\end{equation}
At time $t=t_{\text{on}}=0$, we suddenly switch on the qubit-resonator coupling in \cref{eq:JCmodel} from $0$ to $g$, see \cref{fig:qubitSweetspotAngles}(a), and subsequently allow the system to freely evolve under Hamiltonian dynamics. The goal is to calculate $\ket{\Psi(t)}$ for $t>0$.

For the above setup, it is helpful to simultaneously consider the closely related (classical) Rabi model \footnote{for brevity, we will henceforth omit the ``classical'' label, although it is understood to be implied.}, where the JC resonator is replaced by a classical drive with the same driving frequency $\omega$,
\begin{equation}\label{eq:RabiModel}
    \hat{H}_{\text{R}}=-\frac{\Omega}{2}\hat{S}^z+ \Omega_d e^{-i\omega t}\hat{S}^- + \Omega_d^* e^{i\omega t}\hat{S}^+.
\end{equation}
Applying $\hat{H}_{\text{JC}}$ from \cref{eq:JCmodel} to $\ket{\Psi_0}$ in \cref{eq:qubitCavityInitialState}, we see that, at $t=0$, the Rabi and JC models are equivalent if we replace $\Omega_d\leftrightarrow g\alpha_0$. At later times $t>0$, the dynamics will necessarily diverge, because the Rabi model ignores any back action from the qubit on the photonic drive. As a result, the qubit has nothing to become entangled with, and maintains whatever initial purity it started out with at all times under unitary time evolution, $\frac{d}{dt}\Tr[\hat{\rho}^2_{\text{q}}]=0$. By contrast, in the JC model, the qubit and resonator are able to exchange energy and information, so that the purities of the reduced density matrices of both individual subsystems may fluctuate as a function of time, $\frac{d}{dt}\Tr[\hat{\rho}_{\text{q}}^2],\frac{d}{dt}\Tr[\hat{\rho}_{\text{r}}^2]\neq 0$. Here, $\hat{\rho}_{\text{q}}=\Tr_{\text{r}}[\hat{\rho}]$ and $\hat{\rho}_{\text{r}}=\Tr_{\text{q}}[\hat{\rho}]$ are obtained by tracing out the resonator and qubit degrees of freedom, respectively, from the full-system density matrix $\hat{\rho}$. Nevertheless, the global purity remains constant, $\frac{d}{dt}\Tr[\hat{\rho}^2]=0$, as the system is isolated from the environment. The more physically realistic situation where the system interacts weakly with the environment can be modeled by coupling the qubit and resonator to a Lindblad bath \cite{manzano2020,breuer2002}. In this case, at zero temperature and after a long enough time, the purities of the qubit and resonator trivially tend to 1, as they both lose energy and inevitably end up in their respective ground states. The non-trivial open early- and intermediate-time dynamics will be discussed in a future work \cite{nonDispersiveReadoutFluxonium2026}.  

For both the JC and the Rabi models, calculations are greatly simplified in a frame rotating at angular frequency $\omega$ anticlockwise around the $\vb{e}^z$-axis. Inserting the unitary operators
\begin{equation}\label{eq:unitaryOpsRotFrame}
    \hat{U}_{\text{JC}}(t)=e^{i\omega\left(\hat{a}^{\dagger}\hat{a}-\frac{1}{2}\hat{S}^z\right)t},\quad \hat{U}_{\text{R}}(t)=e^{-\frac{i}{2}\omega\hat{S}^z t},
\end{equation}
into the definitions $\hat{\tilde{H}}_{\text{JC}}=\hat{U}_{\text{JC}}\hat{H}_{\text{JC}}\hat{U}^{\dagger}_{\text{JC}}-i\hat{U}_{\text{JC}}\dot{\hat{U}}^{\dagger}_{\text{JC}}$ and $\hat{\tilde{H}}_{\text{R}}=\hat{U}_{\text{R}}\hat{H}_{\text{R}}\hat{U}^{\dagger}_{\text{R}}-i\hat{U}_{\text{R}}\dot{\hat{U}}^{\dagger}_{\text{R}}$, we find
\begin{equation}\label{eq:HamiltoniansRabiJCrotFrame}
\begin{aligned}
    \hat{\tilde{H}}_{\text{JC}}&=\frac{\Delta}{2}\hat{S}^z+g(\hat{a}\hat{S}^-+\hat{a}^{\dagger}\hat{S}^+),\\
    \hat{\tilde{H}}_{\text{R}}&=\frac{\Delta}{2}\hat{S}^z+\Omega_d\hat{S}^-+\Omega_d^*\hat{S}^+,
\end{aligned}
\end{equation}
where $\Delta=\omega-\Omega$ is the qubit-resonator and qubit-classical drive detuning, respectively. We will now solve for the qubit and resonator dynamics predicted by these two models. We start by reminding ourselves of the trivial dynamics in the Rabi model. We then use the intuition coming from the Rabi model to help us understand the more complicated dynamics in the JC model.

\emph{Rabi model in the rotating frame ---} As $\hat{\tilde{H}}_{\text{R}}$ is now just a time-independent $2\times 2$ matrix, it is straightforward to diagonalize, with eigenvectors $\ket{\pm}$ and eigenvalues $\lambda_{\pm}$ given by
\begin{equation}\label{eq:RabiModelEigenbasis}
    \begin{aligned} \ket{+}&=\cos\left(\frac{\thetaR}{2}\right)\ket{\uparrow}+\sin\left(\frac{\thetaR}{2}\right)e^{i\phiR}\ket{\downarrow}, & \lambda_+&=+\frac{\OmegaR}{2},\\
    \ket{-}&=\sin\left(\frac{\thetaR}{2}\right)\ket{\uparrow}-\cos\left(\frac{\thetaR}{2}\right)e^{i\phiR}\ket{\downarrow}, & \lambda_-&=-\frac{\OmegaR}{2},
    \end{aligned}
\end{equation}
where $\thetaR=\arctan\left(2|\Omega_d|/\Delta\right)$, $\phiR=\arg(\Omega_d)$, and $\OmegaR=\sqrt{\Delta^2+4|\Omega_d|^2}$. For a geometric interpretation, it is useful to define the Rabi drive vector,
\begin{equation}\label{eq:staticRabiVec}
    \OmegaRvec=\begin{pmatrix}
        2\Re[\Omega_d], & 2\Im[\Omega_d], & \Delta
    \end{pmatrix}^T,
\end{equation}
which can alternatively be written as $\OmegaRvec=\OmegaR\begin{pmatrix}\sin(\thetaR)\cos(\phiR),\sin(\thetaR)\sin(\phiR),\cos(\thetaR)\end{pmatrix}^T$. Thus, the length of the Rabi drive vector is just the Rabi frequency, $\OmegaR$, while $\thetaR$ and $\phiR$ describe its polar and azimuthal angles, respectively.  
The qubit wavefunction evolves in time according to
\begin{equation}\label{eq:qubitTimeEvolutionRabiModel}
    \ket{\psi(t)}=c_+e^{-i\lambda_+ t}\ket{+}+c_-e^{-i\lambda_- t}\ket{-},
\end{equation}
where $c_{\pm}$ are complex coefficients that describe the initial state of the qubit, with $|c_+|^2+|c_-|^2=1$ to ensure that the wavefunction is normalized. A useful parametrization of $c_{\pm}$ satisfying this normalization condition is
\begin{equation}\label{eq:cPlusMinusParametrization}
    c_{\pm}=e^{i\Phi_{\pm}}\sqrt{\frac{1}{2}(1\pm r)}, \quad \Delta\Phi=\Phi_+-\Phi_-,
\end{equation}
where $r,\Phi_{\pm}\in\mathbb{R}$. Labeling $\ket{\pm}$ as the North/South Poles in this tilted spherical coordinate system gives a natural geometric interpretation to $r$ and $\Delta\Phi$. Varying $r$ in the range $-1\leq r\leq 1$ displaces the initial qubit state longitudinally along a meridian between the South and North Poles, whereas varying $\Delta\Phi$ in the range $0\leq \Delta\Phi \leq 2\pi$ rotates it around an azimuth; see also the hemispheres on the left side of \cref{fig:stereographicProjectionQNDnessVaryN} for an illustration.
Using \cref{eq:qubitTimeEvolutionRabiModel,eq:staticRabiVec,eq:cPlusMinusParametrization}, it is straightforward to show that the qubit Bloch vector $\langle\hat{S}^i(t)\rangle=\bra{\psi(t)}\hat{S}^i\ket{\psi(t)}$ evolves in time as
\begin{figure}
  \centering
  \includegraphics[width=\linewidth]{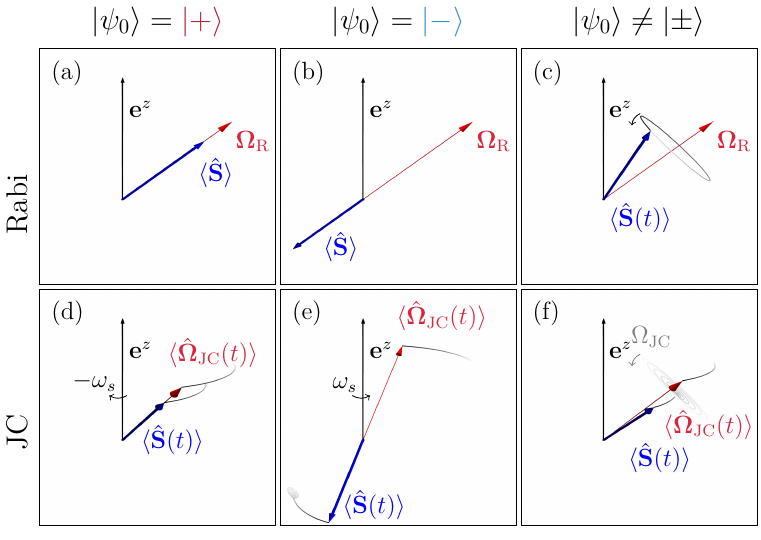}
  \caption{Time traces for $0\leq t\leq 20\, T_{\text{R/JC}}$, with $T_{\text{R/JC}}=2\pi/\Omega_{\text{R/JC}}$, of the resonator static/dynamic drive vector (red) and qubit Bloch vector (blue) in the Rabi (panels (a)--(c)) and JC (panels (d)--(f)) models, in the rotating frame described by the unitary operators $\hat{U}_{\text{JC}}(t)$, $\hat{U}_{\text{R}}(t)$ defined in \cref{eq:unitaryOpsRotFrame}. The qubit is initialized in the $\ket{\pm}$ eigenstates, i.e. parallel or antiparallel to the Rabi vector, in columns 1--2, and in an (unequal) superposition state of $\ket{\pm}$ in column 3. In the Rabi model, $\langle\hat{\vb{S}}\rangle$ always precesses anticlockwise at the Rabi frequency $\OmegaR$ around $\vb{\Omega}_{\text{R}}$, which is fixed [\cref{eq:staticRabiVec,eq:RabiModelBlochVectorTimeEvolution}]. In the JC model, $\hat{\vb{\Omega}}_{\text{R}}$ is promoted to an operator and becomes dynamical [\cref{eq:dynamicalRabiVec}]. Rabi oscillations at frequency $\OmegaJC$ dominate the early-time dynamics of the qubit in the JC model, but quickly decay at rate $\gamma_f$ (see \cref{eq:Timescales} for definitions of $\OmegaJC,\omega_s,\gamma_f$ and $\gamma_s$). For $\gamma_f^{-1}\lesssim t\lesssim \gamma_s^{-1}$, the dominant dynamics in panels (d) and (e) consist of a slow precession of $\langle\hat{\vb{\Omega}}_{\text{JC}}\rangle_\pm$ and $\langle\hat{\vb{S}}\rangle_\pm$ at frequency $\mp\omega_s$ around $\vb{e}^z$ [\cref{eq:semiClassicalApproximationLeadingOrderSolution}]. Meanwhile, in panel (f), $\langle\hat{\vb{\Omega}}_{\text{JC}}\rangle$ and $\langle\hat{\vb{S}}\rangle$ follow elliptical, rather than circular, paths, which result from a superposition of $\langle\hat{\vb{S}}\rangle_\pm$, $\langle\hat{\vb{\Omega}}_{\text{JC}}\rangle_\pm$ [\cref{eq:superpositionSemiClassicalSol}]; see also \cref{fig:JCvsRabiPurityXYplaneTrace}(a)--(b). See App.~\ref{app:videos} in the Supplementary Material for animations of panels (c)--(f). Parameters used for \cref{fig:JCvsRabi,fig:JCvsRabiPurityXYplaneTrace}: $g/2\pi=$\SI{100}{\mega\hertz}, $\Delta/2\pi=$\SI{23}{\mega\hertz}, $\alpha_0=5$.}
\label{fig:JCvsRabi}
\end{figure}
\begin{figure}
  \centering
  \includegraphics[width=\linewidth]{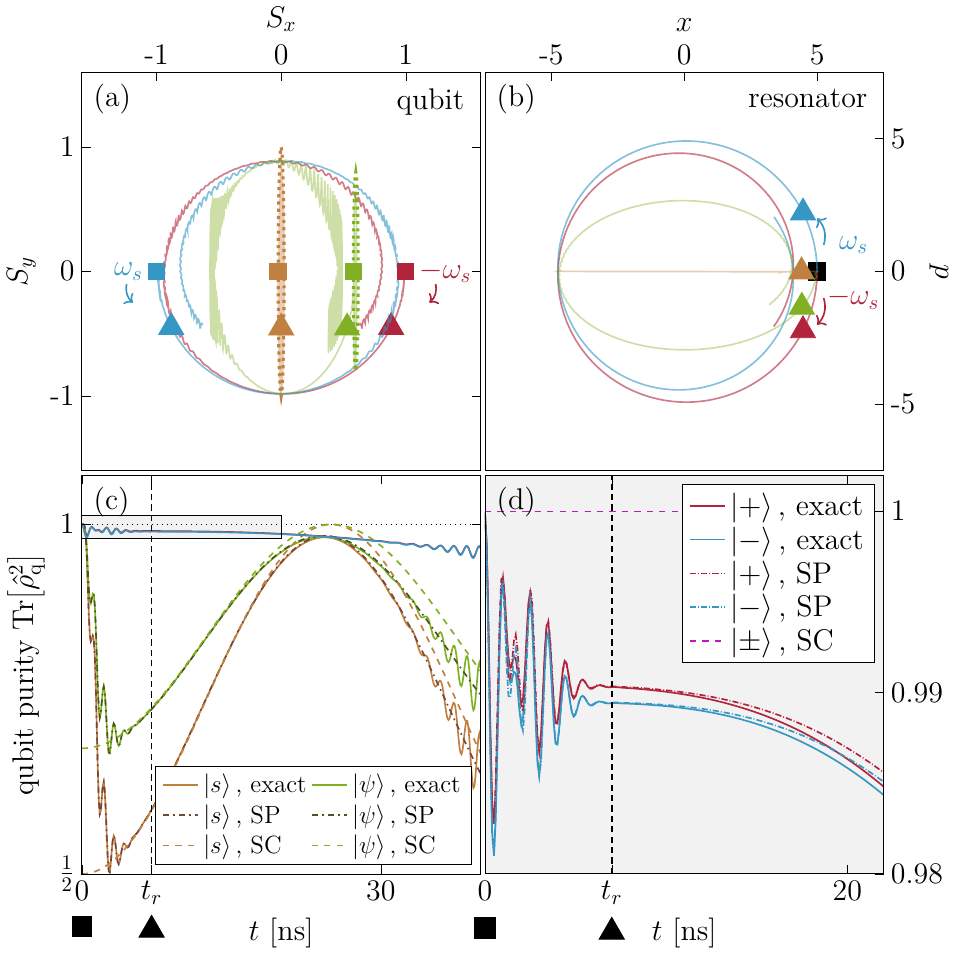}
  \caption{Panels (a)--(b): paths of $\langle\hat{\vb{S}}\rangle$ and $\langle\vb{\hat{\Omega}}_{\text{JC}}\rangle$ projected onto the $S^xS^y$- and $xp$-planes, respectively, for initial qubit states $\ket{\pm}$ (red/blue), $\ket{s}=\frac{1}{\sqrt{2}}(\ket{+}+\ket{-})$ (brown) and $\ket{\psi}=0.89\ket{+}+0.45\ket{-}$ (green), in the rotating frame described by \cref{eq:unitaryOpsRotFrame}. Squares indicate $t=0$, triangles $t_r=2\sqrt{2}\gamma_f^{-1}\approx$\SI{7}{\nano\second}, see also $t$-axis in panels (c)--(d). In the Rabi model, the qubit precesses on a cone, which forms an ellipse upon projection onto the $S^xS^y$-plane (thick dotted brown and green curves for $\ket{s},\ket{\psi}$ in panel (a)). If the qubit is initialized in $\ket{\pm}$, the cone angle shrinks to zero, and the qubit remains stationary (red and blue squares). The resonator is not a dynamical variable and always remains fixed at its initial location (black square in panel (b)). In the JC model, the qubit initially follows Rabi model-like dynamics, but these quickly decay by the time $t\sim t_r$. For $t\gtrsim t_r$, the qubit and resonator both follow elliptical paths for general states $\ket{\psi}$ (thin smooth green lines in panels (a)--(b)). For equal superposition states, such as $\ket{s}$, the ellipses collapse to lines (thin smooth brown lines in panels (a)--(b)). If the qubit is initialized in $\ket{\pm}$, the paths are nearly circular, with opposite directions of rotation $\mp\omega_s$ (thin smooth red and blue lines in panels (a)--(b)). Panel (c): qubit purity, $\Tr[\hat{\rho}_{\text{q}}^2]$, as a function of time. The purity of $\ket{s},\ket{\psi}$ initially drops sharply as the resonator decoheres the qubit. By contrast, the purity for $\ket{\pm}$ only drops proportionally to $1/N$. For times $t_r\lesssim t\lesssim \gamma_s^{-1}$, the purity oscillates at $2\omega_s$, the relative angular frequency between $\langle\hat{\vb{S}}(t)\rangle_\pm$. Panel (d): zoomed in gray rectangular area from panel (c), showing the difference in purity between $\ket{\pm}$ and the comparison between the semiclassical (SC) [\cref{eq:semiclassicalQubitPurity}] and saddlepoint (SP) (\cref{eq:SzSaddlePointGeneral,eq:SPlusSaddlePointGeneral}) approaches.
  }
\label{fig:JCvsRabiPurityXYplaneTrace}
\end{figure}
\begin{equation}\label{eq:RabiModelBlochVectorTimeEvolution}
\begin{aligned}
    &\langle\hat{\vb{S}}(t)\rangle=\mathcal{R}[\thetaR,\phiR].\vb{e}_{\text{c}},\,
    \vb{e}_{\text{c}}=\begin{pmatrix}\sqrt{1-r^2}c(\OmegaR t-\Delta\Phi)\\ \sqrt{1-r^2}s(\OmegaR t-\Delta\Phi)\\ r\end{pmatrix},
\end{aligned}
\end{equation}
where
\begin{equation}\label{eq:rotationalMatrixRabiModel}
\mathcal{R}[\theta,\phi]=\begin{pmatrix}
        -c(\theta)c(\phi) & s(\phi) & s(\theta)c(\phi)\\
         -c(\theta)s(\phi)& -c(\phi) & s(\theta)s(\phi) \\
         s(\theta) &  0 & c(\theta)
    \end{pmatrix},
\end{equation}
and we used the abbreviations $c(x)=\cos(x)$, $s(x)=\sin(x)$. $\vb{e}_\text{c}$ precesses in the anticlockwise direction around the $z$-axis (in the local frame) at the Rabi frequency $\OmegaR$, with conical angle $\arctan(\sqrt{1-r^2}/r)$. $\mathcal{R}[\thetaR,\phiR]$ rotates $\vb{e}_\text{c}$ to make the axis of precession parallel to the Rabi vector, $\mathcal{R}[\thetaR,\phiR].\vb{e}^z=\OmegaRvec/\OmegaR$. In the special limits $r=\pm 1$, when the qubit is initialized in $\ket{\pm}$, the conical angle is zero, and $\langle\hat{\vb{S}}\rangle$ is static and (anti)parallel to $\OmegaRvec$, see \cref{fig:JCvsRabi}(a)--(b). The general case of Rabi oscillations for a qubit \emph{not} initialized in one of the two eigenstates is shown in \cref{fig:JCvsRabi}(c). In this case, the Bloch vector precesses at the Rabi frequency $\Omega_R$ anticlockwise around $\OmegaRvec$, keeping the initial conical angle $\arctan(\sqrt{1-r^2}/r)$ constant throughout.

\emph{JC model in the rotating frame ---}
In the JC model, the resonator is promoted to a quantum harmonic oscillator, with Hilbert space spanned by the Fock eigenstates $\ket{n}$, where $n\in\mathbb{N}$. The Hilbert space of the full qubit-resonator system is thus a tensor product of the Hilbert spaces of the two subsystems, $\mathcal{H}=\mathcal{H}_{\text{qubit}}\otimes\mathcal{H}_{\text{res.}}$. In contrast to the Rabi model, the Hilbert space for the JC model is infinitely large, as $n$ does not have an upper bound. This property makes it difficult to obtain concise analytical expressions for general initial states. However, the coherent envelope of our initial state $\ket{\alpha_0}$, which, for large $N=|\alpha_0|^2$, becomes almost Gaussian-distributed in Fock space, greatly simplifies this task, as it restricts the meaningful $n$ to a window $N-\sqrt{N}\lesssim n\lesssim N+\sqrt{N}$. We can use this, together with the saddlepoint approximation and perturbation theory in $1/\sqrt{N}$, to obtain analytical expressions for \emph{any} observable of interest in the JC model. However, before doing this, we first take a short detour and calculate the dynamics of the Bloch and JC drive vectors in the JC model using a much simpler semiclassical approach. This approximate technique accurately captures the leading-order dynamics --- even in the nondispersive regime --- and builds intuition for the full quantum treatment developed later on.

\textit{Semiclassical solution for the JC model}---
By analogy to \cref{eq:staticRabiVec}, we define the JC drive vector, which is now an operator,
\begin{equation}\label{eq:dynamicalRabiVec}
    \OmegaJCvec=\begin{pmatrix}
        g(\hat{a}+\hat{a}^\dagger), & ig(\hat{a}^{\dagger}-\hat{a}), & \Delta
    \end{pmatrix}^T.
\end{equation}
As the resonator is initialized in a coherent state $\ket{\alpha_0}$, $\langle\OmegaJCvec\rangle=\OmegaRvec$ at $t=0$ if we set $\Omega_d=g\alpha_0$.
Substituting $\hat{\tilde{H}}_{\text{JC}}$, defined in \cref{eq:HamiltoniansRabiJCrotFrame}, into the Heisenberg equation of motion, $\langle\dot{\hat{O}}\rangle=i\langle [\hat{\tilde{H}},\hat{O}]\rangle$, we obtain the following coupled equations of motion for $\hat{O}=\OmegaJCvec,\vb{\hat{S}}$, 
\begin{equation}\label{eq:HeisenbergEoMsJCmodel}
    \begin{aligned}
        \langle\dot{\hat{\vb{\Omega}}}_{\text{JC}}\rangle&=-g^2\vb{e}^z\times \langle\hat{\vb{S}}\rangle, & \langle\dot{\hat{\vb{S}}}\rangle&=\langle \OmegaJCvec \times \hat{\vb{S}}\rangle.
    \end{aligned}
\end{equation}
The first equation contains only expectation values of single operators, so all the terms conveniently decouple. By contrast, the second equation is less pleasant, due to the term on the right which consists of an expectation of a product of operators, $\langle \OmegaJCvec \times \hat{\vb{S}}\rangle$. If we momentarily demote $\OmegaJCvec$ back to a classical drive, $\OmegaJCvec\to\OmegaRvec$, this term decouples to $\OmegaRvec \times \langle\hat{\vb{S}}\rangle$, and we immediately recognize the equation of motion for Larmor precession, with $\langle\hat{\vb{S}}\rangle$ precessing around $\OmegaRvec$ at the Rabi frequency $\OmegaR$, see \cref{fig:JCvsRabi}(c). For the JC model, this term also decouples at time $t=0$, because we initialize the qubit-resonator system in a product state, $\ket{\Psi_0}=\ket{\psi_0}\ket{\alpha_0}$. This is no longer true as soon as $t>0$, as the qubit and resonator become entangled. However, if the amount of entanglement is low, i.e., the purity of both subsystems remains high, $\Tr[\hat{\rho}^2_{\text{q}}]$, $\Tr[\hat{\rho}^2_{\text{r}}]\simeq 1$, we may still approximate $\langle \OmegaJCvec \times \vb{\hat{S}}\rangle\approx \langle \OmegaJCvec \rangle \times \langle \vb{\hat{S}}\rangle$. This assumption is known as the semiclassical approximation. Despite not fully capturing all the details, it predicts the leading order behavior remarkably well.

We now brazenly apply the semiclassical approximation to \cref{eq:HeisenbergEoMsJCmodel}, solving the resulting coupled equations perturbatively in powers of $1/\sqrt{N}$. In analogy to the Rabi frequency $\OmegaR=\sqrt{\Delta^2+4|\Omega_d|^2}$, we define the frequency $\OmegaJC=\sqrt{\Delta^2+4g^2N}$. As $\OmegaJC$ turns out to be the fastest frequency in the problem, we will refer to it from now on as the ``fast'' frequency. Assuming that $\Delta\sim\sqrt{N}$, $g\sim 1$, so that $\OmegaJC\sim\sqrt{N}$, when performing the perturbation theory, we obtain the following two pairs of solutions,
\begin{equation}\label{eq:semiClassicalApproximationLeadingOrderSolution}
    \langle\vb{\hat{S}}\rangle_{\pm}=\frac{\pm\langle\OmegaJCvec\rangle_{\pm}}{|\langle\OmegaJCvec\rangle_{\pm}|},\quad 
    \langle\OmegaJCvec\rangle_{\pm}=\begin{pmatrix}
        2g\sqrt{N}\cos(\phiJC\mp\omega_s t) \\
        2g\sqrt{N}\sin(\phiJC\mp\omega_s t) \\
        \Delta
    \end{pmatrix},
\end{equation}
to leading order in $1/\sqrt{N}$, with $\phiJC=\arg(\alpha_0)$, and a new emergent angular frequency,
\begin{equation}
    \omega_s=g^2/\OmegaJC.
\end{equation}
The first equation in \cref{eq:semiClassicalApproximationLeadingOrderSolution} tells us that the $\langle\vb{\hat{S}}\rangle_{\pm}$ and $\langle\OmegaJCvec\rangle_{\pm}$ vectors are aligned parallel or antiparallel to each other, which closely mirrors the eigenvectors of the Rabi model, see \cref{fig:JCvsRabi}(a)--(b). The second equation provides new information: instead of remaining fixed as in the Rabi model, $\langle\hat{\vb{S}}(t)\rangle_{\pm}$ and $\langle\OmegaJCvec(t)\rangle_{\pm}$ are now both dynamical in time, and precess around the $\vb{e}^z$-axis at angular frequency $\mp\omega_s$, see \cref{fig:JCvsRabi}(d)--(e) and \cref{fig:JCvsRabiPurityXYplaneTrace}(a)--(b). We call the new frequency $\omega_s$ the ``slow'' frequency, because it is suppressed by a factor $\sim N$ relative to the fast frequency $\OmegaJC$. The fact that the direction of precession --- clockwise or anticlockwise around $\vb{e}^z$ --- depends on the relative initial orientation between the drive and Bloch vectors is the key property that allows us to perform qubit readout.

Another way to interpret this behavior is to notice that the system has a global rotational Goldstone mode $\Phi$ around the $\vb{e}^z$-axis, i.e., the transformation $e^{-i\Phi\left(\hat{a}^{\dagger}\hat{a}-\frac{1}{2}\hat{S}^z\right)}\hat{\tilde{H}}_{\text{JC}}e^{i\Phi\left(\hat{a}^{\dagger}\hat{a}-\frac{1}{2}\hat{S}^z\right)}$ leaves $\hat{\tilde{H}}_{\text{JC}}$ unchanged. Precession of $\langle\hat{\vb{S}}(t)\rangle_\pm$, $\langle\OmegaJCvec(t)\rangle_\pm$ at $\pm\omega_s$ can therefore be described as an ``activation'' of the system's rotational Goldstone mode. This effect is generic and will occur in any periodically driven systems with rotational or translational Goldstone modes, as long as the drive breaks enough symmetries. The rate of change of the Goldstone-mode coordinate averaged over one driving period, $\langle\dot{\Phi}\rangle_T$, can be proportional either to the drive amplitude or its power (i.e., the amplitude squared). For example, in our case, $|\omega_s|\sim g^2$ off resonance ($\Delta\gg 4g^2N$), but $|\omega_s|\sim g$ on resonance $(\Delta=0)$. This effect has recently also been predicted and observed in various types of classical driven magnets, giving rise to new dynamical states of matter, such as magnetic Archimedean screws \cite{delser2021,shimizu2023,zhang2025}, skyrmion jellyfish \cite{delser2023}, rotating skyrmion lattices \cite{tengdin2022}, or ultrafast active domain walls \cite{hardt2025}. 

Now, an important observation: the perturbation theory we used to derive \cref{eq:semiClassicalApproximationLeadingOrderSolution} forces the right hand side of the second equation in \cref{eq:HeisenbergEoMsJCmodel} to vanish to leading order in $1/\sqrt{N}$. This means that, to leading order in $1/\sqrt{N}$, the system of equations is linear, so that linear combinations of $\langle \hat{\vb{S}}\rangle_{\pm}$, $\langle\OmegaJCvec\rangle_{\pm}$ are also valid solutions,
\begin{equation}\label{eq:superpositionSemiClassicalSol}
\begin{aligned}
    \langle \hat{\vb{S}} \rangle&=\frac{1}{2}(1+r)\langle \hat{\vb{S}} \rangle_+ + \frac{1}{2}(1-r)\langle \hat{\vb{S}} \rangle_-, \\
    \langle \OmegaJCvec \rangle&=\frac{1}{2}(1+r)\langle \OmegaJCvec \rangle_+ + \frac{1}{2}(1-r)\langle \OmegaJCvec \rangle_-,
\end{aligned}
\end{equation}
where $r$ must be a real number to ensure that $\langle \hat{\vb{S}} \rangle$ and $\langle \OmegaJCvec \rangle$ remain real at all times. In addition, conservation  of probability, $|\langle\hat{\vb{S}}\rangle|\leq 1$, together with $\max(|\langle\hat{\vb{S}}\rangle|^2)=\frac{4g^2N+r^2\Delta^2}{\Delta^2+4g^2N}$ (obtained by setting $\omega_s t=(n+\frac{1}{2})\pi$, $n\in\mathbb{Z}$ in \cref{eq:semiClassicalApproximationLeadingOrderSolution}) places lower and upper bounds on $r$: $-1\leq r\leq 1$. We immediately recognize that the $r$ in \cref{eq:superpositionSemiClassicalSol} is none other than the $r$ in \cref{eq:cPlusMinusParametrization}. Thus, the prefactors of $\langle\hat{\vb{S}} \rangle_{\pm}$ in \cref{eq:superpositionSemiClassicalSol} are simply $|c_{\pm}|^2=\frac{1}{2}(1\pm r)$. 

On the other hand, \cref{eq:superpositionSemiClassicalSol} seems to be missing information about the phase difference $\Delta\Phi$. Where did it go? The clue is in the details. Notice how the fast frequency $\OmegaJC$ is also absent from \cref{eq:superpositionSemiClassicalSol}. It turns out that, for $r\neq\pm 1$, the semiclassical solution for the Bloch vector in fact misses some leading order terms that oscillate at $\OmegaJC$ and rapidly decay with decay rate $\gamma_f=\sqrt{2N}\omega_s$ (derived later using the full quantum approach); see also the early-time gray spiral trace in \cref{fig:JCvsRabi}(f). These terms dominate the early dynamics of the JC model, connecting it to the Rabi model, and contain information about $\Delta\Phi$. 
The fast decay rate dictates the time scale on which a qubit initialized in a general superposition state of $\ket{\pm}$ decoheres, and also corresponds to our definition of the readout time (see \cref{sec:Measurements}),
\begin{equation}
    t_r=2\sqrt{2}\gamma_f^{-1}.
\end{equation}

Neglecting the incorrect early-time ($0<t\lesssim t_r$) prediction for $r\neq\pm 1$, \cref{eq:superpositionSemiClassicalSol} to leading order describes the paths followed by $\langle \hat{\vb{S}} \rangle$ and $\langle \OmegaJCvec \rangle$ remarkably well, up to a time $t_{\text{max}}=\pi\omega_s^{-1}-t_r$, when the fast-oscillating and -decaying terms return. For this reason, $t_{\text{max}}$ is also sometimes referred to as the ``revival'' time \cite{berman2014}. For $r\neq \pm 1$, the magnitudes of the in-plane components, $\langle\hat{\vb{S}}^{\perp}\rangle=(\langle\hat{S}^x\rangle,\langle\hat{S}^y\rangle)^T$ and $\langle\OmegaJCvec^{\perp}\rangle=(\langle\hat{\Omega}_{\text{JC}}^x\rangle,\langle\hat{\Omega}_{\text{JC}}^y\rangle)^T$, oscillate periodically at $2\omega_s$, the relative angular frequency between the $\langle \hat{\vb{S}} \rangle_{+}$,$\langle \OmegaJCvec \rangle_{+}$ and $\langle \hat{\vb{S}} \rangle_{-}$,$\langle \OmegaJCvec \rangle_{-}$ base vectors. This results in elliptical paths, which become nearly circular when $r=\pm 1$, see \cref{fig:JCvsRabiPurityXYplaneTrace}(a)--(b). The paths are not quite circular because of subleading $\mathcal{O}(1/N)$ corrections, which will be discussed in the next part. Note that the ellipse traced in the $S^xS^y$-plane by $\langle\hat{\vb{S}}^{\perp}\rangle$ is rotated by \SI{90}{\degree} relative to the one traced in the $xp$-plane by $\langle\OmegaJCvec^{\perp}\rangle$, due to the opposite initial orientations of $\langle\hat{\vb{S}}\rangle_-$ and $\langle\OmegaJCvec\rangle_-$. The principal axes of the ellipses in \cref{fig:JCvsRabiPurityXYplaneTrace}(a)--(b) align with the $S^xS^y$- or $xp$-axes because we chose $\alpha_0\in\mathbb{R}$. If $\alpha_0$ were instead chosen to be complex, the ellipses would be rotated anticlockwise about the origin by $\phiJC=\arg(\alpha_0)$.

As the $\langle\hat{\vb{S}}^z\rangle$ component remains to leading order fixed, oscillations in the in-plane component $\langle\hat{\vb{S}}^\perp\rangle$ directly lead to oscillations in the length of the Bloch vector, $|\langle\hat{\vb{S}}\rangle|^2$, and therefore also the qubit purity, see green and brown curves in \cref{fig:JCvsRabiPurityXYplaneTrace}(c). Substituting \cref{eq:superpositionSemiClassicalSol} into $\Tr[\hat{\rho}^2_{\text{q}}]=\frac{1}{2}(1+|\langle\hat{\vb{S}}\rangle|^2)$, and using \cref{eq:semiClassicalApproximationLeadingOrderSolution} to evaluate $|\langle\hat{\vb{S}}\rangle_\pm|^2=1$ and $\langle\hat{\vb{S}}\rangle_+\cdot\langle\hat{\vb{S}}\rangle_-=-(\Delta^2+4g^2N\cos(2\omega_s t))/\Omega_{\text{JC}}^2$, the semiclassical approach predicts the leading order purity of the qubit to be
\begin{equation}\label{eq:semiclassicalQubitPurity}
\begin{aligned}
    \Tr[\hat{\rho}^2_{\text{q}}]&=\frac{1}{4}\left[3+r^2-\frac{(1-r^2)}{\OmegaJC^2}(\Delta^2+4g^2N\cos(2\omega_s t))\right]\\
    &\quad +\mathcal{O}(N^{-1/2}),
\end{aligned}
\end{equation}
valid for $t_r\lesssim t\lesssim \pi\omega_s^{-1}$.
\cref{eq:semiclassicalQubitPurity} incorrectly predicts purity less than one at $t=0$ for $r\neq \pm 1$, even though we always initialize the qubit in a pure state. It also ignores the revival at $t_{\text{max}}$ of the same fast-decaying terms that were at the origin of the discrepancy in the $0<t\lesssim t_r$ region. However, in the given time interval $t_r\lesssim t\lesssim t_{\text{max}}$, \cref{eq:semiclassicalQubitPurity} is accurate to leading order in $1/\sqrt{N}$ for all possible initial qubit states parametrized by $r$ and $\Delta\Phi$. For an initial qubit state with $|r|<1$, it predicts oscillations at $2\omega_s$, whereas for $r=\pm1$ --- the ``sweet spot'' orientations --- it predicts $\Tr[\hat{\rho}_{\text{q}}^2]=1$. In reality, the purity of the qubit always decreases below one for $t>0$, even for a qubit initialized in $\ket{\pm}$, as some entanglement with the resonator is unavoidable, see \cref{fig:JCvsRabiPurityXYplaneTrace}(d). We will shortly calculate these subleading order corrections using the full quantum approach.

We have now extracted all the useful information available from the semiclassical solution,
\cref{eq:semiClassicalApproximationLeadingOrderSolution,eq:superpositionSemiClassicalSol,eq:semiclassicalQubitPurity}. Although it is correct to leading order in $1/\sqrt{N}$ for $r=\pm 1$, the semiclassical solution misses the leading order fast-decaying terms for $r\neq \pm 1$. Much more problematically, the semiclassical approach fails as soon as we try to evaluate the subleading $1/\sqrt{N}$ and higher order corrections to $\langle\hat{\vb{S}}\rangle_\pm$. These turn out to be crucial for evaluating the errors in the measurement fidelity and QNDness of the readout scheme in Sec.~\ref{sec:Measurements}. For these reasons, we now move on to the more sophisticated full quantum approach. 

\textit{Full quantum solution for the JC model}---
In the full quantum approach, we do not make any assumptions about the supposedly low degree of entanglement between the qubit and resonator when the qubit is initialized in $\ket{\pm}$. Instead, we directly time-evolve the full-system wavefunction $\ket{\Psi(t)}$, and use that to calculate the expectation value $\bra{\Psi(t)}\hat{O}\ket{\Psi(t)}$ of any desired observable $\hat{O}$. We will specifically calculate analytical expressions for $\hat{O}=\hat{S}^+$, $\hat{S}^z$, $\hat{a}$, $\hat{a}^2$ and $\hat{n}$, where $\hat{n}=\hat{a}^{\dagger}\hat{a}$. $\langle\hat{S}^{+,z}\rangle$ and $\langle\hat{a}\rangle$ will inform us of the corrections to the semiclassical expressions for $\langle\hat{\vb{S}}\rangle$ and $\langle\OmegaJCvec\rangle$ in the previous subsection. As a bonus, using $\langle\hat{a}^2\rangle$ and $\langle \hat{n}\rangle$, we will calculate the amount of work done and the degree of squeezing and spreading in the resonator state, both of which are implicitly assumed to be zero in the semiclassical approach. Note that, while the method and results will be discussed in the main text, some technical details of this, at times challenging, calculation are deferred to Apps.~\ref{app:ObservablesJCmodelExactExpressions}--\ref{app:SaddlePointExpressionsAtSweetSpot}.

We begin with the exact expression for $\ket{\Psi(t)}$. $\hat{\tilde{H}}_{\text{JC}}$ conveniently commutes with the excitation number operator $\hat{n}_{\text{e}}=\hat{a}^{\dagger}\hat{a}-\frac{1}{2}\hat{S}^z$, so that $\ket{m}\ket{\uparrow}$ only couples to $\ket{m-1}\ket{\downarrow}$, where $m\geq 1$. The resulting $2\times 2$ block matrix, and hence also its eigenbasis, is identical to that of the Rabi model, \cref{eq:HamiltoniansRabiJCrotFrame,eq:RabiModelEigenbasis}, if we replace $\Omega_d\to g\sqrt{m}$. We denote the resulting eigenvectors and eigenenergies by $\ket{m,\alpha}$ and $\alpha\lambda_{m}$, respectively, where $\alpha=\pm$, see also \cref{eq:JCmodelEigenbasis} for their precise form. The ground state $\ket{0,\uparrow}=\ket{0}\ket{\uparrow}$, with eigenenergy $\Delta/2$, remains unhybridized. $\ket{\Psi(t)}$ takes the general form
\begin{equation}\label{eq:PsiFullSystemTimeEvolved}
\ket{\Psi(t)}=c_{0,\uparrow}e^{-\frac{i\Delta t}{2}}\ket{0,\uparrow}+\sum_{m=1}^{\infty}\sum_{\alpha=\pm}c_{m,\alpha}e^{-i\alpha\lambda_{m}t}\ket{m,\alpha}.
\end{equation}
The complex coefficients $c_{m,\alpha}=\bra{m,\alpha}\ket{\Psi_0}$ are determined by the initial state of the system, in this case \cref{eq:qubitCavityInitialState}; see \cref{eq:JCModelFullSystemWavefunctionCoeffs} for the resulting expressions.

Next, we substitute \cref{eq:PsiFullSystemTimeEvolved} into $\bra{\Psi(t)}\hat{O}\ket{\Psi(t)}$, resulting in \cref{eq:aExpectValExact,eq:rhoQubitExact,eq:nExpectValExact,eq:aSquaredExpectValExact} for our observables of interest $\langle\hat{a}\rangle$, $\langle\hat{S}^{z,\pm}\rangle$, $\langle\hat{n}\rangle$ and $\langle\hat{a}^2\rangle$, respectively. All these expressions share the same structure, consisting of an infinite sum over the Fock index $m$, coming from the resonator, and a sum over the four possible spin configurations $\alpha=\beta=\pm$ and $\alpha=-\beta=\pm$, coming from the qubit. Although they are exact, these expressions are not yet very useful, because the infinite sum over $m$ obscures any simple dependency that the observables might have on $g,\Delta,N$, and the initial states $\ket{\psi_0}$, $\ket{\alpha_0}$ of the resonator and qubit. To make progress, we first note that the summand approximately follows a Poisson distribution, $e^{-N}N^{m}/m!$, with a global maximum at the mean value $N=|\alpha_0|^2$, due to the coherent envelope of the initial state, \cref{eq:qubitCavityInitialState}. Secondly, the error that arises when we approximate the discrete Riemann sums over $m$ in \cref{eq:aExpectValExact,eq:rhoQubitExact,eq:nExpectValExact,eq:aSquaredExpectValExact} as continuous integrals shrinks exponentially with $N$ for $N\gtrsim 1$, see App.~\ref{app:SaddlePoint}, \cref{eq:eulerMaclaurinError}. Together, these two observations allow us to apply the saddlepoint technique to expand each summand in \cref{eq:aExpectValExact,eq:rhoQubitExact,eq:nExpectValExact,eq:aSquaredExpectValExact} around its global maximum and express it in terms of Gaussian integrals, which are simple to compute.

There is an inherent polynomial error $\mathcal{O}(1/N^k)$, $k\in\mathbb{Z}$ associated with the saddlepoint technique, which can be made arbitrarily small by including higher and higher $k$-moments $x^{2k}e^{-x^2}$ in the integrand. For our purposes, going up to order $\mathcal{O}(1/N)$ is in most cases sufficient to obtain the leading order correction that was inaccessible using the semiclassical technique. Some exceptions to this are quadrature spread in the resonator and the time-dependent decay of qubit QNDness (see Sec.~\ref{sec:Measurements}), both of which require going up to $\mathcal{O}(1/N^2)$ to obtain the leading order correction. In addition, the quadratic-in-time decay of the qubit purity at the slow rate $\gamma_s$ in the $t_r\lesssim t\lesssim t_{\text{max}}$ time window (see downward slopes in $t\gtrsim t_r$ region in \cref{fig:JCvsRabiPurityXYplaneTrace}) is also a strictly $\mathcal{O}(1/N^2)$ effect.

Taking these considerations into account, we apply the saddlepoint approximation to convert the complicated sums \cref{eq:aExpectValExact,eq:rhoQubitExact,eq:nExpectValExact,eq:aSquaredExpectValExact} into the much simpler-to-interpret saddlepoint expressions \cref{eq:SzSaddlePointGeneral,eq:SPlusSaddlePointGeneral,eq:annihilationSaddlePointGeneral,eq:annihilationSquaredSaddlePointGeneral,eq:numberOperatorSaddlePointGeneral}. We observe that the $\alpha=-\beta$ and $\alpha=\beta$ spin index pairings result in qualitatively different terms, with the following emergent oscillation frequencies and decay rates,
\begin{equation}\label{eq:Timescales}
\begin{tabular}{| l | l | l |} 
\hline
& frequency & decay rate \\ [0.5ex] 
 \hline & & \\ [-2ex]
$\alpha=-\beta$ & $\OmegaJC=\sqrt{\Delta^2+4g^2N}$ & $\gamma_f=\sqrt{2N}\omega_s$   \\
$\alpha=\beta$ & $\omega_s=g^2/\OmegaJC$ &  $\gamma_s=\gamma_f \omega_s/\OmegaJC$  \\ 
\hline 
\end{tabular}
\end{equation}
listed in order of decreasing speed. The origin of the frequencies is simple enough to understand: they correspond to the leading order contributions of the prefactor of $it$ in $\exp\left[i(\beta\lambda_{N+1}-\alpha\lambda_{N})t\right]$, the complex exponential factor in the summand of $\bra{\Psi(t)}\hat{O}\ket{\Psi(t)}$ evaluated at the saddle point $m=N$. For the decay rates, an additional step is required. The saddlepoint value is in fact $N$ only to leading order, with imaginary subleading and subsubleading corrections that are linear in $t$, namely $N\left[1-i(\alpha-\beta)\omega_s t-i(\alpha+\beta)\frac{\omega_s^2}{\OmegaJC}t\right]$; see \cref{eq:saddlePointFormula}. Taking these next order corrections into account when evaluating $\exp(i(\beta\lambda_{N+1}-\alpha\lambda_{N})t)$ generates the quadratic in $t$ decay terms in the rightmost column of Table~\ref{eq:Timescales}. 

We recognize the $\alpha=\beta$ contributions in Table~\ref{eq:Timescales}, oscillating at the slow frequency $\omega_s$, from the semiclassical approach. The saddlepoint calculation in addition tells us that these contributions decay at a slow rate $\gamma_s\sim g/N$. On the other hand, the $\alpha=-\beta$ contributions, which oscillate at the fast frequency $\OmegaJC$ and decay at rate $\gamma_f$, both a factor $\sim N$ faster than the $\alpha=\beta$ terms, are a new feature. They solve the conundrum of the incorrect early-time dynamics predicted by the semiclassical expression \cref{eq:superpositionSemiClassicalSol} for $|r|\neq 1$ by modifying $\langle\hat{\vb{S}}\rangle$ to
\begin{equation}\label{eq:fullQuantumSleadingOrder}
\begin{aligned}
\langle\hat{\vb{S}}\rangle &= \mathcal{R}[\theta_+,\phi_+].\vb{e}_{\text{c},+} + \mathcal{R}[\theta_-,\phi_-].\vb{e}_{\text{c},-} + \mathcal{O}(N^{-\frac{1}{2}}),\\
\theta_{\pm}&=\frac{\pi}{2}(1\mp 1)\pm\thetaJC,\,
\phi_{\pm}=\frac{\pi}{2}(1\mp 1)+\phiJC\mp\omega_s t,\\
\vb{e}_{\text{c},\pm}&=\begin{pmatrix}
        \mp \frac{\OmegaJC}{2\Delta}\left(1\mp \frac{\Delta}{\OmegaJC}\right)\sqrt{1-r^2}e^{-\gamma_f^2t^2}\cos(\OmegaJC t-\Delta\Phi) \\
        \pm \frac{1}{2}\left(1\mp \frac{\Delta}{\OmegaJC}\right)\sqrt{1-r^2}e^{-\gamma_f^2t^2}\sin(\OmegaJC t-\Delta\Phi)\\
        \frac{1}{2}(1\pm r)
    \end{pmatrix},
\end{aligned}
\end{equation}
where $\thetaJC=\arctan(2g\sqrt{N}/\Delta)$, $\phiJC=\arg(\alpha_0)$, and $\mathcal{R}[\theta,\phi]$ was defined in \cref{eq:rotationalMatrixRabiModel}. The $\vb{e}_{\text{c},\pm}$ vectors in \cref{eq:fullQuantumSleadingOrder} describe two spinning tops that rotate quickly at $\OmegaJC$ around the $z$-axis in the local frame (which is not the same as the lab frame $\vb{e}^z$-axis). The sense of rotation is clockwise if $\Delta>0$ and anticlockwise if $\Delta<0$. The paths followed by $\vb{e}_{\text{c},\pm}$ are in general elliptical due to the unequal prefactors of the $\cos(\dots)$ and $\sin(\dots)$ time-dependent terms in the $\vb{e}_{\text{c},\pm}^{x,y}$ components. The angle between $\vb{e}_{\text{c},\pm}$ and the $z$-axis is finite when $|r|<1$, but quickly decays to zero at the fast decay rate $\gamma_f$. This results in the spiral trace in \cref{fig:JCvsRabi}(f).

We compare and contrast $\vb{e}_{\text{c},\pm}$ to the $\vb{e}_{\text{c}}$ vector in the Rabi solution, \cref{eq:RabiModelBlochVectorTimeEvolution}. There, the sense of rotation was always anticlockwise, irrespective of the sign of $\Delta$, and the conical angle remained constant over time. The latter effect is a result of the absence of entanglement, and therefore decoherence, between the Bloch vector and the classical drive. When the drive becomes quantized as in the JC model, decoherence cannot be avoided. In fact, it is desirable, because it allows us to use single-shot measurements to distinguish between the $\ket{\pm(t)}$ states resulting from a qubit initialized in a superposition state, see \cref{sec:Measurements} for more details.  

The $\mathcal{R}[\theta_{\pm},\phi_{\pm}]$ rotational matrices initially rotate $\vb{e}_{\text{c},\pm}$, so that at $t=0$, they align with $\langle\vb{S}(0)\rangle_\pm=\bra{\pm}\hat{\vb{S}}\ket{\pm}$, where $\ket{\pm}$ are the same sweet-spot orientations we defined in \cref{eq:RabiModelEigenbasis}, with $\thetaR\to\thetaJC$, $\phiR\to\phiJC$. Using $\mathcal{R}[\theta_-,\phi_-(t=0)]=\mathcal{R}[\theta_+,\phi_+(t=0)]\cdot\text{diag}(1,-1,-1)$, one can verify that the updated Bloch vector for the JC model in \cref{eq:fullQuantumSleadingOrder} (obtained using the full quantum approach) coincides exactly with the Bloch vector in the Rabi model, \cref{eq:RabiModelBlochVectorTimeEvolution}, at $t=0$, if we let $\OmegaR\leftrightarrow \OmegaJC$. This is not surprising, given that the two models are mathematically identical at $t=0$. At finite $t$, $\mathcal{R}[\theta_\pm,\phi_\pm]$ causes each top to slowly precess around the $\vb{e}^z$-axis (now in the lab frame) at the slow frequency $\mp\omega_s$. For $t\gtrsim \gamma_f^{-1}$, we recover the result from the semiclassical approach, \cref{eq:superpositionSemiClassicalSol}. 

The higher order corrections $\mathcal{O}(N^{-k/2})$, with $k\geq 1$, to \cref{eq:fullQuantumSleadingOrder} cannot be nicely written as a superposition of two modes and are listed in \cref{eq:SzSaddlePointGeneral,eq:SPlusSaddlePointGeneral}, up to and including $k=4$ and $k=2$ for the $\alpha=\beta$ and $\alpha=-\beta$ contributions, respectively. Note that they all vanish at $t=0$, where we recover the Rabi result, \cref{eq:RabiModelBlochVectorTimeEvolution} with $\OmegaR\to\OmegaJC$, as required. The leading order $O(N^{-1/2})$ corrections all oscillate and decay at the fast frequency and decay rates $\OmegaJC$ and $\gamma_f$. For a qubit initialized in $\ket{\pm}$, they become the leading order fast-oscillating, fast-decaying terms, as the $\mathcal{O}(1)$ fast-decaying terms vanish from $\langle\hat{\vb{S}}\rangle_\pm$ (to see this, set $r=\pm 1$ in \cref{eq:fullQuantumSleadingOrder}).

\begin{figure}
  \centering
  \includegraphics[width=\linewidth]{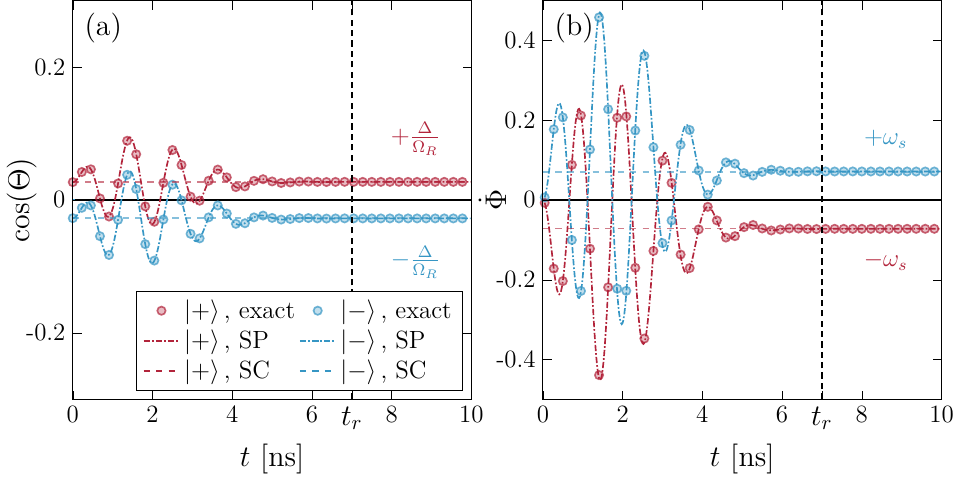}
  \caption{Qubit angles in the early-time window, $0<t\lesssim t_r$, for a qubit initialized in $\ket{\pm}$ (JC model in the rotating frame, as in \cref{fig:JCvsRabi,fig:JCvsRabiPurityXYplaneTrace}). We plot $\cos(\Theta)$ and $\dot{\Phi}$ for simplicity, where $\Theta,\Phi$ are the polar and azimuthal angles of the qubit, respectively. Saddlepoint (SP) and semiclassical (SC) expressions for $\Theta,\Phi$ were obtained by substituting \cref{eq:SzSaddlePointGeneral}, \cref{eq:SPlusSaddlePointGeneral} and \cref{eq:semiClassicalApproximationLeadingOrderSolution} into \cref{eq:ThetaPhiPlusMinusDef}, respectively. Only the SP approach is able to correctly capture the initial $\mathcal{O}(N^{-1})$ oscillations in the exact data for $0<t\lesssim t_r$. At $t\gtrsim t_r$, the exact data, SP and SC approaches match to leading order in $1/\sqrt{N}$. Parameters: $g/2\pi=$\SI{100}{\mega\hertz}, $\Delta/2\pi=$\SI{23}{\mega\hertz}, $\alpha_0=4.4$, with $t_r=$\SI{7}{\nano\second} as in \cref{fig:JCvsRabiPurityXYplaneTrace}.
  }
\label{fig:anglesQubitSweetSpot}
\end{figure}

In \cref{fig:anglesQubitSweetSpot}, we plot $\cos(\Theta)=\langle\hat{S}^z\rangle/|\langle\hat{\vb{S}}\rangle|$ and $\dot{\Phi}$, where $\Phi=\arctan(\langle\hat{S}^y\rangle/\langle\hat{S}^x\rangle)$, as a function of time for a qubit initialized in $\ket{\pm}$. We see that, while the semiclassical approach fails to capture the early-time fast-decaying $\mathcal{O}(1/\sqrt{N})$ oscillations, the saddlepoint approach reproduces them correctly. Beyond $t=t_r$, the leading order corrections are the surviving $\alpha=\beta$, $\mathcal{O}(1/N)$ terms, which oscillate at the slow frequency $\omega_s$, see \cref{eq:qubitDMsweetspotReadoutTime}. These corrections, which are too small to be seen in \cref{fig:anglesQubitSweetSpot}, are nevertheless crucial, as they enter in the calculation of the post-measurement density matrices in \cref{sec:Measurements}.

The qubit purity inherits the $1/\sqrt{N}$ expansion structure of the Bloch vector $\langle\hat{\vb{S}}\rangle$, allowing us to express it as
\begin{equation}\label{eq:purityTaylorExpansion}
\Tr[\hat{\rho}_{\text{q}}^2]=\sum_{k=0}^{\infty} \frac{P^{(\frac{k}{2})}}{N^{\frac{k}{2}}}.
\end{equation}
Using \cref{eq:fullQuantumSleadingOrder}, we calculate the purity of the qubit, which to leading order reads
\begin{equation}\label{eq:purityEarlyTimeWindowJCmodel}
    P^{(0)}=\frac{1}{2}\left[1+r^2+e^{-2\gamma_f^2t^2}(1-r^2)\right],
\end{equation}
in the early-time window $0<t\lesssim t_r$. Note that, to obtain \cref{eq:purityEarlyTimeWindowJCmodel}, we Taylor expanded $\cos(\omega_s t)\approx 1$, $\sin(\omega_s t)\approx 0$, using $\omega_s t_r\sim N^{-1/2}$. The $\mathcal{O}(N^{-1/2})$ and higher order corrections all vanish at $t=0$, so that $\Tr[\hat{\rho}^2_{\text{q}}]$ always starts off at exactly one. At $t= t_r=2\sqrt{2}\gamma_f^{-1}$, $\Tr[\hat{\rho}^2_{\text{q}}]$ decays to $\frac{1}{2}(1+r^2)$, to leading order, see also \cref{fig:JCvsRabiPurityXYplaneTrace}(c). The subleading correction at $t=t_r$ has order $\mathcal{O}(N^{-1})$, as the $\mathcal{O}(N^{-1/2})$ terms all decay at the fast decay rate $\gamma_f$, i.e., $P^{(\frac{1}{2})}(t_r)=0$. The $\mathcal{O}(N^{-1})$ corrections arise because the fast-decaying $\mathcal{O}(N^{-1})$ terms, which at $t=0$ exactly counterbalance their slow-oscillating $\mathcal{O}(N^{-1})$ counterparts, are no longer there at $t=t_r$. To identify the measurement axes, or pointer states, of the qubit, we need to find the two pairs of sweet-spot angles, or, alternatively, $r_{\pm},\Delta\Phi_{\pm}$ using the parametrization in \cref{eq:cPlusMinusParametrization}, that maximize $\Tr[\hat{\rho}^2_{\text{q}}(t_r)]$. Mirroring the expansion of the purity in powers of $1/\sqrt{N}$ in \cref{eq:purityTaylorExpansion}, we assume the following perturbative ansätze for $r_{\pm},\Delta\Phi_{\pm}$,
\begin{equation}
\begin{aligned}
    r_{\pm}&=\sum_{k=0}^{\infty} \frac{r^{(\frac{k}{2})}_{\pm}}{N^{\frac{k}{2}}}, \quad \Delta\Phi_{\pm}=\sum_{k=0}^{\infty} \frac{\Delta\Phi^{(\frac{k}{2})}_{\pm}}{N^{\frac{k}{2}}}.
\end{aligned}
\end{equation}
\begin{figure}
  \centering
  \includegraphics[width=.85\linewidth]{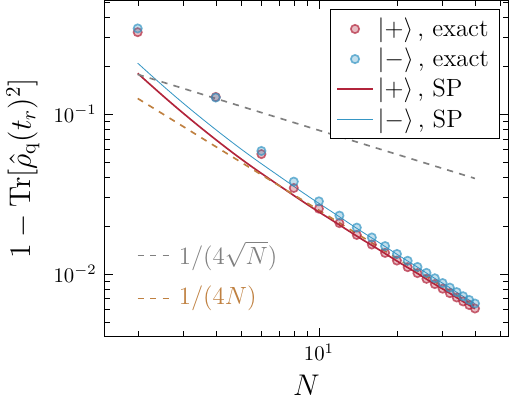}
  \caption{Loss in qubit purity, $1-\Tr[\hat{\rho}_{\text{q}}(t_r)^2]$, at readout time $t_r=$\SI{7}{\nano\second}, for a qubit initialized in sweet-spot states $\ket{\pm}$ $(r=\pm 1)$, as a function of $N$. Dashed lines correspond to the asymptotic $1/(4N)$ line (brown), with a $1/(4\sqrt{N})$ line (gray) added for reference. Parameters: $g/2\pi=$\SI{100}{\mega\hertz}, $\Delta/2\pi=$\SI{23}{\mega\hertz}. Leading order saddlepoint (SP) expression in \cref{eq:QubitPurityReadoutTimeJCmodel}.
  }
\label{fig:qubitPuritySweetSpot}
\end{figure}
Note that all the $P^{(\frac{k}{2})}$ coefficients in \cref{eq:purityTaylorExpansion} are functions of $r^{(\frac{k'}{2})}_\pm$ and $\Delta\Phi^{(\frac{k'}{2})}_{\pm}$, where $k'\leq k$. We now proceed perturbatively, maximizing $P^{(\frac{k}{2})}$ order by order, starting with $k=0$ and increasing $k$ in integer steps. Using \cref{eq:purityEarlyTimeWindowJCmodel}, it is clear that $r^{(0)}=\pm 1$ maximizes $P^{(0)}(r^{(0)})=1$. Note that $\Delta\Phi^{(0)}$ is undefined there, as $r^{(0)}=\pm 1$ corresponds to the two poles of the coordinate system. For the next order, $\mathcal{O}(N^{-1/2})$, using that $P^{(\frac{1}{2})}(t_r)$ vanishes, $r^{(\frac{1}{2})}$ and $\Delta\Phi^{(\frac{1}{2})}$ must also vanish. At order $\mathcal{O}(N^{-1})$, there are two contributions: $\frac{r^{(1)}}{N}\frac{\partial P^{(0)}}{\partial r}|_{r^{(0)}}$ and $P^{(1)}(r^{(0)})$. Now, using that $|r|\leq 1$ and that $r^{(1)}$ dominates all the higher order $r^{(\frac{k}{2})}$, $k\geq 2$ corrections in the $N\to\infty$ limit, we must have $\sign(r^{(1)})=-\sign(r^{(0)})$. But this means that $\frac{r^{(1)}}{N}\frac{\partial P^{(0)}}{\partial r}|_{r^{(0)}}=\frac{r^{(1)}r^{(0)}}{N}$ is always negative, i.e., a finite $r^{(1)}$ always \emph{reduces} the purity at order $\mathcal{O}(N^{-1})$. From this, we conclude that $r^{(1)}=0$ maximizes the purity at order $\mathcal{O}(N^{-1})$. Now, notice that the above argument can be iterated to the next power $k=3$, where there will again only be two $\mathcal{O}(N^{-\frac{3}{2}})$ contributions to the qubit purity: $\frac{r^{(\frac{3}{2})}}{N}\frac{\partial P^{(0)}}{\partial r}|_{r^{(0)}}$ and $P^{(\frac{3}{2})}(r^{(0)})$, as $r^{(\frac{1}{2})}$ and $r^{(1)}$ both vanish. We can repeat this for all consecutive powers, leading us to conclude that $r^{(\frac{k}{2})}=0$ $\forall k\geq 1$ is the optimal choice for maximizing the qubit purity at each order $\mathcal{O}(N^{-\frac{k}{2}})$. Thus, $r_{\pm}=\pm 1$ really do designate the two sweet-spot orientations (with $\Delta\Phi_\pm$ not well defined as these are the two poles in the tilted spherical coordinate system). In \cref{eq:purityJCpostReadoutTime}, we list the resulting expression for qubit purity $\Tr[\hat{\rho}_{\text{q},\pm}^2(t\gtrsim t_r)]$, evaluated at $r_\pm=\pm 1$, correct to order $\mathcal{O}(N^{-2})$. At readout time $t_r$, the leading order deviation of the purity from one is
\begin{equation}\label{eq:QubitPurityReadoutTimeJCmodel}
1-\Tr[\hat{\rho}_{\text{q},\pm}^2(t_r)]=\frac{\omega_s }{\OmegaJC}\left(1\mp \frac{\Delta}{\OmegaJC}\right)^2+\mathcal{O}(N^{-2}).
\end{equation} 
\cref{eq:QubitPurityReadoutTimeJCmodel} was obtained by setting $t=t_r$ in \cref{eq:purityJCpostReadoutTime}, Taylor expanding $\cos(\omega_s t_r)\sim 1$, and keeping only the $\mathcal{O}(N^{-1})$ terms. In \cref{fig:qubitPuritySweetSpot}, we compare \cref{eq:purityJCpostReadoutTime} to the exact qubit purity, obtaining good agreement for $N\gtrsim 5$. Note that the purity is not identical for the $\ket{\pm}$ states, unless $\Delta=0$, i.e., when the qubit and resonator energies are in resonance. In the limit $N\to \infty$, the bracketed part of the first term on the RHS of \cref{eq:QubitPurityReadoutTimeJCmodel} tends to $1$, and the numerical prefactor is dictated by the prefactor $\omega_s/\OmegaJC$, giving $\lim_{N\to \infty}(1-\Tr[\hat{\rho}_{\text{q},\pm}^2(t_r)])=1/(4N)$.

\begin{figure}
  \centering
  \includegraphics[width=\linewidth]{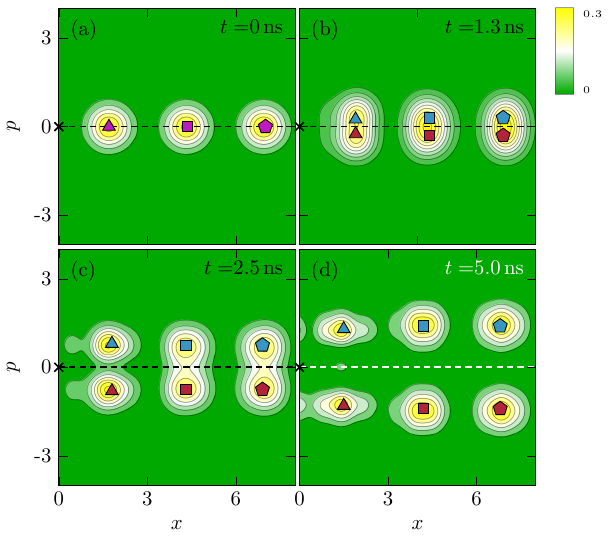}
  \caption{Evolution of the resonator's phase space combined marginal probability distribution, $P_{\pm}(x,p)=|R_{\pm}(x)|^2|R_{\pm}(p)|^2$, in the early-time window $0\leq t\lesssim t_r$. All data are for the initial state \cref{eq:qubitCavityInitialState}, in six different configurations: $\ket{\psi}=\ket{\pm}$ and $\alpha_0=1.7,4.4,7.1$, indicated by a triangle, square, and pentagon, respectively. The polygons track the centre of each marginal probability distribution, $\langle\hat{a}(t)\rangle$, over time. Panel (a): $\ket{R_{\pm}}=\ket{\alpha_0}$, the probability distributions are indistinguishable and perfectly Gaussian, $\Delta x=\Delta p=1/\sqrt{2}$. Panels (b)--(c): $P_{\pm}(x,p)$ start moving down/up at angular frequency $\mp\omega_s$ relative to $(0,0)$ (indicated by a black cross) and deforming. Panel (d): at readout time $t_r\approx 4g^{-1}$, $P_{\pm}(x,p)$ become distinguishable. They are squeezed and stretched compared to panel (a), but the total quadrature spread decreases with $N$ as $\zeta\sim \mathcal{O}(1/N)$, see \cref{eq:squeezingParameterReadoutTime}. Parameters $g,\Delta$ as in \cref{fig:JCvsRabiPurityXYplaneTrace}.
  }
\label{fig:wignerPlotsVaryNandTime}
\end{figure}

Let us now turn our attention to the resonator. The semiclassical approach gave us the correct leading order $\mathcal{O}(N^{1/2})$ contribution to the JC drive vector $\langle\OmegaJCvec\rangle$, see \cref{eq:semiClassicalApproximationLeadingOrderSolution,eq:superpositionSemiClassicalSol}. The in-plane components of $\langle\OmegaJCvec\rangle$ are straightforwardly related to $\langle\hat{a}\rangle$ via \cref{eq:dynamicalRabiVec}, giving
\begin{equation}\label{eq:aExpectValLeadingOrder}
    \langle\hat{a}\rangle=\frac{1}{2}e^{i\phiJC}\sqrt{N}\left[(1+r)e^{-i\omega_s t}+(1-r)e^{i\omega_s t}\right]+\mathcal{O}(N^{-\frac{1}{2}}),
\end{equation}
to leading order. The next order corrections can be evaluated using the saddlepoint approach and are given in \cref{eq:annihilationSaddlePointGeneral}, down to $\mathcal{O}(N^{-3/2})$ for the slow terms and $\mathcal{O}(N^{-1/2})$ for the fast-decaying terms, respectively. Note the absence of any fast-decaying terms in \cref{eq:aExpectValLeadingOrder}, in contrast to what we saw in the leading order terms for the qubit Bloch vector, \cref{eq:fullQuantumSleadingOrder}. For $\langle\hat{a}\rangle$, the fast-decaying terms first appear only at order $\mathcal{O}(N^{-1/2})$. This makes sense if one carefully considers the first of the two equations in \cref{eq:HeisenbergEoMsJCmodel}. On the left side of this equation, the time derivatives of the $\mathcal{O}(N^{-\frac{k}{2}})$ fast-decaying contributions have order $\mathcal{O}(N^{-\frac{k-1}{2}})$ or $\mathcal{O}(N^{-\frac{k}{2}})$, depending on whether the time derivative hits an $\OmegaJC t$ or a $\gamma_f^2t^2$ term, respectively. Meanwhile, the right side of the equation is $\mathcal{O}(N^0)$, which enforces $k\geq 1$, i.e., the leading fast-decaying contribution can be at most $\mathcal{O}(N^{-1/2})$. The absence of $\mathcal{O}(N^{\frac{1}{2},1})$ fast-decaying terms leads to the orbits of the resonator being less ``wiggly'' than those of the qubit, even when the qubit is initialized in a superposition of its two sweet-spot states, $\ket{\psi_0}=c_+\ket{+}+c_-\ket{-}$; compare Figs.~\ref{fig:JCvsRabiPurityXYplaneTrace}(a) and (b). 

\begin{figure}
  \centering
  \includegraphics[width=.65\linewidth]{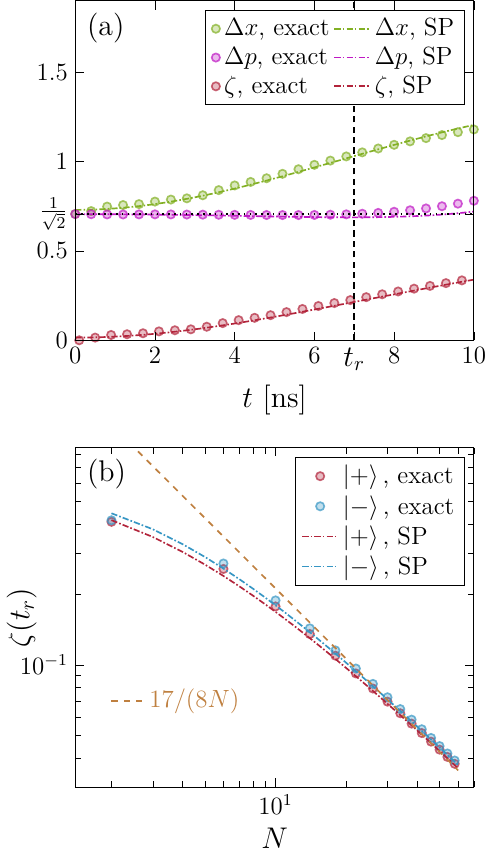}
  \caption{Panel (a): Quadrature spread of the resonator state as a function of time, for a qubit initialized in $\ket{+}$ with $N=7.1$. Comparison of exact data with the saddlepoint approach, \cref{eq:sumDeltaXsquaredDeltaPsquared,eq:differenceDeltaXsquaredDeltaPsquared}. The resonator state initially gets stretched in the $x$-direction and slightly squeezed, before also getting stretched, in the $p$-direction. Panel (b): total quadrature spread parameter $\zeta$ for a qubit initialized in $\ket{\pm}$, \cref{eq:squeezingParameter}, as function of $N$. The asymptotic brown dashed line $17/(8N)$ is obtained by taking the large $N$ limit of \cref{eq:squeezingParameterReadoutTime}. Parameters: $g/2\pi=$\SI{100}{\mega\hertz}, $\Delta/2\pi=$\SI{23}{\mega\hertz}, $\phiJC=0$.
  }
\label{fig:squeezingPanels}
\end{figure}

For qubit readout, we also care about how the probability distribution of the resonator is distributed in phase space. The center of the probability distribution is given by $\langle\hat{a}\rangle=\langle\hat{x}\rangle+i \langle\hat{p}\rangle$, where $\hat{x}=\sqrt{\frac{m\omega}{2\hbar}}\hat{X}$ and $\hat{p}=\frac{1}{\sqrt{2m\omega\hbar}}\hat{P}$ are the dimensionless position and momentum coordinates, respectively. But what about its shape---does going into the nondispersive regime have any surprising effects resulting in quadrature spreading or nonlinearities? Tracing over the degrees of freedom of the qubit returns the density matrix of the resonator, $\hat{\rho}_{\text{r}}=\Tr_{\text{q}}[\ket{\Psi}\bra{\Psi}]$. The probability densities in the $x$- and $p$-representations are then given by $|R(x)|^2=\bra{x}\hat{\rho}_{\text{r}}\ket{x}$ and $|R(p)|^2=\bra{p}\hat{\rho}_{\text{r}}\ket{p}$, where $\ket{x}$ and $\ket{p}$ are the eigenstates of the $\hat{x}$ and $\hat{p}$ operators, respectively. We can also express them in the simple harmonic oscillator (SHO) basis by inserting two resolutions of the identity, $\mathcal{I}=\sum_m\ket{m}\bra{m}$, before and after $\hat{\rho}_{\text{r}}$, giving
\begin{equation}\label{eq:probabilityDistributionCavity}
    \begin{aligned}
        |R(x)|^2&=\sum_{m,n}\psi_m(x)\psi^*_n(x)\bra{m}\hat{\rho}_{\text{r}}\ket{n},\\
        |R(p)|^2&=\sum_{m,n}\psi_m(p)\psi^*_n(p)\bra{m}\hat{\rho}_{\text{r}}\ket{n},
    \end{aligned}
\end{equation}
where $\ket{m}$ are the eigenstates of $\hat{H}_{\text{SHO}}=\omega \hat{a}^{\dagger}\hat{a}$. $\psi_m(x)=\bra{x}\ket{m}$ and $\psi_m(p)=\bra{p}\ket{m}$ are the wavefunctions of the SHO eigenstates in the position and momentum representations, respectively \footnote{They take the form $\psi_m(x)=(2^m m!)^{-\frac{1}{2}}\pi^{-\frac{1}{4}} e^{-\frac{x^2}{2}}H_m(x)$, $\psi_m(p)=(-i)^m(2^m m!)^{-\frac{1}{2}}\pi^{-\frac{1}{4}} e^{-\frac{p^2}{2}}H_m(p)$, where the Hermite polynomials are defined as $H_n(x)=(-1)^ne^{x^2}\frac{d^n}{dx^n}e^{-x^2}$}. 
At $t=0$, we always have $\hat{\rho}_{\text{r}}=\ket{\alpha_0}\bra{\alpha_0}$, since we initialize the resonator in a coherent state. Using 
\begin{equation}
    \begin{aligned}
        \bra{x}\ket{\alpha_0}&=\pi^{-\frac{1}{4}}e^{-\Im[\alpha_0]^2/2}\exp\left(-\frac{1}{2}(x-\alpha_0)^2\right),\\
        \bra{p}\ket{\alpha_0}&=\pi^{-\frac{1}{4}}e^{-\Re[\alpha_0]^2/2}\exp\left(-\frac{1}{2}(p+i\alpha_0)^2\right)
    \end{aligned}
\end{equation}
\footnote{To see this, consider the definition $\hat{a}\ket{\alpha_0}=\alpha_0\ket{\alpha_0}$. In, e.g., the $x$-representation, this takes the form $\left(x+\frac{\partial}{\partial x}\right)R(x)=\alpha_0 R(x)$, which can be trivially solved to give the stated result.}, we conclude that the initial probability is always Gaussian-distributed and centered at $\Re[\alpha_0]$ and $\Im[\alpha_0]$ in the $x$- and $p$-representations, respectively. To visualize this, in \cref{fig:wignerPlotsVaryNandTime}(a), we plot the combined marginal distribution $P(x,p)=|R(x)|^2|R(p)|^2$ for three different values $\alpha_0=1.7,4.4$ and $7.1$. As expected, the probability distributions are centered at $\alpha_0$ and perfectly circular (i.e., Gaussian). Note that $P(x,p)$ is not the same object as the Wigner function, but regardless of this fact captures the separation of the pointer states.

For a reliable readout, it is important that the probability distributions not deform too much by the readout time $t_r$, remaining as close to Gaussian as possible. This is because the width of the distribution in the $p$-direction enters the signal-to-noise formula (SNR) of the measurement, see \cref{eq:SNRdef} in Sec.~\ref{sec:Measurements}. To check this, we use the definitions in \cref{eq:probabilityDistributionCavity} to plot the evolution of the resonator probability distribution $P(x,p)$ in the early-time window $0<t\lesssim t_r$, \cref{fig:wignerPlotsVaryNandTime}(b)--(d). For each of the different $\alpha_0$ values, we initialize the qubit in the corresponding $\ket{\pm}$ state, following \cref{eq:RabiModelEigenbasis}, and replacing $\thetaR\to\thetaJC$, $\phiR\to\phiJC$. As $t$ increases, each probability distribution becomes slightly stretched or squeezed in the $x$- and $p$-directions due to the interaction with the qubit, but overall their shapes remain broadly circular. The probability distributions corresponding to $\ket{\pm}$ separate at the rate
\begin{equation}\label{eq:DeltaaSeparationRate}
    \Delta\dot{a}=\frac{d}{dt}\left|\langle\hat{a}\rangle_+-\langle\hat{a}\rangle_-\right|=2\sqrt{N}\omega_s\cos(\omega_s t) + \mathcal{O}(N^{-1/2}).
\end{equation}
In the nondispersive regime, $\Delta^2\ll 4g^2N$, the separation rate is independent of $N$, with $\Delta \dot{a}(0)\sim g$. 

On the other hand, inspection of \cref{fig:wignerPlotsVaryNandTime}(d) suggests that the deformation of the probability distribution appears to become less severe with increasing $N$. One simple way to quantify the degree of deformation is through the variances $\Delta x,\Delta p$. To this end, we calculate $\langle\hat{a}^2\rangle$ and $\langle\hat{n}\rangle$ using the saddlepoint approach, including corrections down to $\mathcal{O}(N^{-1})$ and $\mathcal{O}(N^{0})$ for the slow and fast-decaying terms, respectively, giving \cref{eq:annihilationSquaredSaddlePointGeneral,eq:numberOperatorSaddlePointGeneral}. The resulting variances $\Delta x_{\pm}(t_r)$ and $\Delta p_{\pm}(t_r)$, evaluated at readout time $t_r$ for a system with the qubit initialized in $\ket{\pm}$, can be obtained from \cref{eq:sumDeltaXsquaredDeltaPsquared,eq:differenceDeltaXsquaredDeltaPsquared}. To quantify the overall spread of the probability distribution, we define the total quadrature spread parameter 
\begin{equation}
    \zeta=\frac{1}{2}\log(\Delta x^2+\Delta p^2).
\end{equation}
In an ideal readout experiment, the resonator state remains perfectly coherent, such that $\Delta x=\Delta p=1/\sqrt{2}$, resulting in $\zeta=0$. We now wish to calculate the leading order imperfection in $\zeta$ resulting from our nondispersive readout scheme. Since the Heisenberg uncertainty principle requires $\Delta x,\Delta p\geq 1/\sqrt{2}$, $\zeta$ is always positive. We once again proceed perturbatively, calculating $\zeta$ in powers of $1/\sqrt{N}$. Interestingly, to obtain a sufficiently accurate $\zeta$, it is crucial not just to stop at the subleading order $\mathcal{O}(N^0)$ corrections, but to also include the subsubleading $\mathcal{O}(N^{-1})$ terms, see definition in \cref{eq:differenceDeltaXsquaredDeltaPsquared}. This is because the $\mathcal{O}(N^0)$ terms in fact become $\mathcal{O}(N^{-1})$ as $N$ becomes large, as they are proportional to $\sin^2(\omega_s t_r)$, which Taylor expands to $ \omega_s^2t_r^2\sim \mathcal{O}(N^{-1})$. Thus, the leading order correction is in fact $\mathcal{O}(N^{-1})$, not $\mathcal{O}(N^{0})$, and all other terms of that order also need to be included to allow for accurate comparison with the exact data.

In \cref{fig:squeezingPanels}(a), we plot $\Delta x$, $\Delta p$ and $\zeta$ as a function of time, for a system initialized in $\ket{\Psi_0}=\ket{+}\ket{\alpha_0}$, with $\alpha_0=7.1$. We compare the exact data with the saddlepoint formulas \cref{eq:sumDeltaXsquaredDeltaPsquared,eq:differenceDeltaXsquaredDeltaPsquared}, and obtain a very good match, even in the $0<t\lesssim t_r$ region. This is a pleasant surprise, given that Eqs.~(\ref{eq:sumDeltaXsquaredDeltaPsquared}) and (\ref{eq:differenceDeltaXsquaredDeltaPsquared}) ignore the fast-decaying early-time contributions, which we know vanish for $r=\pm 1$ at order $\mathcal{O}(N^{(1,\frac{1}{2},0)})$ from \cref{eq:annihilationSaddlePointGeneral,eq:annihilationSquaredSaddlePointGeneral,eq:numberOperatorSaddlePointGeneral}, but which we did not check were small at order $\mathcal{O}(N^{(-\frac{1}{2},-1)})$. Although there is initially some squeezing in the $p$-direction at very early times, overall the probability distribution always spreads, as confirmed by $\zeta$ increasing over time.

To conclude this section, we investigate how the degree of quadrature spread of the resonator state at readout time varies with $N$. We plot $\zeta(t_r)$ as a function of $N$ in \cref{fig:squeezingPanels}(b). The negative gradient confirms the qualitative observation from \cref{fig:wignerPlotsVaryNandTime}(d), namely, that the degree of spreading decreases with increasing $N$. In the limit of large $N$, $\zeta(t_r)$ can be Taylor expanded to give
\begin{equation}\label{eq:squeezingParameterReadoutTime}
   \begin{aligned}
   \zeta(t_r)&=\frac{1}{2}\frac{\omega_s}{\OmegaJC}\left(1\mp\frac{\Delta}{\OmegaJC}\right)\left(4N\omega_s^2 t_r^2+\left(1\mp\frac{\Delta}{\OmegaJC}\right)\right)\\
   &\quad + \mathcal{O}(N^{-2}),
   \end{aligned}
\end{equation}
i.e., $\zeta(t_r)\sim\mathcal{O}(N^{-1})$, just like $1-\Tr[\hat{\rho}_{\text{q}}^2(t_r)]$ in \cref{eq:QubitPurityReadoutTimeJCmodel}. The $1/N$ deviation from the ideal value $1$ for figures of merit in the nondispersive model is a common feature, and will reappear regularly throughout the paper. In the limit of large $N$ limit, the terms in the second set of brackets simplify to $17$, where we used $\lim_{N\to\infty}4N\omega_s^2t^2=16$, giving $\lim_{N\to\infty}\zeta(t_r)=17/(8N)$ (dashed brown reference line in \cref{fig:squeezingPanels}(b)). 

This concludes our discussion of the qubit and resonator dynamics in the JC model. In Sec.~\ref{sec:SecondDrive}, we will investigate what happens when we add a classical drive to the qubit-resonator system, thereby combining the JC and Rabi models covered in detail in this section, and show how this can benefit the readout process.

\section{Two drives}\label{sec:SecondDrive}

So far, we have looked at the Rabi and JC models separately, but what if we combine the two? In a physical system, this would correspond to a qubit connected to a resonator, with an additional classical drive $\hat{H}_d(t)=\Omega_d e^{-i\omega_d t}\hat{S}^-+\text{h.c.}$ applied to the qubit, for example via a time-dependent magnetic flux. The additional classical drive gives us two extra degrees of freedom to experiment with: the ratio of the resonator and classical drive frequencies, $\omega/\omega_d$, and the phase difference between the drive amplitudes.

\begin{figure}
  \centering
  \includegraphics[width=.75\linewidth]{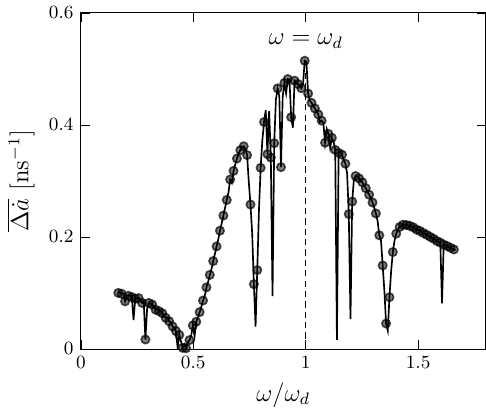}
  \caption{Average separation rate $\overline{\Delta\dot{a}}$ as a function of the frequency ratio $\omega/\omega_d$. Data obtained by averaging $\overline{\Delta\dot{a}}$ over one common time period $T_c=mT_r=nT_d$, where $m,n\in\mathbb{Z}$, $T_r=2\pi/\omega_r$, and $T_d=2\pi/\omega_d$. The maximum $\overline{\Delta\dot{a}}$ occurs when the frequencies are identical, $\omega=\omega_d$. Parameters: $|\Omega_r|=$\SI{169}{\mega\hertz}, $|\Omega_d|=$\SI{84}{\mega\hertz}, with the phase difference $\arg(\Omega_d/\Omega_r)$ optimized for each data point $\omega/\omega_d$ to maximize $\overline{\Delta \dot{a}}$. The data was generated numerically using QuTiP \cite{qutip2026} (see \cite{zenodoDataset2026} for the detailed implementation).
  }
\label{fig:varyRatioOfDriveFrequencies}
\end{figure}

For simplicity, we will set the frequencies of the resonator and classical drive equal to each other, $\omega_d=\omega$. The choice of the symmetric point $\omega_d=\omega$ is further justified by the fact that it maximizes the average rate $\overline{\Delta\dot{a}}$ at which the $P_{\pm}(x,p)$ probability distributions separate; see \cref{fig:varyRatioOfDriveFrequencies} for a numerical demonstration of this result. To generate \cref{fig:varyRatioOfDriveFrequencies}, the resonator was replaced by another classical drive of the form $\hat{H}_{\text{res}}=\Omega_r e^{-i\omega t}\hat{S}^-+h.c.$, and the ratio $\omega/\omega_d$ between them was varied, while keeping the amplitudes of each drive constant. The phase difference $\varphi$ between two drives was numerically optimized to maximize the displacement rate over one common period, $\overline{\Delta\dot{a}}$; see also \cite{zenodoDataset2026} for the detailed technical implementation. This maximizes the rate of growth of the signal-to-noise ratio (SNR), defined in \cref{eq:SNRdef}, and therefore minimizes the readout time.

Choosing the drives to be on resonance still leaves the choice of the phase difference between their amplitudes. The phase difference enters the Floquet eigenenergies $\epsilon$ of the system, allowing energy to be pumped into or out of the resonator at a rate $\sim \nabla_{\varphi}\epsilon$ \cite{psaroudaki2023,yinan2025}. Might this be harnessed in any way to improve the readout rate or qubit purity at readout time? We answer this question in this section.

Since the frequencies of the two drives are equal, we may use the same rotating frame as in Sec.~\ref{sec:JCmodelDynamics}. Applying $\hat{U}_{\text{JC}}$ from \cref{eq:unitaryOpsRotFrame}, the two-drive JC+Rabi Hamiltonian takes the form
\begin{equation}\label{eq:JCplusRabiModel}
    \hat{\tilde{H}}_{\text{JC+R}}=\frac{\Delta}{2}\hat{S}^z+g\left((\hat{a}+s)\hat{S}^-+(\hat{a}^{\dagger}+s^*)\hat{S}^+\right),
\end{equation}
where $s=\Omega_d/g$ is a complex dimensionless parameter measuring the strength of the classical drive. The phase difference between the two drive amplitudes is then defined as
\begin{equation}\label{eq:phaseDifferenceBetweenTwoDrives}
    \varphi=\arg(s/\alpha_0).
\end{equation}
To bring \cref{eq:JCplusRabiModel} to a more familiar form, we apply a displacement transformation on the resonator state, leaving the qubit unaffected. Defining the photonic displacement operator $\hat{D}(s)=e^{s\hat{a}^{\dagger}-s^*\hat{a}}$, the displaced Hamiltonian reads $\hat{\bar{H}}_{\text{JC+R}}=\hat{D}(s)\hat{\tilde{H}}_{\text{JC+R}}\hat{D}^{\dagger}(s)$. Using $\hat{D}(s)\hat{a}\hat{D}^{\dagger}(s)=\hat{a}-s$, $\hat{D}(s)\hat{a}^{\dagger}\hat{D}^{\dagger}(s)=\hat{a}^{\dagger}-s^*$, we see that the displaced two-drive Hamiltonian is identical to the original JC Hamiltonian in the rotating frame, $\hat{\bar{H}}_{\text{JC+R}}=\hat{\tilde{H}}_{\text{JC}}$. This fact makes the calculation of observables for the two-drive model defined in \cref{eq:JCplusRabiModel} remarkably straightforward. In the rotating frame, the eigenstates take the form $\hat{D}^{\dagger}(s)\ket{m,\alpha}$, where $\ket{m,\alpha}$ are the original JC eigenstates defined in \cref{eq:JCmodelEigenbasis}. The wavefunction for the full system therefore takes the form
\begin{equation}\label{eq:psiTwoDrives}
\begin{aligned}
    \ket{\Psi(t)}_{\text{JC+R}}&=\hat{D}^{\dagger}(s)\bigg(\bar{c}_{0,\uparrow}e^{-\frac{i\Delta t}{2}}\ket{0,\uparrow}\\
    &+\sum_{m=1}^{\infty}\sum_{\alpha=\pm}\bar{c}_{m,\alpha}e^{-i\alpha\lambda_m t}\ket{m,\alpha}\bigg),
\end{aligned}
\end{equation}
where the $\bar{c}_{m,\alpha}$ coefficients are given by
\begin{equation}
    \bar{c}_{m,\alpha}=\bra{m,\alpha}\hat{D}(s)\ket{\Psi(0)}.
\end{equation}
Substituting \cref{eq:qubitCavityInitialState} into the above and using $\hat{D}(s)\ket{\alpha_0}=\ket{\alpha_0+s}$, we conclude that all the $\bar{c}_{m,\alpha}$ coefficients can be obtained from $c_{m,\alpha}$ defined in \cref{eq:JCModelFullSystemWavefunctionCoeffs} by sending $\alpha_0\to\alpha_0+s$. This motivates us to define a new quantity, the effective two-drive amplitude
\begin{equation}\label{eq:effectiveTwoDriveAmplitude}
    \alpha_{\text{eff}}=\alpha_0+s.
\end{equation}
whose norm squared defines an effective photon number, 
\begin{equation}\label{eq:effectiveTwoDrivePhotonNumber}
   N_{\text{eff}}=|\alpha_{\text{eff}}|^2=N+|s|^2+2\sqrt{N}|s|\cos(\varphi),
\end{equation}
for the two-drive system. We also combine the Rabi and JC drive vectors defined in \cref{eq:staticRabiVec,eq:dynamicalRabiVec} into a single two-drive vector by sending $\hat{a}\to\hat{a}+s$ in \cref{eq:dynamicalRabiVec},
\begin{equation}\label{eq:JCRabiDriveVec}
    \hat{\vb{\Omega}}_{\text{JC+R}}=\begin{pmatrix}
        g(\hat{a}+\hat{a}^{\dagger}+s+s^*), & ig(\hat{a}^{\dagger}-\hat{a}+s^*-s) ,& \Delta 
    \end{pmatrix}^T.
\end{equation}
Note that at $t=0$, $\langle\hat{\vb{\Omega}}_{\text{JC+R}}(0)\rangle=\langle\hat{\vb{\Omega}}_{\text{JC}}(0)\rangle_{N\to N_{\text{eff}}}$. 

Next, we consider the effect that the addition of the classical drive has on the expectation values of observables in the two-drive system, $\langle\hat{O}\rangle_{\text{JC+R}}=\bra{\Psi}_{\text{JC+R}}\hat{O}\ket{\Psi}_{\text{JC+R}}$. For observables that \emph{do not} involve a photonic creation or annihilation operator, such as $\hat{S}^i$, no further work is required, since $\hat{D}(s)\hat{O}\hat{D}^{\dagger}(s)=\hat{O}$. We can simply reuse the saddlepoint formulas \cref{eq:SzSaddlePointGeneral,eq:SPlusSaddlePointGeneral}, sending $\sqrt{N}\to \sqrt{N}+|s|e^{i\varphi}$ to account for the addition of the classical drive $\hat{H}_d(t)$. Observables that \emph{do} involve a photonic operator require the additional step of evaluating $\hat{D}(s)\hat{O}\hat{D}^{\dagger}(s)$. This produces some extra terms compared to the JC model saddlepoint formulas, see \cref{eq:resonatorObservablesTwoDrivesSP} in App.~\ref{app:SaddlePointExpressionsSweetspot} for the resulting expressions for $\langle\hat{a}\rangle$, $\langle\hat{a}^2\rangle$ and $\langle\hat{n}\rangle$. 
\begin{figure*}
  \centering
  \includegraphics[width=.95\linewidth]{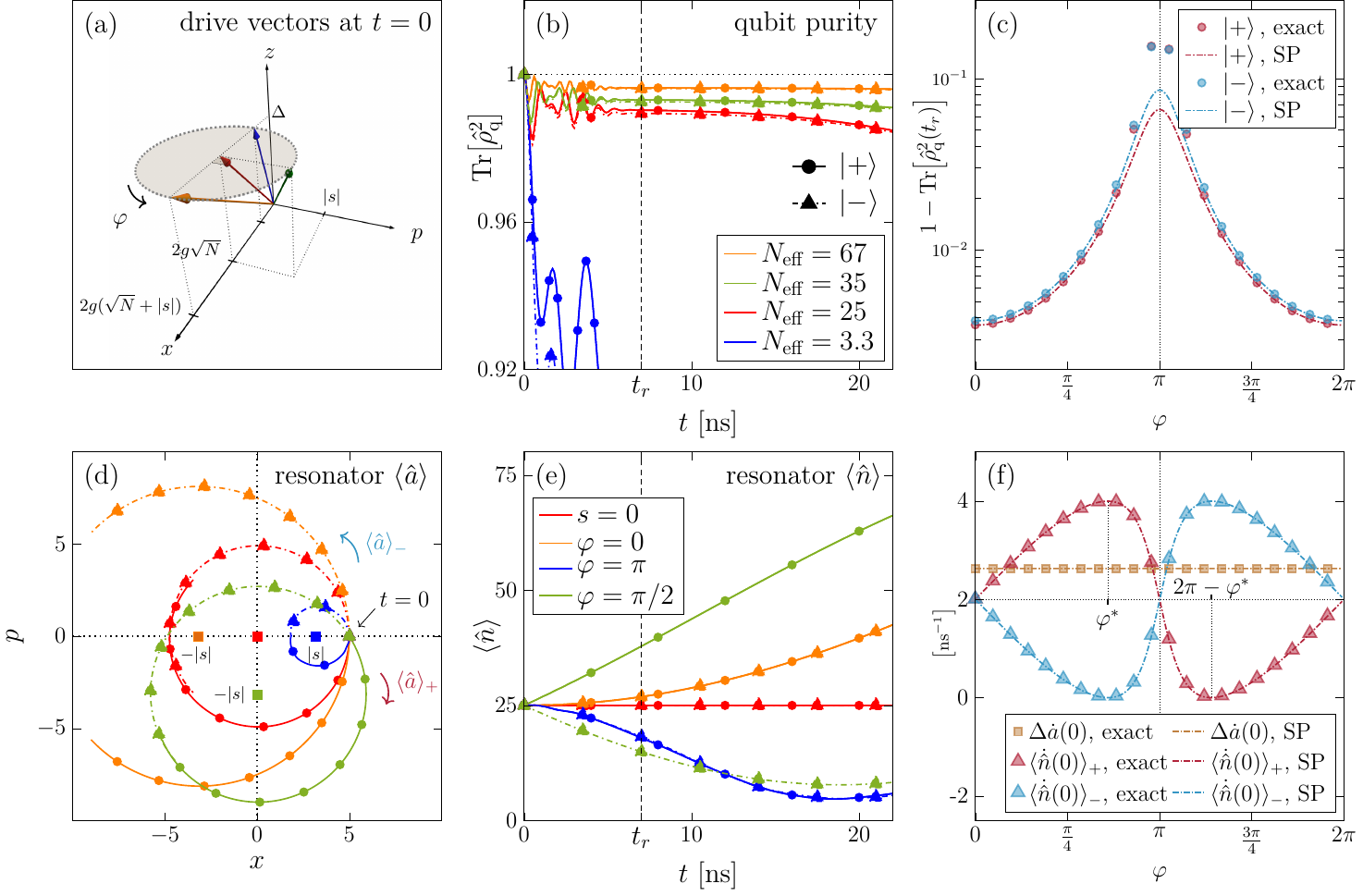}
  \caption{The two-drive model $\hat{H}_{\text{JC+R}}$, \cref{eq:JCplusRabiModel}, in the rotating frame described by $\hat{U}_{\text{JC}}$, defined in \cref{eq:unitaryOpsRotFrame}. For all data shown in (a)--(f), the resonator was initialized in $\ket{\alpha_0}$, and the qubit in $\ket{+}$ (circles) or $\ket{-}$ (triangles), with $\ket{\pm}$ defined as in \cref{eq:RabiModelEigenbasis}, with $\thetaR\to\theta_{\text{JC+R}}$ and $\phiR\to\phi_{\text{JC+R}}$. The amplitudes $|\alpha_0|=5$ and $|s|=3.18$ are kept constant, and the phase difference $\varphi=\arg(s/\alpha_0)$ between the two drives is varied. The triangles and circles denote exact data, while the smooth and dashdotted lines refer to the corresponding SP formulas. In panels (a),(b),(d) and (e), we show data for three discrete values of $\varphi$, obtained by varying the phase of the classical drive while keeping $\alpha_0\in\mathbb{R}$: $s=3.18,\varphi=0$ (yellow), $s=3.18i,\varphi=\pi/2$ (green), $s=-3.18,\varphi=\pi$ (blue), and a reference setting with $|s|=0$ (red). In panels (c) and (f), we scan over all values $\varphi$ in the range $\varphi\in [0,2\pi)$. Panel (a): two-drive vectors $\langle\OmegaJCRvec\rangle$, see \cref{eq:JCRabiDriveVec} for definition,  at $t=0$. The magnitude of $\langle\OmegaJCRvec\rangle$ is maximized when $\varphi=0$. (Note the axes are not to scale: the $z$-axis has been stretched for visual clarity). Panel (b): qubit purity $\Tr[\hat{\rho}^2_{\text{q}}(t)]$ as a function of time. The larger $\Neff$, the closer the purity stays to one. We used the saddlepoint (SP) expressions in \cref{eq:SzSaddlePointGeneral,eq:SPlusSaddlePointGeneral} with $N\to\Neff$. Panel (c): deviation of the qubit purity from one at readout time as a function of $\varphi$. We used the SP expression in \cref{eq:QubitPurityReadoutTimeJCmodel} with $N\to N_{\text{eff}}$.
  $\Tr[\hat{\rho}^2_{\text{q}}(t_r)]$ is proportional to $\bar{\omega}_s/\bar{\Omega}_{\text{JC}}\sim 1/\Neff$. Hence, using $\Neff\sim\cos(\varphi)$ [\cref{eq:effectiveTwoDrivePhotonNumber}], the best purity is achieved at $\varphi=0$. The discrepancy between the exact data and the SP prediction in the $\varphi\sim\pi$ region is caused by low values of $N_{\text{eff}}\sim 3$, where the SP approximation breaks down. Panel (d): paths traced by $\langle\hat{a}(t)\rangle_\pm$ in the $xp$-plane (compare to \cref{fig:JCvsRabiPurityXYplaneTrace}(c)). Squares indicate the centers of the (to leading order) circular orbits traced out by $\langle\hat{a}\rangle_{+}$ (circles, precessing clockwise) and $\langle\hat{a}\rangle_{-}$ (triangles, precessing anticlockwise). SP expression in \cref{eq:amplitudeTwoDrivesSP}. Panel (e): mean photon number $\langle\hat{n}(t)\rangle$ in the resonator as a function of time. SP expression in \cref{eq:photonNumberTwoDrivesSP}. Panel (f): pointer state separation rate, $\Delta \dot{a}(0)$, and initial growth/decay rate of resonator photon number, $\langle \dot{n}(0)\rangle_\pm$, SP expressions in \cref{eq:DeltaaSeparationRate} and \cref{eq:photonNumberTwoDrivesSP}, respectively. $\langle \dot{n}(0)\rangle_\pm$ is minimized at $\varphi=0$. Parameters: $g/2\pi=$\SI{100}{\mega\hertz}, $\Delta/2\pi=$\SI{23}{\mega\hertz}, $\alpha_0=$\SI{5}{} and $|\Omega_d|/2\pi=$\SI{318}{\mega\hertz}.
}
\label{fig:twoDrivesPanels}
\end{figure*}

Let us investigate the dynamics of the two-drive system when we vary the phase difference $\varphi$ between the two drives, keeping the amplitudes $\sqrt{N}=5$ and $|s|=3.18$ of the two drives constant. In \cref{fig:twoDrivesPanels}(a)--(b), (d)--(e), we consider four different settings: in-phase drives with $\varphi=0$ (orange); drives with $\varphi=\pi/2$ (green); out-of-phase drives with $\varphi=\pi$ (blue); and, for reference, a system where the classical drive is turned off, $s=0$ (red). For each drive setting, we use initial states $\ket{\Psi_0}=\ket{\pm}\ket{\alpha_0}$, replacing $\phiR\to\phi_{\text{JC+R}}=\arg(\alpha_{\text{eff}})$ and $\thetaR\to\theta_{\text{JC+R}}=\arctan(2g\sqrt{\Neff}/\Delta)$ in the definition of $\ket{\pm}$ in \cref{eq:RabiModelEigenbasis}. In \cref{fig:twoDrivesPanels}(a), we show the two-drive vectors $\langle\hat{\vb{\Omega}}_{\text{JC+R}}\rangle$, defined in \cref{eq:JCRabiDriveVec}, at $t=0$, for each of the four drive settings. Using \cref{eq:effectiveTwoDriveAmplitude}, the corresponding effective photon numbers are $N_{\text{eff}}=|\alpha_\textrm{eff}|^2=67,35,3.3$ and $25$, respectively (also listed in the legend of \cref{fig:twoDrivesPanels}(b)).

Let us first consider the qubit purity at readout time, $\Tr[\hat{\rho}^2_{\text{q}}(t_r)]$. In \cref{fig:twoDrivesPanels}(b), we see that $\Tr[\hat{\rho}^2_{\text{q}}(t_r)]$ correlates with the inverse of $N_{\text{eff}}$. This makes sense if we consider \cref{eq:QubitPurityReadoutTimeJCmodel}. Sending $N\to N_{\text{eff}}$, the formula predicts that $1-\Tr[\hat{\rho}^2_{\text{q}}(t_r)]$ scales with the inverse of the length of the two-drive vector, $|\langle\hat{\vb{\Omega}}_{\text{JC+R}}\rangle |=\bar{\Omega}_{\text{JC}}=\sqrt{\Delta^2+4g^2N_{\text{eff}}}$. Thus, the highest and lowest $\Tr[\hat{\rho}^2_{\text{q}}(t_r)]$ are achieved by maximizing and minimizing $N_{\text{eff}}$, which occurs at $\varphi=0$ and $\varphi=\pi$, respectively; see \cref{fig:twoDrivesPanels}(a) and (c). 

Next, we consider the amplitude $\langle\hat{a}\rangle_{\pm}$ of the resonator state in the two-drive system. The saddlepoint expression for $\langle\hat{a}\rangle_{\pm}$ is given in \cref{eq:resonatorObservablesTwoDrivesSP}. The leading order in $1/\sqrt{\Neff}$ contribution reads
\begin{equation}\label{eq:amplitudeTwoDrivesSP}
    \begin{aligned}
        \langle\hat{a}\rangle_{\pm}&=(\alpha_0+s)e^{\mp i \bar{\omega}_s t}-s+\mathcal{O}(\Neff^{-\frac{1}{2}}),
    \end{aligned}
\end{equation}
where $\bar{\omega}_s=g^2/\sqrt{\Delta^2+4g^2N_{\text{eff}}}$. \cref{eq:amplitudeTwoDrivesSP} predicts that, to leading order, $\langle\hat{a}\rangle_{\pm}$ once again follows a circular path, but with the center shifted from the origin to $-s$ in the $xp$-plane, see also \cref{fig:twoDrivesPanels}(d). In addition, the radius of the circle is rescaled from $\sqrt{N}$ to $\sqrt{N_{\text{eff}}}$, and the angular frequency from $\omega_s$ to $\bar{\omega}_s$. Since the radius of the path and the angular frequency $\bar{\omega}_s$ are rescaled in the same way, $N\to N_{\text{eff}}$, the separation rate $\Delta\dot{a}$, defined in \cref{eq:DeltaaSeparationRate}, is unaffected to leading order in $1/\sqrt{\Neff}$, as long as the system remains in the nondispersive regime. We test this claim by plotting $\Delta\dot{a}(0)$ as a function of $\varphi$; see the brown data in \cref{fig:twoDrivesPanels}(f). As expected, we obtain a constant value $g$, independent of $\varphi$, to leading order in $1/\sqrt{\Neff}$.

Finally, we look at the mean number of photons in the resonator, $\langle\hat{n}\rangle_{\pm}$, as a function of time. In the JC model, this quantity remained constant and to leading order equal to the initial number of photons $N$ in the resonator, see \cref{eq:numberOperatorSaddlePointGeneral}. However, with the addition of the second drive, $\langle\hat{n}\rangle_{\pm}$ starts to fluctuate in time,
\begin{equation}\label{eq:photonNumberTwoDrivesSP}
    \begin{aligned}
        \langle\hat{n}&\rangle_{\pm}=N+2|s|^2(1-\cos(\bar{\omega}_s t))\\
        &+2\sqrt{N}|s|\left(\cos(\varphi)-\cos(\varphi\pm\bar{\omega}_s t)\right)+\mathcal{O}(\Neff^{-1}),
    \end{aligned}
\end{equation}
as energy is pumped into or out of the resonator by the classical drive $\hat{H}_d(t)$. We could also have obtained \cref{eq:photonNumberTwoDrivesSP} by taking the absolute value squared of \cref{eq:amplitudeTwoDrivesSP}, since $\langle\hat{n}\rangle_{\pm}=|\langle\hat{a}\rangle_{\pm}|^2$ to leading order in $1/\sqrt{\Neff}$. We verify \cref{eq:photonNumberTwoDrivesSP} by plotting $\langle\hat{n}(t)\rangle_\pm$ as a function of time for the same four two-drive settings in \cref{fig:twoDrivesPanels}(e), obtaining an excellent match between the exact data and the saddlepoint expression.

Note that in \cref{eq:photonNumberTwoDrivesSP}, terms are grouped so as to make it obvious that we always have $\langle\hat{n}(0)\rangle_{\pm}=N$, independent of both $\varphi$ and $|s|$. On the other hand, the initial power $\dot{W}(0)=\omega\langle\dot{\hat{n}}(0)\rangle$ absorbed or emitted by the resonator, \emph{does} depends on both the initial orientation of the qubit and the phase difference $\varphi$ between the two drives. Using \cref{eq:photonNumberTwoDrivesSP}, the initial rate of change of the number of photons in the resonator is given by
\begin{equation}\label{eq:rateOfChangeResonatorPhotons}
    \langle\dot{\hat{n}}(0)\rangle_\pm=\pm 2\sqrt{N}|s|\bar{\omega}_s\sin(\varphi).
\end{equation}
We can use \cref{eq:rateOfChangeResonatorPhotons} to identify which values of $\varphi$ maximize and minimize $\langle\dot{\hat{n}}(0)\rangle_\pm$. Recalling that $\bar{\omega}_s$ has a $\varphi$-dependence via $N_{\text{eff}}$, see \cref{eq:effectiveTwoDrivePhotonNumber}, \cref{eq:rateOfChangeResonatorPhotons} has two turning points $\varphi^*$ and $2\pi-\varphi^*$ satisfying $\partial_\varphi \langle\dot{\hat{n}}(0)\rangle_\pm=0$ in the range $0\leq \varphi^*< 2\pi$, with
\begin{equation}\label{eq:turningPointsNdot}
\begin{aligned}
    \varphi^*&=\arccos(\sqrt{b^2-1}-b),\\
    \text{where }b&=\frac{1}{2}\left(\frac{\Delta^2}{4g^2\sqrt{N}|s|}+\frac{\sqrt{N}}{|s|}+\frac{|s|}{\sqrt{N}}\right).
\end{aligned}
\end{equation}
The minimum possible value of $b$ is $1$, occurring at $\Delta=0$, $\sqrt{N}=s$. This puts $\varphi^*$ in the range $\frac{\pi}{2}\leq\varphi^*\leq \pi$. Substituting $\varphi^*$ back into \cref{eq:rateOfChangeResonatorPhotons}, we conclude that $\max(\langle\dot{\hat{n}}(0)\rangle_+)$ and $\min(\langle\dot{\hat{n}}(0)\rangle_+)$ occur at $\varphi^*$ and $2\pi-\varphi^*$, respectively, and vice versa for $\max(\langle\dot{\hat{n}}(0)\rangle_-)$ and $\min(\langle\dot{\hat{n}}(0)\rangle_-)$. We verify this by plotting $\langle\dot{\hat{n}}(0)\rangle_\pm$ in \cref{fig:twoDrivesPanels}(f) (red and blue data), and find that their turning points correctly follow our prediction.

We contrast this observation with the deviation of the qubit purity at readout time from the ideal value one, $1-\Tr[\hat{\rho}^2_{\text{q}}(t_r)]$, plotted in \cref{fig:twoDrivesPanels}(c). This quantity is always minimized at $\varphi=0$, which never coincides with $\varphi^*$. In fact, at $\varphi=0$, $\langle\dot{\hat{n}}(0)\rangle=0$; see \cref{eq:rateOfChangeResonatorPhotons} and \cref{fig:twoDrivesPanels}(f). Thus, maximizing $\Tr[\hat{\rho}^2_{\text{q}}(t_r)]$ necessarily minimizes $|\langle\dot{\hat{n}}(0)\rangle|$.

For a geometric interpretation of why this must be the case, consider the circular paths traced by $\langle\hat{a}(t)\rangle_\pm$ in \cref{fig:twoDrivesPanels}(d), described to leading order by \cref{eq:amplitudeTwoDrivesSP}. The optimal setting $\varphi=0$ corresponds to maximizing the radius of the circular path. In this setting, $\alpha_0\parallel s$, so that the line connecting the center of the circle at $-s$ with the starting point of $\langle\hat{a}(t)\rangle_\pm$ at $\alpha_0$ always goes through the origin of the $xp$-plane.
Meanwhile, the circle of constant $N$ is always centered at the origin, with radius given by $N=|\langle\hat{a}(0)\rangle_\pm|^2$. Hence, the two circles---one describing the trajectory of $\langle\hat{a}(t)\rangle_\pm$, the other defined by constant $N$---touch tangentially at $\langle\hat{a}(0)\rangle_\pm$. This guarantees that $\langle\dot{\hat{n}}\rangle_\pm=0$ when $\varphi=0$.

The main conclusion of this section is that adding a second classical drive improves qubit readout, provided that the relative phase $\varphi$ is chosen such that $N_{\text{eff}}>N$. The optimal choice is the one that maximizes $N_{\text{eff}}$, i.e., $\varphi=0$ (see \cref{eq:effectiveTwoDriveAmplitude}). This setting maximizes the length of the two-drive vector $\langle\OmegaJCRvec\rangle$ at $t=0$, see \cref{fig:twoDrivesPanels}(a), and therefore minimizes the deviation of the qubit purity from one at readout time, as predicted by \cref{eq:QubitPurityReadoutTimeJCmodel} with $\Omega_{\text{JC}}\to\bar{\Omega}_{\text{JC}}$. Maximizing $\Neff$ also minimizes the measurement errors of the readout scheme, namely the errors in the measurement fidelity and post-measurement QNDness of the qubit, as will be shown in \cref{sec:Measurements}. In the dispersive regime, it additionally reduces the readout time, since $t_r\sim 2\abs{\Delta}/(g^2 \sqrt{N_{\text{eff}}})$ is inversely proportional to $N_{\text{eff}}$. Note that this effect is absent to leading order in the nondispersive regime, since the readout time $t_r\sim 4g^{-1}$ does not depend on $N_{\text{eff}}$.

We now proceed to the final section of the paper, related to the measurement process. We will show how projective measurements of the resonator state provide us with information on the initial state of the qubit. We will also demonstrate how, in the presence of a high enough number $N_{\text{eff}}$ of effective photons, the measurement is QND, even in the nondispersive regime.

\section{Measurement, Fidelity and QNDness}
\label{sec:Measurements}

\begin{figure*}
  \centering
  \includegraphics[width=.85\linewidth]{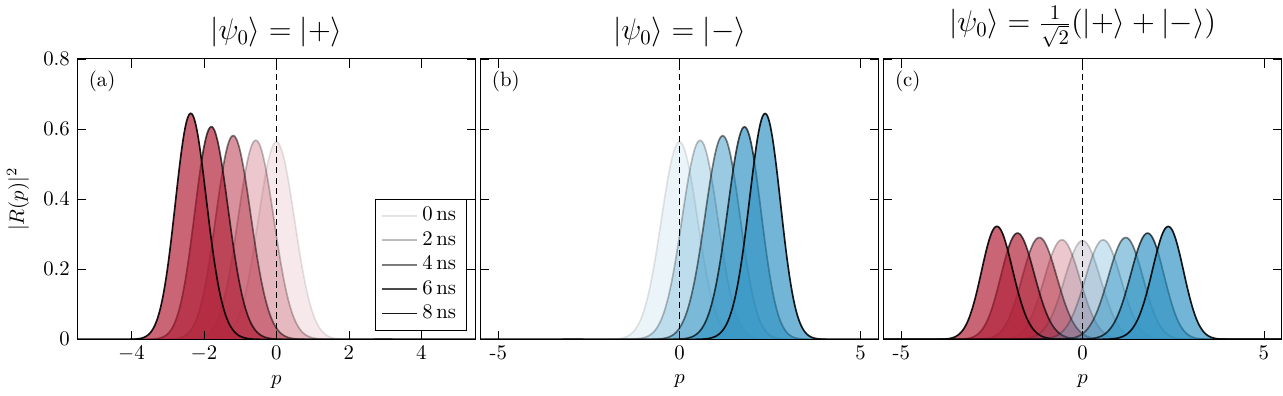}
  \caption{Early-time snapshots of the probability distributions $|R_{\pm}(p,t)|^2$, where $R_{\pm}(p,t)=(\bra{p}\otimes\bra{\pm(t)})\ket{\Psi}$, for three different initial qubit states $\ket{\psi_0}$: (a) $\ket{+}$, (b) $\ket{-}$, and (c) $\frac{1}{\sqrt{2}}(\ket{+}+\ket{-})$. $|R_{+}(p,t)|^2$ (red) and $|R_{-}(p,t)|^2$ (blue) both start as Gaussians centered at $p=0$, but over time move towards negative and positive values of $p$, respectively. 
  }
\label{fig:polarizationCavityMomentumProbabilityDistribution}
\end{figure*}

So far, we have discussed the dynamics of the qubit and resonator states in great detail, but what about the measurement process itself? In the bang-bang protocol, the qubit-resonator coupling is switched off at time $t_\textrm{off}$, and the resonator is subsequently measured. Typically, homodyne measurements are used to achieve this. These monitor some quadrature of the resonator field, $\hat{Y}=\frac{1}{2}(e^{-i\theta}\hat{a}+e^{i\theta}\hat{a}^{\dagger})$, where $\theta$ determines the measurement axis in the $xp$-plane. The optimal choice of $\theta$ aligns the measurement axis with the initial direction of separation of $\langle \hat{a}\rangle_{\pm}$, allowing the fastest distinction between the two corresponding photonic probability distributions. Instead of modeling the full homodyne measurement, here we employ a proxy: binary projective measurements of the appropriately chosen quadrature that allow us to make analytical progress. While this approach does not resolve the time it takes to perform the projective measurements, we expect it to closely reproduce the fidelity and QNDness of homodyne readout.

Without loss of generality, we can set both $\alpha_0$ and $\Omega_d$ to be real. Note that we implicitly assumed $\alpha_0$ and $\Omega_d$ have the same phase, $\arg(\Omega_d/\alpha_0)=0$, as this ensures optimal qubit purity at readout time for two given drive amplitudes $|\alpha_0|,|\Omega_d|$, see Sec.~\ref{sec:SecondDrive}, \cref{fig:twoDrivesPanels}(c). With this choice, $\langle\hat{a}\rangle_\pm$ initially separate in the $\pm p$-directions, see \cref{fig:wignerPlotsVaryNandTime}, so we set $\theta=\pi/2$, giving $\hat{Y}=\hat{p}$. The measurement operator $\hat{Y}$ therefore projects the resonator component of $\ket{\Psi(t)}$ onto an eigenstate of $\hat{p}$, $\ket{p}$, with probability $|\bra{p}\ket{\Psi}|^2$, and measurement outcome (eigenvalue) $p$. Repeating the measurement on the same initial state $\ket{\Psi_0}$ many times allows us to statistically reconstruct the probability distribution of outcomes, $|R(p)|^2$; see \cref{fig:polarizationCavityMomentumProbabilityDistribution} for some examples. Since the $|R(p)|^2$ distribution is uniquely \footnote{up to not knowing $\Delta\Phi$. To find out $\Delta \Phi$, a second set of experiments where the qubit is initially rotated in the plane perpendicular to the $\langle\hat{\vb{S}}(0)\rangle_+=\bra{+}\hat{\vb{S}}\ket{+}$ (or equivalently $\langle\hat{\vb{S}}(0)\rangle_-$) axis is required, as explained later in the text.} linked to the initial state of the qubit, we can therefore deduce the state of the qubit via measurements performed on the resonator.

A commonly used metric to estimate the readout time is the signal-to-noise ratio (SNR), for which we adopt the following definition 
\begin{equation}\label{eq:SNRdef}
    \text{SNR}(t)=\int_0^t\dd{t'}\sqrt{\frac{|\dot{\bar{p}}_+-\dot{\bar{p}}_-|^2}{\Delta p^2_++\Delta p^2_-}},
\end{equation}
where $\bar{p}_{\pm}=\bra{\Psi_{\pm}}\hat{p}\ket{\Psi_{\pm}}$ is the mean momentum, and $\Delta p_{\pm}^2=\bra{\Psi_{\pm}}\hat{p}^2\ket{\Psi_{\pm}}-\bar{p}_{\pm}^2$ is the spread of the probability distribution $|R(p)|^2$ in momentum space. Note that \cref{eq:SNRdef} differs slightly from another definition of SNR frequently encountered in the literature, which depends on the damping rate in the resonator \footnote{If we include the damping rate $\kappa$, the rate of information collected by the classical transmission line connected to the resonator is proportional to $\kappa$, and the SNR is defined as $\text{SNR}^2=\kappa\int_0^t\dd{t}'\frac{|\bar{p}_+-\bar{p}_-|^2}{\Delta p_+^2+\Delta p_-^2}$}. Here, we chose the simpler definition in \cref{eq:SNRdef}, as the analysis in this paper is performed in the absence of damping, but we will refer to the other definition in future work \cite{nonDispersiveReadoutFluxonium2026}.

The SNR quantifies how quickly the probability distributions of the resonator state become distinguishable for two systems where the qubit is initialized in $\ket{+}$ and $\ket{-}$, respectively. The readout time $t_r$ is defined as the time it takes for $\text{SNR}(t)\geq 4$. Substituting \cref{eq:amplitudeTwoDrivesSP} into $\bar{p}_{\pm}=\Im[\langle\hat{a}\rangle_{\pm}]$ yields $|\dot{\bar{p}}_+-\dot{\bar{p}}_-|=2\sqrt{N_{\text{eff}}}\bar{\omega}_s\cos(\bar{\omega}_s t)$, to leading order in $1/\sqrt{N_{\text{eff}}}$. Previously, we also showed that $\Delta p_{\pm}^2=1/2+2N_{\text{eff}}\bar{\omega}_s/\Omega_{\text{JC+R}}\sin^2(\bar{\omega}_s t)(1\mp\frac{\Delta}{\Omega_{\text{JC+R}}})^2$ to leading order, see \cref{eq:sumDeltaXsquaredDeltaPsquared,eq:differenceDeltaXsquaredDeltaPsquared} with $N\to N_{\text{eff}}$. Inserting these into \cref{eq:SNRdef}, Taylor expanding $\sin(\bar{\omega}_s t_r)\approx \bar{\omega}_s t_r$, and keeping only the leading order terms yields
\begin{equation}\label{eq:readoutTimeDef}
    t_r\approx\frac{2}{\sqrt{N_{\text{eff}}}\bar{\omega}_s}=2\sqrt{2}\bar{\gamma}_f^{-1},
\end{equation}
where $\bar{\gamma}_f$ is just $\gamma_f$ in \cref{eq:Timescales} with $N\to N_{\text{eff}}$. Using \cref{eq:readoutTimeDef}, the readout time takes the values
\begin{equation}
    t_r=\begin{cases}
    2(\chi\sqrt{N})^{-1}, & \Delta^2\gg 4g^2N,\\
    4g^{-1}, & \Delta^2\ll 4g^2N,
    \end{cases}
\end{equation}
where $\chi=g^2/\Delta$ is the dispersive shift, in the dispersive and nondispersive limits, respectively. Since in the dispersive limit, $2(\chi\sqrt{N})^{-1}\gg 4g\sqrt{N}/(g^2\sqrt{N})=4g^{-1}$, $t_r$ is much larger, and correspondingly readout is much slower, than in the nondispersive limit. In the nondispersive limit, $t_r$ is determined only by $g$, which can take values as high as $2\pi \times$\SI{900}{\mega\hertz} \cite{bosman2017}, although we used a more typical value of $2\pi \times$\SI{100}{\mega\hertz} \cite{houck2008} throughout this paper. This gives readout times on the order of 5--\SI{10}{\nano\second}, excluding the time required to unitarily prepare the qubit in the sweet-spot basis, projectively measure the resonator state after $g$ is switched off, and perform the unitary reset. Accounting for these additional times, we still expect a significant speed-up over reported state-of-the-art dispersive readout times which are typically $>$\SI{50}{\nano\second} \cite{stefanski2024,stefanski2411}. 

A short readout time alone is not sufficient: the measurement should also have a high measurement fidelity and ideally be QND, so that the qubit can be reliably reset (by fast unitaries, rather than cooling) and reused after the measurement. We remark that more sophisticated methods have recently been proposed that use full tomography to more accurately characterize fidelity, QNDness and an additional quantity known as destructiveness \cite{pereira2022}. However, in this work we restrict ourselves to two metrics: first, the Bhattacharyya fidelity, which measures how well the readout statistics realize the ideal Born distribution, and second, a definition of QNDness that accounts for the fact that our protocol includes a unitary reset operation. In dispersive readout, the measurement inherently satisfies these criteria --- but what happens as we move into the nondispersive regime? Let us first review what precisely constitutes a high-fidelity and QND measurement, and why dispersive readout satisfies these requirements, before proceeding to the less explored case of our bang-bang nondispersive readout protocol. 

As we set $\alpha_0,\Omega_d\in\mathbb{R}$, the initial photonic probability distribution $|R(p,0)|^2$ in momentum space is always centered on the $x$-axis. Over time, the probabilities $|R_{\pm}(p,t)|^2$ associated with initial qubit states $\ket{\pm}$ move to negative/positive values of $p$, see \cref{fig:polarizationCavityMomentumProbabilityDistribution}. Note that this remains true when the qubit is initialized in some superposition state,
\begin{equation}\label{eq:qubitInitialSuperpositionState}
    \ket{\psi_0}=c_+\ket{+}+c_-\ket{-};
\end{equation}
see, e.g., \cref{fig:polarizationCavityMomentumProbabilityDistribution}(c), where $c_\pm=1/\sqrt{2}$. This is due to linearity of unitary time evolution in quantum mechanics. For $t\gtrsim t_r$, the $|R_{\pm}(p,t)|^2$ probability distributions appear well-confined to negative/positive values of $p$, so we expect $\sign(p)$ to be a reliable indicator of the state of the qubit after the measurement of the resonator state. 

We now put these intuitive statements onto a more quantitative footing. Following the approach in, e.g., \cite{girvin2014}, we define the lower/upper half-plane projection operators
\begin{equation}
\begin{aligned}
\hat{P}_{<}&=\int_{-\infty}^0\dd{p}\ket{p}\bra{p}, &\hat{P}_{>}&=\int_0^{\infty}\dd{p}\ket{p}\bra{p}.
\end{aligned}
\end{equation}
$\hat{P}_{\gtrless}$ projects the resonator part of the full-system wavefunction $\ket{\Psi(t)}$ onto a momentum eigenstate $\ket{p}$ in the upper or lower momentum half-plane, and returns $p$ as the measured value. The probability of measuring $p\gtrless 0$ is $P_{\gtrless}=|\hat{P}_{\gtrless}\ket{\Psi}|^2=\bra{\Psi}\hat{P}_{\gtrless}\ket{\Psi}$, where we used $\hat{P}_{\gtrless}^2=\hat{P}_{\gtrless}$. The probability that the qubit is in the state $\ket{\sigma}$ after measuring $p\gtrless 0$ in the resonator is $P(\ket{\sigma}\,\cap\gtrless)=|\bra{\sigma}\hat{P}_{\gtrless}\ket{\Psi}|^2$.

We define two post-measurement density matrices for the qubit, corresponding to the two measurement outcomes $p\gtrless 0$ on the resonator state,
\begin{equation}\label{eq:postMeasurementDensityMatrix}
\begin{aligned}
    \hat{\rho}_{\text{q},\gtrless}&=\Tr_{\text{r}}[\hat{P}_{\gtrless}\ket{\Psi}\bra{\Psi}\hat{P}_{\gtrless}].
\end{aligned}
\end{equation}
The probability of measuring $p\gtrless 0$ is directly given by the trace of $\hat{\rho}_{\text{q},\gtrless}$, $P_{\gtrless}=\Tr[\hat{\rho}_{\text{q},\gtrless}]$. The probabilities $P(\ket{\sigma_{\parallel,\perp}}\,\cap\gtrless)$ are given by the diagonal entries of $\hat{\rho}_{\text{q},\gtrless}$ in some orthonormal basis $\{\ket{\sigma_\parallel},\ket{\sigma_\perp}\}$, where $\braket{\sigma_{i}}{\sigma_j}=\delta_{ij}$, with $i,j=\{\parallel,\perp\}$.

Using \cref{eq:postMeasurementDensityMatrix}, we define two figures of merit that characterize how close the readout scheme is to being ideal: measurement fidelity (here, Bhattacharyya fidelity) and QNDness. Fidelity is given by
\begin{equation}\label{eq:fidelityDef}
    \mathcal{F}=|c_{+}|\sqrt{P_{<}}+|c_{-}|\sqrt{P_{>}},
\end{equation}
and QNDness by 
\begin{equation}\label{eq:QNDnessDef}
    \text{QNDness}=P(\ket{+(t)}\,\cap <)+P(\ket{-(t)}\,\cap >).
\end{equation}
Fidelity compares how close the measured probabilities $\{P_<,P_>\}$ are to the ideal values $\{|c_+|^2,|c_-|^2\}$ describing the initial qubit state. The Cauchy-Schwarz inequality ensures that $\mathcal{F}\leq \sqrt{(|c_+|^2+|c_-|^2)(P_<+P_>)}=1$, with equality occurring only when $P_{\gtrless}=|c_{\mp}|^2$. The QNDness goes one step further and also quantifies the degree of definiteness in the qubit state after the resonator has been projectively measured. In the ideal case, measuring $p\gtrless 0$ projects the qubit into the states $\ket{\mp(t)}$ with probability $100\%$, so that $P(\ket{\mp(t)}\,\cap <)=P_\gtrless$, and $\text{QNDness}=1$. In reality, there is a finite probability that the qubit ends up in the wrong state. This means that, after the unitary reset operation, we cannot be $100\%$ certain that the qubit is back in its correct initial state. The QNDness quantifies this uncertainty. We remark that, in contrast to some common definitions of fidelity and QNDness in the literature, which only consider the computational states $\ket{\pm}$, \cref{eq:fidelityDef,eq:QNDnessDef} apply to a qubit in \emph{any} general superposition state, as defined in \cref{eq:qubitInitialSuperpositionState}.


Since the ideal values of fidelity and QNDness are one, the corresponding errors are defined as
\begin{equation}
\begin{aligned}
        \text{error}(\mathcal{F})&=1-\mathcal{F},\\
        \text{error}(\text{QNDness})&=1-\text{QNDness}.
\end{aligned}
\end{equation}

As a small aside, we note that the physical measurement we are modeling gives us access to $r_{\text{meas.}}=P_<-P_>$, which is defined in analogy to $r=|c_+|^2-|c_-|^2$. However, it does not provide us with any information on the phase difference, $\Delta\Phi=\arg(c_+/c_-)$. This is not surprising, as the best we can hope for is to precisely measure the projection along a single spin axis at a time---in this case, one of the (time-dependent) $\langle\hat{\vb{S}}(t)\rangle_\pm$-axes. Perfect knowledge of the qubit component along one axis necessarily means zero knowledge of the qubit components in the plane perpendicular to it, by the Heisenberg uncertainty principle and the fact that different spin components do not commute, $[\hat{S}^i,\hat{S}^j]=i\epsilon_{ijk}\hat{S}^k$. An additional measurement is therefore required to determine $\Delta \Phi$. The most common approach is to first apply a unitary transformation to rotate the qubit about one of the axes in the plane perpendicular to the initial two-drive vector $\langle\vb{\Omega}_{\text{JC+R}}\rangle$ (which is (anti)parallel to $\langle\hat{\vb{S}}(0)\rangle_\pm$), before repeating the experiment. This technique, known as quantum tomography \cite{altepeter2004}, allows us to distinguish between states of constant $r$ but different $\Delta\Phi$.

\textit{Fidelity and QNDness in the dispersive model---}Let us first calculate the post-measurement density matrices, fidelity and QNDness for the diagonal dispersive Hamiltonian in \cref{eq:dispersiveHamiltonian}, with the classical drive $\hat{H}_d(t)$ turned off, $\Omega_d=0$. In the rotating frame defined by \cref{eq:unitaryOpsRotFrame}, the time-evolved full-system wavefunction is 
\begin{equation}
\begin{aligned}
    \ket{\Psi_{\text{disp.}}(t)}&=\hat{U}_{\text{disp.}}(t)\ket{\Psi(0)},\quad \text{with}\\
    \hat{U}_{\text{disp.}}(t)&=e^{-i\hat{\tilde{H}}_{\text{disp.}} t}, \, \hat{\tilde{H}}_{\text{disp.}}=\hat{U}_{\text{JC}}\hat{H}_{\text{disp.}}\hat{U}_{\text{JC}}^{\dagger},
\end{aligned}
\end{equation}
and $\hat{\tilde{H}}_{\text{disp.}}$ defined in \cref{eq:dispersiveHamiltonian}. Parametrizing the initial state as $\ket{\Psi(0)}=(c_{\uparrow}\ket{\uparrow}+c_{\downarrow}\ket{\downarrow})\ket{\alpha_0}$, the time-evolved wavefunction in the dispersive limit reads
\begin{equation}\label{eq:timeEvolvedWavefunctionDispersiveLimit}
\begin{aligned}
    \ket{\Psi_{\text{disp.}}(t)}&=c_{\uparrow}(t)\ket{\uparrow}\ket{\alpha_{\uparrow}}+c_{\downarrow}(t)\ket{\downarrow}\ket{\alpha_{\downarrow}},
\end{aligned}
\end{equation}
where $c_{\uparrow/\downarrow}(t)=e^{\mp\frac{i(\Delta+\chi)}{2}t}c_{\uparrow/\downarrow}$ and $\alpha_{\uparrow/\downarrow}=\alpha_0e^{\mp i\chi t}$. \cref{eq:timeEvolvedWavefunctionDispersiveLimit} shows that the initial coherent resonator state $\ket{\alpha_0}$ splits into two coherent states $\ket{\alpha_{\uparrow/\downarrow}}=\ket{\alpha_0e^{\mp i\chi t}}$, which rotate in opposite directions due to their opposite time-dependent phases $\pm\chi t$.

For later comparison with the post-measurement density matrices, we first evaluate the reduced qubit density matrix for \cref{eq:timeEvolvedWavefunctionDispersiveLimit}. Tracing out the resonator degrees of freedom, $\hat{\rho}_{\text{q,disp.}}=\Tr_{\text{r}}[\ket{\Psi_{\text{disp.}}(t)}\bra{\Psi_{\text{disp.}}(t)}]$, and expressing the resulting reduced density matrix in the $\{\uparrow,\downarrow\}$ basis, we obtain
\begin{equation}
    \rho_{\text{q,disp.}}=\begin{pmatrix}
        |c_{\uparrow}|^2 & c_{\uparrow}(t)c^*_{\downarrow}(t)\bra{\alpha_{\downarrow}}\ket{\alpha_{\uparrow}} \\
        c^*_{\uparrow}(t)c_{\downarrow}(t)\bra{\alpha_{\uparrow}}\ket{\alpha_{\downarrow}} & |c_{\downarrow}|^2
    \end{pmatrix}.
\end{equation}
After a time $t_r\sim 2(\chi\sqrt{N})^{-1}$, most of the probability distribution associated with $\ket{\alpha_{\uparrow}}$/$\ket{\alpha_{\downarrow}}$ moves to the lower/upper half-planes in $xp$-space, with only an exponentially small tail in the ``wrong'' half-plane. In this limit, the $\bra{\alpha_{\uparrow}}\ket{\alpha_\downarrow}$ overlaps in the off-diagonal elements vanish, and $\rho_{\text{q,disp.}}$ tends to a diagonal matrix, $\frac{1}{2}(1+r\sigma^z)$. The qubit purity after switching off the coupling, $\Tr[\rho^2_{\text{q,disp.}}(t_\textrm{off} \gtrsim t_r)]\simeq\frac{1}{2}(1+r^2)$, therefore depends on $r$.

Next, we compare this with the post-measurement density matrices. Substituting \cref{eq:timeEvolvedWavefunctionDispersiveLimit} into \cref{eq:postMeasurementDensityMatrix} and expressing $\hat{\rho}_{\text{q,disp.,}\gtrless}$ in the $\{\ket{\uparrow},\ket{\downarrow}\}$ basis gives
\begin{equation}\label{eq:postMeasurementDensityMatrixDispersive}
    \rho_{\text{q,disp.,}\gtrless}=\begin{pmatrix}
    |c_{\uparrow}|^2\bra{\alpha_{\uparrow}}\hat{P}_{\gtrless}\ket{\alpha_{\uparrow}} &  c_{\uparrow}(t)c^*_{\downarrow}(t)\bra{\alpha_{\downarrow}}\hat{P}_{\gtrless}\ket{\alpha_{\uparrow}} \\
       c^*_{\uparrow}(t)c_{\downarrow}(t)\bra{\alpha_{\uparrow}}\hat{P}_{\gtrless}\ket{\alpha_{\downarrow}}  & |c_{\downarrow}|^2\bra{\alpha_{\downarrow}}\hat{P}_{\gtrless}\ket{\alpha_{\downarrow}}
    \end{pmatrix}.
\end{equation}
The $\bra{\alpha_{\uparrow}}\hat{P}_{\gtrless}\ket{\alpha_{\uparrow}}$ and $\bra{\alpha_{\downarrow}}\hat{P}_{\gtrless}\ket{\alpha_{\uparrow}}$ overlaps are straightforward to evaluate analytically, see \cref{eq:halfPlaneOverlapsDispersiveLimit} for the exact expressions. Substituting these into \cref{eq:postMeasurementDensityMatrixDispersive}, we see that $\rho_{\text{q,disp.,}\gtrless}=\frac{1}{2}\rho_{\text{q,disp.}}$ when $t_\textrm{off} =n\pi \chi^{-1},n\in\mathbb{Z}$. In the $t_\textrm{off} \gtrsim t_r$-limit, on the other hand, $\tilde{\rho}_{\text{q,disp,}\gtrless}(t_\textrm{off} \gtrsim t_r)\simeq\frac{1}{2}(1\mp\sigma^z)$. Hence, both post-measurement density matrices $\tilde{\rho}_{\text{q,disp,}\gtrless}(t_\textrm{off} \gtrsim t_r)$ are pure, and independent of $r$. We contrast this with the reduced density matrix for the qubit, $\rho_{\text{q,disp.}}(t_\textrm{off} \gtrsim t_r)$, whose purity was $r$-dependent. This is not surprising: having access to additional information about the system via the sign of the measured momentum $p$ improves our knowledge of the qubit state.

Next, we use \cref{eq:halfPlaneOverlapsDispersiveLimit} to evaluate the measurement fidelity and QNDness in the dispersive model. Inserting \cref{eq:halfPlaneOverlapsDispersiveLimit} into \cref{eq:postMeasurementDensityMatrixDispersive}, and using the definitions \cref{eq:fidelityDef,eq:QNDnessDef} with $c_{\pm}\to c_{\uparrow/\downarrow}$, we obtain
\begin{align}
        &\mathcal{F}_{\text{disp.}}=\frac{1}{2}\bigg(\sqrt{1+r}\sqrt{1+f r}+\sqrt{1-r}\sqrt{1-f r}\bigg),\label{eq:fidelityDispersive}\\
        &\text{QNDness}_{\text{disp.}}=\frac{1}{2}(1+f),\label{eq:QNDnessDispersive}
\end{align}
where $f=\erf(\sqrt{N}\sin(\chi t))$. Note that, while $\rho_{\text{q,disp.,}\gtrless}$ contains information about $\Delta\Phi$ via $c_{\uparrow}c_{\downarrow}^*$ in the off-diagonal terms, neither $\mathcal{F}_{\text{disp.}}$ nor $\text{QNDness}_{\text{disp.}}$ have any knowledge of $\Delta\Phi$, with only $r$ entering \cref{eq:fidelityDispersive,eq:QNDnessDispersive}.

\begin{figure*}
  \centering
  \includegraphics[width=.8\linewidth]{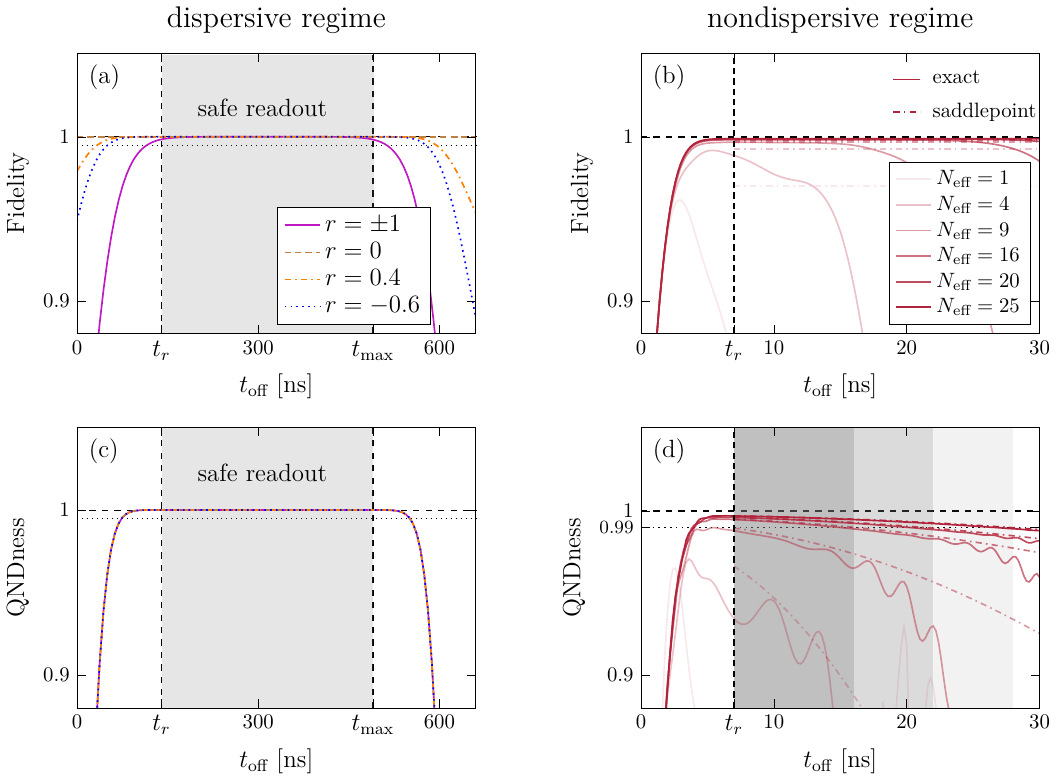}
  \caption{Time traces for fidelity and QNDness as a function of the switch-off time $t_{\text{off}}$ in the dispersive [\cref{eq:dispersiveHamiltonian}, panels (a), (c)] and nondispersive two-drive [\cref{eq:JCplusRabiModel}, panels (b), (d)] models, respectively. Parameters for panels (a),(c): $\chi/(2\pi)=$\SI{0.8}{\mega\hertz}, $\sqrt{N}=3$, for several different $r$. We directly plot the analytical formulas for fidelity [\cref{eq:fidelityDispersive}] and QNDness [\cref{eq:QNDnessDispersive}]. The dotted line denotes $\mathcal{F}=\text{QNDness}=0.995$. Parameters for panels (b),(d): $g/(2\pi)=$\SI{100}{\mega\hertz}, $\Delta/(2\pi)=$\SI{23}{\mega\hertz}, varying $\Neff$ with the qubit always initialized in $\ket{+}$ $(r=1)$. The exact data are plotted as solid lines, while the saddlepoint approximation formulas in \cref{eq:fidelityErrorNonDispersive,eq:qndnessErrorNonDispersive} (valid in the region $t_{\text{off}}\geq t_r$) are shown as dashdotted lines. The gray rectangular regions in panel (b) denote the regions of safe readout where QNDness remains $>0.99$. For the parameters chosen here, these exist for $\Neff\geq \Neff^*\approx 11 $ (see also \cref{fig:stereographicProjectionQNDnessVaryN,fig:fidelityQNDnessVaryRandDeltaPhi}), i.e., for the three largest values $\Neff=16,20,25$ (shaded dark gray, gray, and light gray, respectively). The positions of the right edges of the rectangles were calculated using \cref{eq:tcritQNDness}, with $\epsilon=0.01$.
  }
\label{fig:fidelityQNDnessFourPanels}
\end{figure*}

In \cref{fig:fidelityQNDnessFourPanels}(a) and (c), we plot \cref{eq:fidelityDispersive,eq:QNDnessDispersive}, respectively, as a function of switch-off time $t_\textrm{off}$ for several different values of $r$. At $t=0$, $f=0$, giving $\mathcal{F}_{\text{disp.}}(0)=\frac{1}{\sqrt{2}}(\sqrt{1+r}+\sqrt{1-r})\leq 1$ and $\text{QNDness}_{\text{disp.}}(0)=\frac{1}{2}$. Hence, the initial value of the fidelity varies as a function of $r$. The initial QNDness is always $\frac{1}{2}$, which makes sense because the center of the initial resonator coherent state always lies on the $x$-axis, such that exactly half of the probability distribution associated with each qubit eigenstate $\ket{\pm}$ lies in the ``wrong'' half-plane. As $t_\textrm{off}$ increases, the fidelity increases towards the ideal value $1$, or stays constant if $r=0$ (qubit initialized in the $S^xS^y$-plane). Meanwhile, the QNDness increases towards $1$ identically for all values of $r$.

The special value $r=0$, where $\mathcal{F}_{\text{disp.}}$ remains equal to the ideal value one at all times, is essentially a happy accident. To properly benchmark our readout scheme, we should instead use the values of $r$ that minimize $\mathcal{F}_{\text{disp.}}$. These are given by $r=\pm1$; see \cref{eq:fidelityDispersiveWorst} for the corresponding expression for minimal fidelity and a derivation of this result. In the $t_\textrm{off} \gtrsim t_r$ limit, \cref{eq:fidelityDispersiveWorst,eq:QNDnessDispersive} become very close to $1$, and we can use the large $z$-expansion of $\erf(z)$ \cite{erfcFormula} to calculate the errors $\text{error}(\text{QNDness}_{\text{disp.}})$ and $\max_{r}(\text{error}(\mathcal{F}_{\text{disp.}}))$. The leading-order error in the QNDness is twice as large the error in the fidelity, and given by
\begin{equation}\label{eq:errorFidelityQNDnessDispersive}
    \begin{aligned}
        \text{error}(\text{QNDness}_{\text{disp.}}(t_\textrm{off}))\simeq\frac{1}{2}\frac{e^{-A(t_\textrm{off})^2}}{\sqrt{\pi A(t_\textrm{off})^2}},
    \end{aligned}
\end{equation}
where $A(t)=\sqrt{N}\sin(\chi t)$. \cref{eq:errorFidelityQNDnessDispersive} takes its smallest value $e^{-N}/\sqrt{\pi N}$ at the maximum value of $A$, $A(\frac{\pi}{2}\chi^{-1})=\sqrt{N}$. However, there is no need to wait this long to achieve a satisfactory readout. Defining the acceptable fidelity and QNDness for safe readout to be $0.995$, or $99.5\%$ of the ideal value 1, and inverting \cref{eq:errorFidelityQNDnessDispersive}, we find that we need to wait a time $t_{\text{min}}=\chi^{-1}\arcsin(2N^{-1/2})\simeq t_r$. Waiting \emph{too} long to perform the measurement is also dangerous, as after time $t_\textrm{off} =\frac{\pi}{2}\chi^{-1}$, the probability distributions $|R_{\pm}(p,t)|^2$ start to rotate back towards the $x$-axis, eventually crossing over into the ``wrong'' half-planes at $t_\textrm{off} =\pi\chi^{-1}$. This results in dips in $\mathcal{F}_{\text{disp.}}$ and $\text{QNDness}_{\text{disp.}}$, which become problematic after $t_{\text{max}}\simeq\pi\chi^{-1}-t_r$. Thus, safe readout can be performed in the time interval $t_r\leq t_\textrm{off} \leq t_{\text{max}}$ indicated by the gray rectangles in Fig.~\ref{fig:fidelityQNDnessFourPanels}(a),(c).

\textit{Fidelity and QNDness in the nondispersive model---}Having gained some intuition from the dispersive model, we proceed to the calculation of post-measurement density matrices, fidelity, and QNDness in the nondispersive model. Similarly to \cref{eq:timeEvolvedWavefunctionDispersiveLimit}, we should express the full-system wavefunction $\ket{\Psi(t)}$ in a basis that maximizes the purity of the reduced density matrix $\hat{\rho}_{\text{q}}(t)$ when the qubit is initialized in one of the basis states. In the dispersive model, the $\{\uparrow,\downarrow\}$ basis ideally satisfies this requirement, since $\Tr[\hat{\rho}^2_{\text{q,disp.}}(t)]=1\,\forall t$ when the qubit is initialized in $\ket{\uparrow}$ or $\ket{\downarrow}$. In the nondispersive model, the analysis of the qubit-resonator dynamics in Sec.~\ref{sec:JCmodelDynamics} revealed that $\hat{\rho}_{\text{q}}(t)$ can never remain completely pure at finite $t$. However, initializing the qubit in one of the two $\ket{\pm}$ states defined in \cref{eq:RabiModelEigenbasis} (with $\thetaR\to\theta_{\text{JC+R}}$, $\phiR\to\phi_{\text{JC+R}}$) results in $\Tr[\hat{\rho}^2_{q,\pm}(t_r)]$ within $\mathcal{O}(\Neff^{-1})$ of the ideal value $1$, see \cref{eq:QubitPurityReadoutTimeJCmodel}. Unlike the static $\ket{\uparrow,\downarrow}$ basis states in the dispersive model, the $\ket{\pm_{\parallel}(t)}$ states in the nondispersive model do not stay constant over time. Instead, they point along the time-dependent Bloch vector in a system initialized with the qubit in $\ket{\pm}$: $\bra{\pm_{\parallel}(t)}\hat{\sigma}^i\ket{\pm_{\parallel}(t)}=\langle\hat{S}^i(t)\rangle_\pm/|\langle\hat{\vb{S}}(t)\rangle_\pm|$. The normalization on the right hand side accounts for the fact that the length of $\langle\hat{\vb{S}}(t)\rangle_\pm$ decreases over time; see \cref{fig:JCvsRabiPurityXYplaneTrace}(d). We parametrize $\ket{\pm(t)}$ as
\begin{equation}\label{eq:timeDependenceNonDispersiveBasisStates}
    \ket{\pm_{\parallel}(t)}=\ket{\Theta_\pm(t),\Phi_\pm(t)},
\end{equation}
where $\ket{\Theta,\Phi}=\cos(\Theta/2)\ket{\uparrow}+\sin(\Theta/2)e^{i\Phi}\ket{\downarrow}$. The time-dependent polar and azimuthal angles are given by 
\begin{equation}\label{eq:ThetaPhiPlusMinusDef}
    \begin{aligned}
\Theta_\pm(t)&=\arctan(\sqrt{\langle\hat{S}^x(t)\rangle_\pm^2+\langle\hat{S}^y(t)\rangle_\pm^2}/\langle\hat{S}^z(t)\rangle_\pm),\\       \Phi_\pm(t)&=\arctan(\langle\hat{S}^y(t)\rangle_\pm/\langle\hat{S}^x(t)\rangle_\pm),
    \end{aligned}
\end{equation}
see also \cref{fig:anglesQubitSweetSpot} for plots of $\cos(\Theta_{\pm}(t))$ and $\dot{\Phi}_\pm(t)$ as a function of time. Although $\ket{\pm_{\parallel}(t)}$ are orthogonal to each other at $t=0$,
they stop being orthogonal as $t$ increases, because $\dot{\Phi}_\pm\simeq \mp\omega_s$ have opposite signs. More precisely, their inner product oscillates at the slow frequency $\tilde{\omega}_s$, $\bra{-_{\parallel}(t)}\ket{+_{\parallel}(t)}=i\sin(\tilde{\omega}_s t)e^{-i\tilde{\omega}_s t}\sin(\theta_{\text{JC+R}})$, to leading order in $1/\sqrt{\Neff}$. Thus, we need to define two separate orthonormal bases for the two measurement outcomes $p\gtrless 0$: $\{\ket{\mp_{\parallel}(t)},\ket{\mp_{\perp}(t)}\}$. By construction, the perpendicular states $\ket{\pm_{\perp}(t)}=\ket{\pi-\Theta_\pm(t),\pi+\Phi_\pm(t)}$ are orthogonal to $\ket{\pm_{\parallel}(t)}$ at all times, $\bra{\pm_{\perp}(t)}\ket{\pm_{\parallel}(t)}=0$.
Expressed in terms of these basis vectors, $\ket{\Psi(t)}$ reads
\begin{equation}\label{eq:timeEvolvedWavefunctionNonDispersiveLimit}
\begin{aligned}
    \ket{\Psi(t)}&=c_+\ket{+,R_+(t)}+c_-\ket{-,R_-(t)},
\end{aligned}
\end{equation}
where
\begin{equation}\label{eq:ResonatorPMstate}
    \begin{aligned}
        \ket{\pm,R_{\pm}(t)}=\ket{\pm_{\parallel}(t)}\ket{R_{\pm,\parallel}(t)}+\ket{\pm_\perp(t)}\ket{R_{\pm,\perp}(t)}.
    \end{aligned}
\end{equation}
The resonator wavefunctions $\ket{R_{\pm,\parallel,\perp}(t)}$ are obtained by setting $c_+$ or $c_-$ to zero and projecting $\bra{\pm_{\parallel,\perp}(t)}$ onto $\ket{\Psi(t)}$ in \cref{eq:timeEvolvedWavefunctionNonDispersiveLimit}; see \cref{eq:resonatorPlusMinusParallelPerpKets} for the resulting exact expressions. Physically, $\ket{\pm_{\parallel}(t)}$ corresponds to the ``right'' qubit state (the one in which we expect the qubit to end up if the projective measurement of the resonator state is ideal), while $\ket{\pm_{\perp}(t)}$ corresponds to the ``wrong'' state (the one leading to loss in QNDness). Taylor expanding $\ket{R_{\pm,\parallel,\perp}(t)}$ around the saddlepoint value $m=\Neff$, we obtain
\begin{equation}\label{eq:TaylorExpandedResonatorStates}
\begin{aligned}
    \ket{R_{\pm,\parallel}(t)}&=e^{\mp i\bar{\Omega}_{\text{JC}}t/2}\ket{\alpha_{\pm}(t)} + \mathcal{O}(\Neff^{-1/2}),\\
    \ket{R_{\pm,\perp}(t)}&=\mathcal{O}(\Neff^{-1/2}),
\end{aligned}
\end{equation}
where $\alpha_{\pm}(t)=\sqrt{\Neff}e^{\pm i\bar{\omega}_s t}$ is the time-dependent amplitude of the almost-coherent resonator state. One can also verify \cref{eq:TaylorExpandedResonatorStates} by checking that the Schr{\"o}dinger equation $i\partial_t\ket{\Psi(t)}=\hat{\tilde{H}}_{\text{JC+R}}\ket{\Psi(t)}$ is solved by
\begin{equation}
    \ket{\Psi}=e^{\mp i\bar{\Omega}_{\text{JC}}t/2}\ket{\pm_{\parallel}}\ket{\alpha_\pm(t)},
\end{equation}
 to leading order in $1/\sqrt{\Neff}$, with eigenenergy $\pm\frac{1}{2}\bar{\Omega}_{\text{JC}}$. Since $\ket{R_{\pm,\perp}(t)}$ is suppressed by a factor $1/\sqrt{\Neff}$ relative to $\ket{R_{\pm,\parallel}(t)}$, in the large-$\Neff$ limit we obtain a perfect product state $\lim_{\Neff\to\infty}\ket{\pm,R_{\pm}(t)}=\ket{\pm_{\parallel}(t)}\ket{R_{\pm,\parallel}(t)}$.

Let us compare the expansions of $\ket{\Psi(t)}$ in the dispersive and nondispersive models, \cref{eq:timeEvolvedWavefunctionDispersiveLimit} and \cref{eq:timeEvolvedWavefunctionNonDispersiveLimit}, respectively. In the dispersive model, $\ket{\pm,R_{\pm}(t)}=e^{\mp i\tilde{\Delta}t/2}\ket{\uparrow/\downarrow}\ket{\alpha_{\uparrow/\downarrow}}$ remains a product state at all times. In contrast, in the nondispersive model, at finite $t$ and moderately large $\Neff$, it becomes an entangled state, due to the unavoidable presence of the parasitic $\ket{\pm_\perp(t)}\ket{R_{\pm,\perp}(t)}$ terms. This is reflected in the fact that, for $t>0$, the purity $\Tr[\hat{\rho}^2_{\text{q}}]$ inevitably drops below 1 in the nondispersive model, even when the qubit is initialized in $\ket{\pm}$, as was shown in \cref{fig:JCvsRabiPurityXYplaneTrace}(d). 

We now evaluate the post-measurement density matrices $\hat{\rho}_{\text{q,}\gtrless}$ in the nondispersive model at the switch-off time $t_\textrm{off}$. We use the same definition as in \cref{eq:postMeasurementDensityMatrix}, except that we express $\hat{\rho}_{\text{q,}\gtrless}$ in the time-dependent qubit bases $\{\ket{\mp_{\parallel}(t)},\ket{\mp_\perp(t)}\}$ instead of the fixed basis $\{\uparrow,\downarrow\}$. The resulting exact components of $\rho_{\text{q,}\gtrless}$, are listed in \cref{eq:overlapHalfPlaneRmatComponents,eq:overlapHalfPlaneRmatComponentsBreakdown}. Remarkably, the saddlepoint approach can also be applied in this case to derive analytical expressions for all these quantities, see App.~\ref{app:PostMeasurementDensityMatricesSPapproach} for technical details. The saddlepoint expressions for the components of the post-measurement density matrices $\rho_{\text{q},\gtrless}$ in the two-drive nondispersive model are given in \cref{eq:PostMeasurementDensityMatrixComponentsSaddlepoint}.

To leading order $\mathcal{O}(1)$, the post-measurement density matrices $\rho_{\text{q},\gtrless}=\frac{1}{4}(1\mp r)(1+\sigma^z)$ are diagonal and pure. In fact, the leading order contributions correspond exactly to what we would expect for an ideal readout scheme, $\rho_{\text{q},\gtrless,\text{ideal}}=\frac{1}{2}|c|_\mp^2(1+\sigma^z)$. The higher order $\mathcal{O}(\Neff^{-k})$, $k\geq \frac{1}{2}$ terms describe the imperfections in our nondispersive readout scheme, with the leading order imperfection being of order $\mathcal{O}(\Neff^{-1})$ (brown terms in \cref{eq:PostMeasurementDensityMatrixComponentsSaddlepoint}).

 \begin{figure}
  \centering
  \includegraphics[width=.9\linewidth]{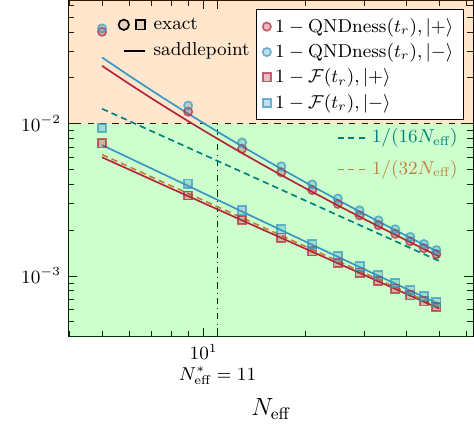}
  \caption{Deviation in QNDness and fidelity from the ideal value $1$ choosing the switch-off time to equal the readout time, $t_\textrm{off} = t_r=$\SI{7}{\nano\second} in the nondispersive two-drive model \cref{eq:JCplusRabiModel}, for a qubit initialized in $\ket{\pm}$, as a function of $\Neff$ (defined in \cref{eq:effectiveTwoDrivePhotonNumber}). Polygons denote exact data, while smooth lines are plotted using the saddlepoint expressions in \cref{eq:qndnessErrorNonDispersive,eq:fidelityErrorNonDispersive}. The dashed sloped lines denote the asymptotic errors in $\text{QNDness}(t_r)$, $1/(16\Neff)$ (teal), and $\mathcal{F}(t_r)$, $1/(32\Neff)$ (brown). The dashed horizontal line delimits regions where the error is larger (orange) or smaller (green) than $1\%$. The dashdotted vertical lines shows the critical value $\Neff^*$ at which the error in QNDness$(t_r)$ drops below $1\%$ for these parameters (see also \cref{fig:stereographicProjectionQNDnessVaryN}(c)). Parameters: $g/2\pi=$\SI{100}{\mega\hertz}, $\Delta/2\pi=$\SI{23}{\mega\hertz}. }
\label{fig:fidelityQNDnessLossReadoutTimeVaryN}
\end{figure}

\begin{figure*}
  \centering
  \includegraphics[width=.85\linewidth]{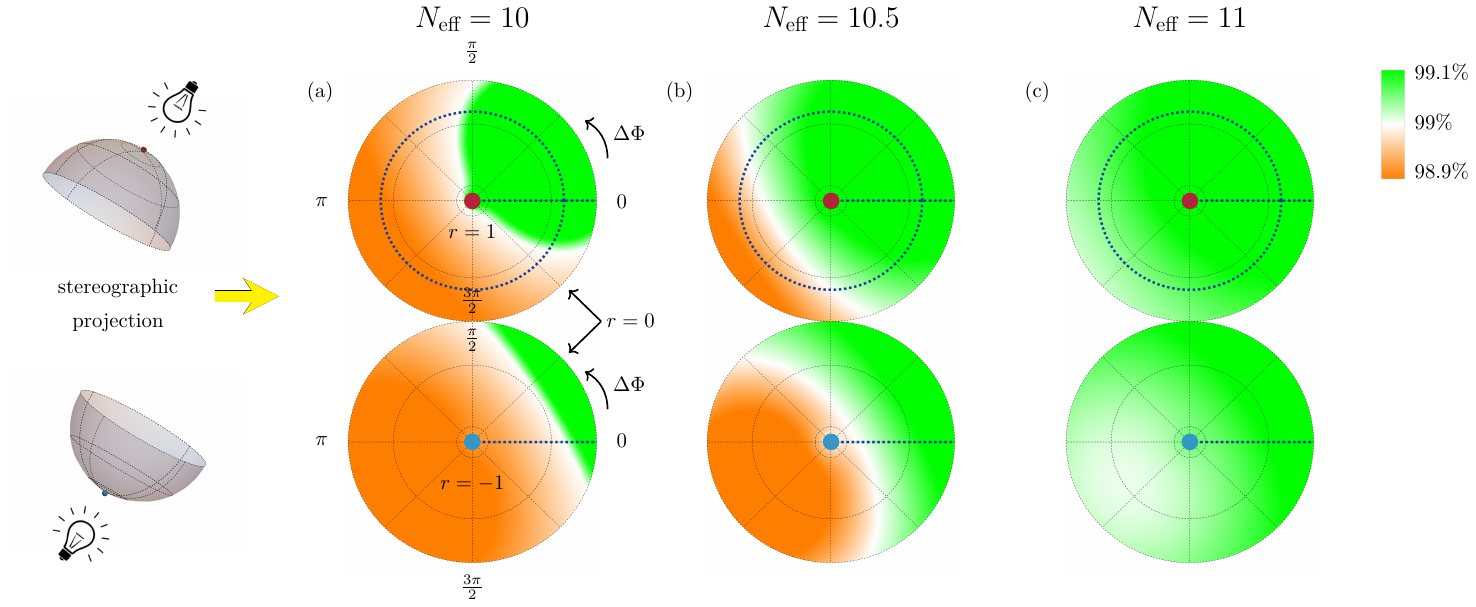}
  \caption{QNDness$(t_r)$ for all possible initial qubit orientations $\ket{\psi_0}=c_+\ket{+}+c_-\ket{-}$ stereographically projected onto the 2D plane. The upper and lower rows are the stereographically projected Bloch hemispheres contained within $0\leq r\leq 1$ and $-1\leq r\leq 0$, respectively. Red and blue dots indicate a qubit initialized in the $\ket{\pm}$ ($r=\pm 1$) states, respectively. Dotted circles are lines of constant $r$; dotted radial lines are lines of constant $\Delta\Phi$. Stereographic projection of a sphere onto a plane necessarily contains some redundancies; in this case, the redundant points are those on the circumferences of the circles in the upper and lower rows. Increasing $N_{\text{eff}}$ improves the QNDness$(t_r)$, with $99\%$ cleared by all initial qubit orientations at $N_{\text{eff}}^*\approx 11$ for the given parameters. Parameters $g,\Delta$ as in \cref{fig:fidelityQNDnessLossReadoutTimeVaryN}. See also \cref{fig:stereographicProjectionFidelityVaryN} for stereographic projection plots of $\mathcal{F}(t_r)$.
  }
\label{fig:stereographicProjectionQNDnessVaryN}
\end{figure*}

 \begin{figure*}
  \centering
  \includegraphics[width=.6\linewidth]{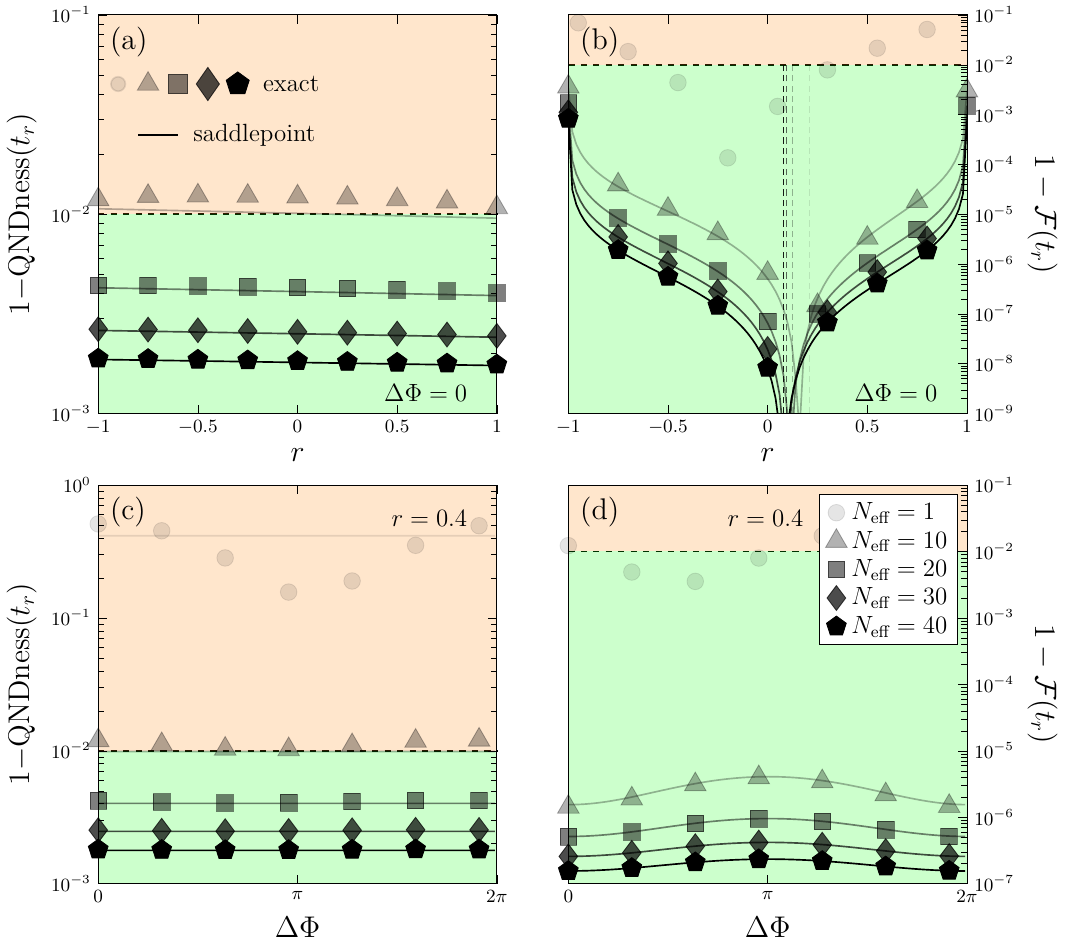}
  \caption{Errors in QNDness and fidelity at readout time, $1-\text{QNDness}(t_r)$ and $1-\mathcal{F}(t_r)$, as a function of $r$, $\Delta\Phi$ (defined in \cref{eq:cPlusMinusParametrization}), for several $\Neff$. Polygons show exact data, solid lines refer to the saddlepoint formulas \cref{eq:qndnessErrorNonDispersive,eq:fidelityErrorNonDispersive}. Panels (a)--(b): dataset described by the flat blue dotted lines in \cref{fig:stereographicProjectionQNDnessVaryN}, where $\Delta\Phi=0$ is kept constant and $r$ is varied in the range $-1\leq r\leq +1$. The dashed vertical lines in panel (b) correspond to $r^*$, defined in \cref{eq:rStar}, where $1-\mathcal{F}(t_r)$ vanishes to leading order $\mathcal{O}(1/\Neff)$. Panels (c)--(d): dataset described by the circular dashed blue paths in \cref{fig:stereographicProjectionQNDnessVaryN}, where $r=0.4$ is kept constant and $0\leq \Delta\Phi\leq 2\pi$ is varied. Note the minimum fidelity is $10^{2}$ larger than in panel (b) because the data range does not include $r^*$. Parameters $g,\Delta$ as in \cref{fig:fidelityQNDnessLossReadoutTimeVaryN}.
  }
\label{fig:fidelityQNDnessVaryRandDeltaPhi}
\end{figure*}

We evaluate the fidelity and QNDness in the nondispersive, two-drive model by substituting \cref{eq:PostMeasurementDensityMatrixComponentsSaddlepoint} into the definitions \cref{eq:fidelityDef,eq:QNDnessDef}. The resulting saddlepoint expressions are given in \cref{eq:fidelityErrorNonDispersive,eq:qndnessErrorNonDispersive}, accurate to $\mathcal{O}(\Neff^{-2})$. In the large $\Neff$-limit, the leading order error in the QNDness at readout time is 
\begin{equation}\label{eq:errorNonDispersive}
\begin{aligned}
    \text{max}_{r,\Delta\Phi} (\text{error}(\text{QNDness}(t_r)))&=\frac{1}{4}\frac{\bar{\omega}_s}{\bar{\Omega}_{\text{JC}}}\left(1+\frac{|\Delta|}{\bar{\Omega}_{\text{JC}}}\right)^2 \\
    &+\mathcal{O}(\Neff^{-2}),
\end{aligned}
\end{equation} 
once again exactly twice as large as the maximum error in the fidelity, $\max_{r,\Delta\Phi}(\text{error}(\mathcal{F}(t_r)))$ (first line in \cref{eq:fidelityErrorNonDispersive}), as well as one quarter of the leading order loss in qubit purity (\cref{eq:errorNonDispersive} with the sign chosen to match $\sign(\Delta)$). Here, $\max_{r,\Delta\Phi}(\dots)$ stands for finding the maximum error over all possible initial qubit states parametrized by $r,\Delta\Phi$. This corresponds to $r=-\sign(\Delta)$ for both the fidelity and the QNDness in the large $\Neff$-limit, but takes other values of $r,\Delta \Phi$ for intermediate $\Neff$; see \cref{fig:fidelityQNDnessVaryRandDeltaPhi}(a)--(d).

In both the dispersive and the nondispersive readout schemes, the error decreases with increasing (effective) photon number, scaling as $e^{-N}/\sqrt{N}$ and $1/\Neff$; see \cref{eq:errorFidelityQNDnessDispersive,eq:errorNonDispersive}, respectively. This stems from the fact that a higher number of initial (effective) photons makes the system more classical, hence reducing the amount of entanglement built up between the qubit and resonator. Since entanglement leads to leakage into the parasitic orthogonal states $\ket{\pm_{\perp}(t)}$, minimizing the entanglement also minimizes the measurement error of the readout scheme. Although $e^{-N}/\sqrt{N}$ decays faster than $1/\Neff$, very good QNDness values can still be achieved in the nondispersive readout scheme at experimentally realistic values of $\Neff$. For example, using the typical parameters $g/2\pi=$\SI{100}{\mega\hertz}, $\Delta=$\SI{23}{\mega\hertz} we adopted throughout this paper, $\text{QNDness}(t_r)>99\%$ is cleared for all initial qubit orientations for $\Neff\approx 11$; see \cref{fig:stereographicProjectionQNDnessVaryN}.

In \cref{fig:fidelityQNDnessFourPanels}(b),(d), we plot the fidelity and QNDness of the nondispersive two-drive model, \cref{eq:JCplusRabiModel}, as a function of the switch-off time $t_\textrm{off}$, for a qubit always initialized in the $\ket{+}$ state. We plot the exact data as solid lines and the saddlepoint formulas for fidelity and QNDness, \cref{eq:fidelityErrorNonDispersive,eq:qndnessErrorNonDispersive}, evaluated at and after $t_\textrm{off}=t_r$, as dashdotted lines. The readout time $t_r\approx $\SI{7}{\nano\second} is more than 10 times faster than the \SI{130}{\nano\second} value in dispersive readout, see \cref{fig:fidelityQNDnessFourPanels}(a),(c). Increasing the effective number of photons $N_{\text{eff}}$, either by populating the resonator with more photons or by applying a stronger classical drive, improves both $\mathcal{F}(t_r)$ and $\text{QNDness}(t_r)$.

One notable difference between the dispersive and nondispersive models arises in the QNDness at $t_\textrm{off}\gtrsim t_r$: while this stays essentially constant until $t_\textrm{off}\simeq t_{\text{max}}$ in the dispersive model, see \cref{fig:fidelityQNDnessFourPanels}(c), in the nondispersive model it worsens over time at an increasing rate, see \cref{fig:fidelityQNDnessFourPanels}(d). This effect is not described by the leading order $\mathcal{O}(\Neff^{-1})$ terms in the saddlepoint solution, which are constant (brown terms in \cref{eq:qndnessErrorNonDispersive}). Rather, it is captured at $\mathcal{O}(\Neff^{-2})$ by the $-4\Neff^2\bar{\omega}_s^5t^2/\bar{\Omega}_{\text{JC}}^3$ term, which grows quadratically in time (first orange term in \cref{eq:qndnessErrorNonDispersive}). As a result, the QNDness drops below an acceptable error threshold $\epsilon$ after a critical time
\begin{equation}\label{eq:tcritQNDness}
    t_{\text{crit}}=\left[\frac{\bar{\Omega}_{\text{JC}}^3}{4\Neff^2\bar{\omega}_s^5}\left(\epsilon-\frac{1}{4}\frac{\bar{\omega}_s}{\bar{\Omega}_{\text{JC}}}\left(1+\frac{|\Delta|}{\bar{\Omega}_{\text{JC}}}\right)^2\right)\right]^{1/2},
\end{equation}
where we used the leading order $\mathcal{O}(\Neff^{-1})$ error stated in \cref{eq:errorNonDispersive}. $t_{\text{crit}}$ scales as $\Neff$, whereas $t_{\text{max}}$ (the time it takes for $|R_\pm(p)|^2$ to complete half a revolution and cross into the ``wrong'' momentum half planes) scales as $\sqrt{\Neff}$. Hence, in the large $\Neff$ limit, the drop in QNDness due to the $-8\Neff^2\bar{\omega}_s^5/\bar{\Omega}_{\text{JC}}^3 t^2$ terms becomes negligible on the relevant time scale, and the measurement is QND.

The saddlepoint formula for the fidelity in the nondispersive model, \cref{eq:fidelityErrorNonDispersive}, on the other hand, contains only constant $\mathcal{O}(\Neff^{-1})$ and $\mathcal{O}(\Neff^{-2})$ terms. Hence, $\mathcal{F}(t\gtrsim t_r)$ stays essentially flat until $t_{\text{max}}$ in the nondispersive regime, just like in the dispersive one, see \cref{fig:fidelityQNDnessFourPanels}(a),(b). Note that $t_{\text{max}}=\pi\bar{\omega}_s^{-1}-t_r$ grows like $\sqrt{\Neff}$, due to $\bar{\omega}_s^{-1}\sim \sqrt{\Neff}$. This explains why in \cref{fig:fidelityQNDnessFourPanels}(b), the fidelity stays high and constant for longer at larger values of $\Neff$.


Next, we focus our attention on the errors in the QNDness and fidelity of our nondispersive readout scheme, setting the switch-off time equal to the readout time $t_\textrm{off} = t_r$. In \cref{fig:fidelityQNDnessLossReadoutTimeVaryN}, we plot $1-\text{QNDness}(t_r)$ and $1-\mathcal{F}(t_r)$ for a qubit initialized in $\ket{\pm}$ as a function of $\Neff$. We see that the saddlepoint expressions (\cref{eq:qndnessErrorNonDispersive,eq:fidelityErrorNonDispersive} with $r=\pm 1$, solid lines) are a good approximation to the exact data (circles and squares) for $\Neff\gtrsim 7$. Both errors scale as $1/\Neff$ at large $\Neff$, replicating the same scaling we saw previously in the loss of qubit purity; see \cref{fig:qubitPuritySweetSpot} and \cref{eq:QubitPurityReadoutTimeJCmodel}. 

For qubit readout, we care not only about the errors in QNDness$(t_r)$ and $\mathcal{F}(t_r)$ for a qubit initialized in $\ket{\pm}$, but also about the errors for any superposition state, $\ket{\psi_0}=c_+\ket{+}+c_-\ket{-}$. To visualize these, we perform a stereographic projection of the Bloch sphere onto the 2D plane. Using $\ket{\pm}$ $(r=\pm 1)$ as the North and South poles, we split the Bloch sphere in half along the equator ($r=0$), resulting in two hemispheres. We project each hemisphere onto a disk, mapping the $\ket{\pm}$ point to the center of the disk and the points on the equator to its circumference. \cref{fig:stereographicProjectionQNDnessVaryN} shows the stereographically projected QNDness$(t_r)$, for three increasing values $\Neff=10,10.5,11$. We set the acceptable QNDness at readout time to be $99\%$. The white regions denote the threshold, while the orange and green regions indicate values below and above the threshold, respectively. For the parameters used, a critical value $\Neff^*=11$ is required to clear the threshold for all initial qubit states. 

The equivalent stereographic projection plots for $\mathcal{F}(t_r)$ are shown in \cref{fig:stereographicProjectionFidelityVaryN}. Note the different distribution of orange and green regions there compared to \cref{fig:stereographicProjectionQNDnessVaryN}: the worst fidelity is concentrated near the poles, whereas the worst QNDness occurs in a region closer to the equator. In addition, a lower $\Neff^*=5$ (roughly a factor of two smaller, reflecting the scaling of the errors; see text under \cref{eq:errorNonDispersive}) is required for $\mathcal{F}(t_r)$ to cross the $99\%$ threshold for all initial $c_{\pm}$, or equivalently, $r$ and $\Delta\Phi$. 

Finally, \cref{fig:fidelityQNDnessVaryRandDeltaPhi} show scans of QNDness$(t_r)$, $\mathcal{F}(t_r)$ as functions of $r$ (keeping $\Delta\Phi$ fixed) and $\Delta\Phi$ (keeping $r$ fixed), corresponding to the dark blue circles (upper row) and lines (both rows) in \cref{fig:stereographicProjectionQNDnessVaryN,fig:stereographicProjectionFidelityVaryN}, respectively. There, we also explicitly demonstrate the excellent agreement between the exact data and the saddlepoint formulas \cref{eq:fidelityErrorNonDispersive,eq:qndnessErrorNonDispersive} for all $r,\Delta\phi$ in the $\Neff\gtrsim 7$-limit.

\section{Conclusion and outlook}
\label{sec:conclusion}

In this work, we introduced a quenched ``bang-bang'' readout protocol. It is based on a resonator preloaded with photons, and sudden switch-on and switch-off of the qubit-resonator coupling $g$. We showed that the bang-bang protocol performs a fast and QND readout of a qubit in the nondispersive regime ($4g^2N/\Delta^2\gg 1$), notably without adiabatically ramping to a dressed basis. Using typical values $g/2\pi\sim$\SI{100}{\mega\hertz} and $\Delta/2\pi\sim$\SI{23}{\mega\hertz}, the readout time $t_r$ is expected to be less than \SI{10}{\nano\second}, with QNDness$(t_r)$ reaching values above $99\%$ for an effective photon number $\Neff\gtrsim 11$. 
A key difference to dispersive readout and protocols that employ adiabatic ramps is that the qubit state after measurement is not in a computational state $\ket{\uparrow,\downarrow}$, but rather in a time-dependent superposition state $\ket{\pm(t)}$.
Crucially, the qubit post-measurement is in a close-to-definite state, as captured by the high values of the QNDness. Thus, the qubit can be reset and reused in other operations with a high degree of precision. In the present case, the reset operation  to leading order consists of a unitary rotation about the $S^z$-axis, clockwise or anticlockwise depending on the measurement outcome in the resonator, by an angle $\Phi_\pm$. Timescales for such rotations are typically on the order of \SI{10}{\nano\second} \cite{nguyen2022}, i.e., on the same order of magnitude as $t_r$. 
 
We developed a full analytical theory based on a combination of the saddlepoint approximation for large $N$ and perturbation theory in $1/\sqrt{N}$ to model the dynamics of the qubit-resonator system and calculate the measurement errors of the readout scheme in the JC model. We verified the analytical results numerically through simulations of the unitary time evolution of the qubit-resonator system. We found that it is almost always sufficient to stop at order $1/N$ to obtain asymptotically correct analytical expressions for important quantities such as loss in qubit purity and errors in measurement fidelity and QNDness. Some features, such as total quadrature spread of the coherent state in the resonator and the slow $\sim t^2$ decay in qubit purity and QNDness, require going up to order $N^{-2}$. 

We also found that driving the qubit with a classical field in the appropriate way (i.e., matching the phase of the coherent state populating the resonator) is analogous to increasing the effective number of photons in the system. Thus, the measurement errors of the readout scheme can be suppressed either by populating the resonator with a coherent state of larger amplitude or by driving the qubit with a classical field. 

While providing a theoretical framework for understanding nondispersive qubit readout, our work does not yet capture all the experimental complexities encountered in real superconducting circuit systems. Within the two-level qubit approximation, the JC and Rabi models used here do not capture the counter-rotating (CR) terms which automatically result when coupling a qubit capacitively to a resonator or to an external flux drive. It has recently been shown that, for a classical drive, these effects can be mitigated by carefully engineering the drive using a combination of capacitative and flux drives \cite{rower2024}. However, it is less obvious how to do this in a single readout resonator. Moreover, while we expect that dispensing with adiabatic ramps will avoid resonances and alleviate the problem of MIST, we have so far not shown that resonances can be fully avoided after the coupling quench. This requires a careful study that goes beyond the two-level qubit approximation for realistic quantum devices.

\begin{acknowledgments}
The authors acknowledge discussions with M. Lin and M. Kolodrubetz. NdS thanks A. Almanakly, W. Oliver, I. Esin, M. Rizzi, S. Trebst, C. Artiaco, and A. Rosch for thought-provoking questions and comments. NdS is supported by a DFG Walter Benjamin stipend, project number 527528104. GR acknowledges support from the AFOSR MURI program under agreement number FA9550-22-1-0339, as well as to the Simons Foundation and the Aspen Center for Theoretical Physics. The authors acknowledge support from the Institute for Quantum Information and Matter, an NSF Physics Frontiers Center (NSF Grant PHY-2317110).

\end{acknowledgments}

\section*{Author Contributions}
NdS developed the analytical theory and wrote the manuscript. YC and JS performed numerical simulations. All authors contributed to the conception of the project.

\bibliography{Library}

\clearpage 

\appendix

\onecolumngrid

\begin{table}\label{table:Timescales}
\begin{tabular}{| l | l | l | l | l |} 
\hline
Section & Symbol & Description & Definition & Units \\ [0.5ex] 
 \hline & & \\ [-2ex]
II & $g$ & qubit-resonator coupling & --- & \SI{}{\giga\hertz} \\
  & $\Omega$ & qubit frequency & --- & \SI{}{\giga\hertz}  \\ 
  & $\omega$ & cavity frequency & --- & \SI{}{\giga\hertz}  \\ 
  & $\Delta$ & qubit-resonator detuning & $\omega-\Omega$ & \SI{}{\giga\hertz}  \\
  & $\alpha_0$ & amplitude of coherent state in resonator at $t=0$ & --- & ---\\
  & $N$ & number of photons in resonator at $t=0$ & $|\alpha_0|^2$ & ---\\
  & $\Omega_d$ & amplitude of classical driving field & --- & \SI{}{\giga\hertz}\\
  & $\hat{H}_{\text{JC}}$ & Jaynes-Cummings (JC) Hamiltonian  & \cref{eq:JCmodel} & \SI{}{\giga\hertz}\\
  & $\hat{H}_{\text{disp.}}$ & dispersive Hamiltonian  & \cref{eq:dispersiveHamiltonian} & \SI{}{\giga\hertz}\\
  & $\hat{H}_{\text{R}}$ & Rabi Hamiltonian  & \cref{eq:RabiModel} & \SI{}{\giga\hertz}\\
  & $\hat{\tilde{H}}_{\text{JC}}$ & JC Hamiltonian in the rotating frame  & \cref{eq:HamiltoniansRabiJCrotFrame}, first line & \SI{}{\giga\hertz}\\
  & $\hat{\tilde{H}}_{\text{R}}$ & Rabi Hamiltonian in the rotating frame & \cref{eq:HamiltoniansRabiJCrotFrame}, second line & \SI{}{\giga\hertz}\\
  & $\thetaR$ &  sweet-spot polar angle of qubit in Rabi model & $\arctan(2|\Omega_d|/\Delta)$ &  ---\\
  & $\phiR$ &   sweet-spot azimuthal angle of qubit in Rabi model & $\arg(\Omega_d)$ & ---\\
  & $\thetaJC$ & sweet-spot polar angle of qubit in JC model & $\arctan(2g\sqrt{N}/\Delta)$ & --- \\
  & $\phiR$ & sweet-spot azimuthal angle of qubit in JC model & $\arg(\alpha_0)$ & --- \\
  & $\OmegaR$ & Rabi frequency & $\sqrt{\Delta^2+4|\Omega_d|^2}$ & \SI{}{\giga\hertz}\\
  & $\OmegaJC$ & fast frequency in JC model & $\sqrt{\Delta^2+4g^2N}$ & \SI{}{\giga\hertz}\\
  & $\omega_s$ & slow frequency in JC model & $g^2/\OmegaJC$ & \SI{}{\giga\hertz}\\
  & $\gamma_f$ & fast decay rate in JC model & $\sqrt{2N}\omega_s$ & \SI{}{\giga\hertz}\\
  & $\gamma_s$ & slow decay rate in JC model & $\gamma_f \omega_s/\OmegaJC$ & \SI{}{\giga\hertz}\\
  & $\zeta(t_r)$ & total quadrature spread of resonator state at readout time in JC model & \cref{eq:squeezingParameterReadoutTime} & ---\\
\hline 
III & $\hat{H}_{\text{JC+R}}$ & two-drive Hamiltonian in the rotating frame & \cref{eq:JCplusRabiModel} & \SI{}{\giga\hertz}\\
  & $\alpha_{\text{eff}}$ & effective two-drive amplitude in two-drive model & $\alpha_0+\Omega_d/g$ & ---\\
  & $\Neff$ & effective number of photons in two-drive model & $|\alpha_{\text{eff}}|^2$ & ---\\
  & $\theta_{\text{JC+R}}$ & sweet-spot polar angle of qubit in two-drive model & $\arctan(2g\sqrt{\Neff}/\Delta)$ & --- \\
  & $\phi_{\text{JC+R}}$ & sweet-spot azimuthal angle of qubit in two-drive model & $\arg(\alpha_{\text{eff}})$ & --- \\
  & $\bar{\Omega}_{\text{JC}}$ & fast frequency in two-drive model & $\sqrt{\Delta^2+4g^2\Neff}$ & \SI{}{\giga\hertz}\\
  & $\bar{\omega}_s$ & slow frequency in two-drive model & $g^2/\bar{\Omega}_{\text{JC}}$ & \SI{}{\giga\hertz}\\
  & $\bar{\gamma}_f$ & fast decay rate in two-drive model & $\sqrt{2\Neff}\bar{\omega}_s$ & \SI{}{\giga\hertz}\\
  & $\bar{\gamma}_s$ & slow decay rate in two-drive model & $\bar{\gamma}_f \bar{\omega}_s/\bar{\Omega}_{\text{JC}}$ & \SI{}{\giga\hertz}\\
\hline
IV  & SNR$(t)$ & signal-to-noise ratio & \cref{eq:SNRdef} & \SI{}{\giga\hertz} \\
  & $t_r$ & readout time & \cref{eq:readoutTimeDef} & \SI{}{\nano\second} \\
  & $t_{\text{max}}$ & maximum time for safe readout & $\pi\bar{\omega}_s^{-1}-t_r$ & \SI{}{\nano\second} \\
  & $\mathcal{F}(t)$ & measurement (Bhattacharyya) fidelity & \cref{eq:fidelityDef} & ---\\
  & QNDness$(t)$ & --- & \cref{eq:QNDnessDef} & ---\\
\hline
\end{tabular}
\caption{Table of symbols, definitions and units, in roughly chronological order of appearance in the main text.}
\end{table}

Metadata for the paper is available at Zenodo in \cite{zenodoDataset2026}. This includes animations of the qubit-resonator dynamics and the measurement process, described in more detail in \cref{app:videos}. It also includes a Mathematica notebook \texttt{paperSaddlePointJCmodelPublicationFinal.nb} with the implementation of the exact and saddlepoint expressions for qubit and resonator dynamics, as well measurement quantities, such as post-measurement density matrices, fidelity, and QNDness, covered in detail in \cref{app:ObservablesJCmodelExactExpressions,app:SaddlePoint,app:SaddlePointExpressions,app:SaddlePointExpressionsAtSweetSpot,app:SaddlePointExpressionsSweetspot,app:measurementDispersiveModel,app:measurementNondispersiveModel}. The python code used to generate \cref{fig:varyRatioOfDriveFrequencies} is provided in \texttt{Displacement\_JC\_model\_frequency\_sweep.ipynb}.

\section{Supplementary videos}\label{app:videos}
\begin{figure*}
  \centering
  \includegraphics[width=.76\linewidth]{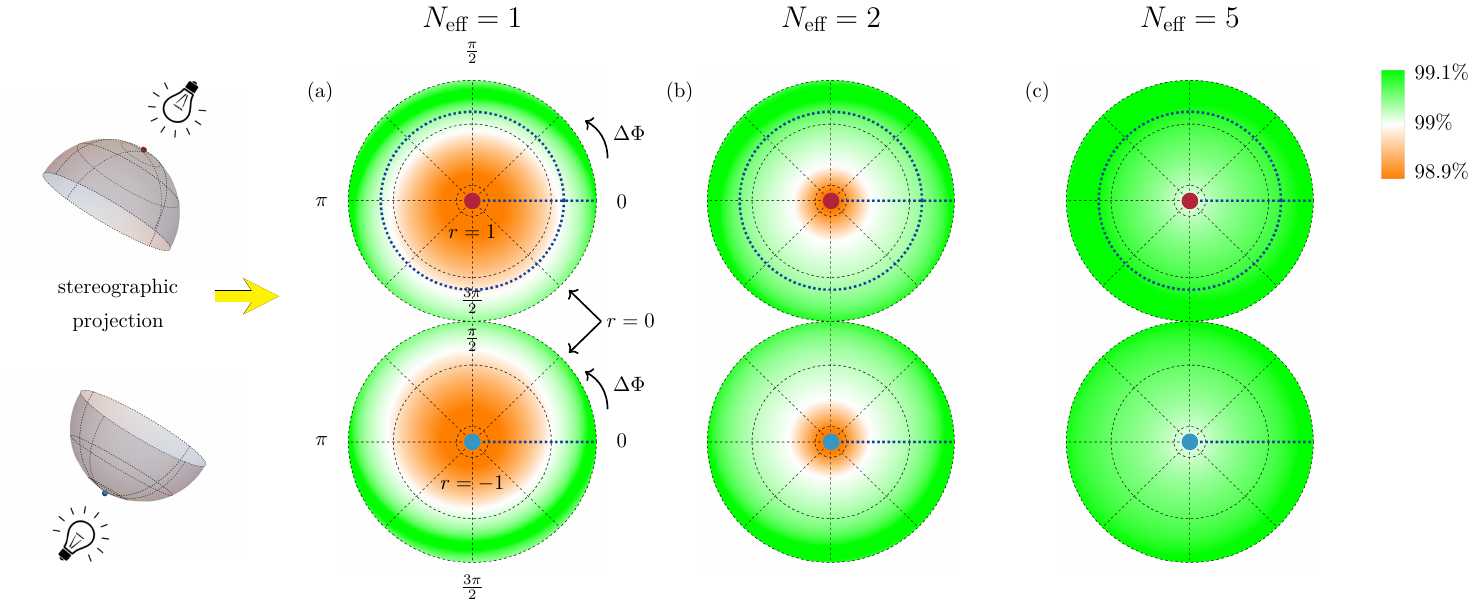}
  \caption{Fidelity $\mathcal{F}(t_r)$ for all possible initial qubit orientations $c_+\ket{+}+c_-\ket{-}$ stereographically projected onto the 2D plane. The worst fidelity always occurs at the poles, $r=\pm 1$. $99\%$ is cleared by every initial qubit orientation at $N_{\text{eff}}\approx 5$ for the given parameters, roughly a factor of four less than the critical $N_{\text{eff}}$ for QNDness$(t_r)$ in \cref{fig:stereographicProjectionQNDnessVaryN}. Parameters $g,\Delta$ as in \cref{fig:fidelityQNDnessFourPanels}.
  }
\label{fig:stereographicProjectionFidelityVaryN}
\end{figure*}

The \texttt{Rabi\_JC\_dynamics} folder contains animations of the Rabi and JC model dynamics in the rotating frame, \cref{fig:JCvsRabi}(c)--(f) (recall that both the qubit Bloch vector and the drive vector describing the resonator remain static in panels (a)--(b)). \texttt{panel\_c.mp4} shows the qubit Bloch vector performing Rabi oscillations at the fast Rabi frequency $\OmegaR$ on a cone around the drive vector (Rabi model). \texttt{panel\_d.mp4} and \texttt{panel\_e.mp4} show the slow clockwise/anticlockwise precessions at $\omega_s$ around the $\vb{e}^z$-axis of both the qubit and the (now dynamical) drive vectors in the JC model, for a qubit initialized in the sweetspot states $\ket{\pm}$, respectively. \texttt{panel\_f.mp4} shows the JC model dynamics for a qubit initialized in a superposition of $\ket{+}$ and $\ket{-}$: the qubit initially does Rabi-like oscillations around the drive vectors, but these quickly decay to reveal a linear combination of slow precessions at $\pm\omega_s$ around the $\vb{e}^z$-axis. In the \texttt{measurements} folder, we show how the fidelity and QNDness of the nondispersive readout scheme evolve over time. \texttt{probabilityFidelityQNDnessVsTimePlusInitialState.mp4} and \texttt{probabilityFidelityQNDnessVsTimeSuperpositionInitialState.mp4} are animated versions of a combination of \cref{fig:polarizationCavityMomentumProbabilityDistribution} and \cref{fig:fidelityQNDnessFourPanels}, for a qubit initialized in $\ket{+}$ and $\frac{1}{\sqrt{2}}(\ket{+}+\ket{-})$, respectively. \texttt{qndnessStereographicVaryt.mp4} and \texttt{fidelityStereographicVaryt.mp4} are animations of \cref{fig:stereographicProjectionQNDnessVaryN,fig:stereographicProjectionFidelityVaryN} in the time window $0\leq t\leq $\SI{20}{\nano\second} for $\Neff=14$ and $\Neff=4$, respectively. The QNDness always starts at exactly $1/2$, independent of the initial qubit orientation, since the qubit is initialized in a pure state, resulting in two fully orange disks after stereographic projection. QNDness$(t)$ steadily improves during the early-time dynamics up to a value $>99\%$ for all initial qubit states at $t=t_r$. At some critical time $t_{\text{crit}}\sim N$ ($>t_r$), QNDness drops below the acceptable value again, meaning that one should decouple the resonator from the qubit before the color switches to orange if the measurement is to remain QND. For the fidelity, different initial qubit orientations on the Bloch spheres result in different $\mathcal{F}(0)$, in the range $\frac{1}{\sqrt{2}}\leq \mathcal{F}(0)\leq 1$. As the resonator probability distributions $|R_\pm(p)|^2$ start to move left/right of the $p=0$ line, the fidelity increases for all initial qubit states. At $t=t_r$, $\mathcal{F}(t)$ reaches a plateau close to one, staying constant until $t_{\text{max}}$, rather than decreasing quadratically in time like the QNDness. Eventually, at $t=t_{\text{max}}$, $|R_\pm(p)|^2$ recross the $p=0$ line into the wrong half-plane and $\mathcal{F}(t)$, also drops below the acceptable value.

\section{Observables in the JC model: exact expressions}\label{app:ObservablesJCmodelExactExpressions}
In this appendix, we give exact expressions for $\langle\hat{a}(t)\rangle$, $\rho_{\text{q}}(t)$, $\langle\hat{n}(t)\rangle$ and $\langle\hat{a}^2(t)\rangle$ in the JC model for a system with initial state \cref{eq:qubitCavityInitialState}. 

We begin by recalling the standard derivation of the JC eigenbasis, using $\ket{m}\ket{\uparrow}$, $\ket{m}\ket{\downarrow}$ as our basis states. Here, $m\in\mathbb{N}$ labels the $m^{\text{th}}$ Fock state of the resonator, with $\hat{n}\ket{m}=m\ket{m}$, where $\hat{n}=\hat{a}^{\dagger}\hat{a}$, while $\uparrow,\downarrow$ describe the qubit, with $\hat{S}^z\ket{\uparrow}=\ket{\uparrow}$, $\hat{S}^z\ket{\downarrow}=-\ket{\downarrow}$. The ground state of the JC model is $\ket{0,\uparrow}=\ket{0}\ket{\uparrow}$, with eigenenergy $\lambda_{0,\uparrow}=\frac{\Delta}{2}$. For the excited states, recall that $\hat{H}_{\text{JC}}$ commutes with the number of excitations, $\hat{n}_e=\hat{a}^{\dagger}\hat{a}-\frac{1}{2}\hat{S}^z$. This means that $\hat{H}_{\text{JC}}$ only couples together $\ket{m}\ket{\uparrow}$ and $\ket{m-1}\ket{\downarrow}$, and the Hamiltonian matrix for the excited states may be written as a tensor product of $2\times 2$ blocks, $H_m=\frac{\Delta}{2}S^z+g\sqrt{m}S^x$. Diagonalizing $H_m$, we find the following pairs of excited states
\begin{equation}\label{eq:JCmodelEigenbasis}
    \begin{aligned}
        \ket{m,+}&=c(\theta_m)\ket{m}\ket{\uparrow}+s(\theta_m)\ket{m-1}\ket{\downarrow},& \lambda_{m,+}=+\lambda_m,\\
        \ket{m,-}&=s(\theta_m)\ket{m}\ket{\uparrow}-c(\theta_m)\ket{m-1}\ket{\downarrow},& \lambda_{m,-}=-\lambda_m,
    \end{aligned}
\end{equation}
for $m\geq 1$, with $c(\theta_m)=\cos(\theta_m/2)$, $s(\theta_m)=\sin(\theta_m/2)$, $\theta_m=\arctan(2g\sqrt{m}/\Delta)$, and $\lambda_m=\frac{1}{2}\sqrt{\Delta^2+4g^2m}$. Notice that the JC (\cref{eq:JCmodelEigenbasis}) and Rabi (\cref{eq:RabiModelEigenbasis}) eigenbases are identical if we swap $\Omega_d\leftrightarrow g\sqrt{m}$. 

We may express the time-evolved wavefunction $\ket{\Psi(t)}$ for the full system in terms of the JC eigenstates,
\begin{equation}\label{eq:psiFullSystemTimeEvolution}
\ket{\Psi(t)}=c_{0,\uparrow}e^{-\frac{i\Delta t}{2}}\ket{0,\uparrow}+\sum_{m=1}^{\infty}\sum_{\alpha=\pm}c_{m,\alpha}e^{-i\alpha\lambda_{m}t}\ket{m,\alpha},
\end{equation}
with $c_{m,\alpha}=\bra{m,\alpha}\ket{\Psi_0}$, where $\ket{\Psi_0}$ is the full-system wavefunction at time $t=0$, defined in \cref{eq:qubitCavityInitialState}. The most convenient parametrization for the initial qubit state $\ket{\psi_0}$ turns out to be in terms of the eigenstates of the Rabi model, $\ket{\psi_0}=c_+\ket{+}+c_-\ket{-}$, if we replace $|\Omega_d|\to g\sqrt{N}$ in $\ket{\pm}$, and define the JC angles $\thetaJC=\arctan(2g\sqrt{N}/\Delta)$, $\phiJC=\arg(\alpha_0)$. This should not come as too much of a surprise, given the similarities between the two models. Using \cref{eq:JCmodelEigenbasis} and the definitions of $\ket{\pm}$ in \cref{eq:RabiModelEigenbasis}, we find
\begin{equation}\label{eq:JCModelFullSystemWavefunctionCoeffs}
    \begin{aligned}
        c_{0,\uparrow}&=e^{-\frac{N}{2}}\left(c_+c(\thetaJC)+c_-s(\thetaJC)\right),\\
        c_{m,+}&=e^{im\phiJC}\frac{e^{-\frac{N}{2}}N^{\frac{m}{2}}}{\sqrt{m!}}\Big[c_+\left(c(\theta_m)c(\thetaJC)+\frac{\sqrt{m}}{\sqrt{N}}s(\theta_m)s(\thetaJC)\right)+c_-\left(c(\theta_m)s(\thetaJC)-\frac{\sqrt{m}}{\sqrt{N}}s(\theta_m)c(\thetaJC)\right)\Big],\\
        c_{m,-}&=e^{im\phiJC}\frac{e^{-\frac{N}{2}}N^{\frac{m}{2}}}{\sqrt{m!}}\Big[c_+\left(s(\theta_m)c(\thetaJC)-\frac{\sqrt{m}}{\sqrt{N}}c(\theta_m)s(\thetaJC)\right)+c_-\left(s(\theta_m)s(\thetaJC)+\frac{\sqrt{m}}{\sqrt{N}}c(\theta_m)c(\thetaJC)\right)\Big],
    \end{aligned}
\end{equation}
where we replaced $\alpha_0\to\sqrt{N}e^{i\phiJC}$. For comparison with the Rabi model, the quantities of interest are $\langle\hat{a}\rangle=\bra{\Psi(t)}\hat{a}\ket{\Psi(t)}$ and $\hat{\rho}_{\text{q}}=\Tr_{\text{r}}[\ket{\Psi(t)}\bra{\Psi(t)}]$. We have
\begin{equation}\label{eq:aExpectValExact}
    \begin{aligned}
        \langle\hat{a}(t)\rangle&=c^*_{0,\uparrow}e^{\frac{i\Delta t}{2}}\left(c(\theta_1)c_{1,+}e^{-i\lambda_1t}+s(\theta_1)c_{1,-}e^{i\lambda_1t}\right)\\
        &\quad +\sum_{l=1}^{\infty} e^{-i(\lambda_{l+1}-\lambda_l)t}\left(\sqrt{\frac{(l+1)!}{l!}}c(\theta_l)c(\theta_{l+1})+\sqrt{\frac{l!}{(l-1)!}}s(\theta_l)s(\theta_{l+1})\right)c^*_{l,+}c_{l+1,+}\\
        &\qquad + e^{i(\lambda_{l+1}-\lambda_l)t}\left(\sqrt{\frac{(l+1)!}{l!}}s(\theta_l)s(\theta_{l+1})+\sqrt{\frac{l!}{(l-1)!}}c(\theta_l)c(\theta_{l+1})\right)c^*_{l,-}c_{l+1,-}\\
        &\qquad +e^{-i(\lambda_{l+1}+\lambda_l)t}\left(\sqrt{\frac{(l+1)!}{l!}}s(\theta_l)c(\theta_{l+1})-\sqrt{\frac{l!}{(l-1)!}}c(\theta_l)s(\theta_{l+1})\right)c^*_{l,-}c_{l+1,+}\\
        &\qquad + e^{i(\lambda_{l+1}+\lambda_l)t}\left(\sqrt{\frac{(l+1)!}{l!}}c(\theta_l)s(\theta_{l+1})-\sqrt{\frac{l!}{(l-1)!}}s(\theta_l)c(\theta_{l+1})\right)c^*_{l,+}c_{l+1,-}.
    \end{aligned}
\end{equation}
Next, we calculate $\hat{\rho}_{\text{q}}$ in the $\{\ket{\uparrow},\ket{\downarrow}\}$-basis. It is sufficient to calculate just the $[\rho_{\text{q}}]_{\uparrow\uparrow}=\bra{\uparrow}\hat{\rho}_{\text{q}}\ket{\uparrow}$ and $[\rho_{\text{q}}]_{\downarrow\uparrow}=\bra{\downarrow}\hat{\rho}_{\text{q}}\ket{\uparrow}$ components; the other two automatically follow by $\Tr[\hat{\rho}_{\text{q}}]=1$ and $\hat{\rho}_{\text{q}}^\dagger=\hat{\rho}_{\text{q}}$. We find
\begin{equation}\label{eq:rhoQubitExact}
    \begin{aligned}
[\rho_{\text{q}}]_{\uparrow\uparrow}&=|c_{0,\uparrow}|^2+\sum_{l=1}^{\infty}c^2(\theta_l)|c_{l,+}|^2+s^2(\theta_l)|c_{l,-}|^2+s(\theta_l)c(\theta_l)\left(e^{-2i\lambda_l t} c^*_{l,-}c_{l,+} + e^{2i\lambda_l t}c^*_{l,+}c_{l,-}\right),\\
        [\rho_{\text{q}}]_{\downarrow\uparrow}&=\left(c_{1,+}s(\theta_1)e^{-i\lambda_1 t}-c_{1,-}c(\theta_1)e^{i\lambda_1 t}\right)c_{0,\uparrow}^*e^{\frac{i\Delta t}{2}}\\
        &+\sum_{l=1}^{\infty}e^{-i(\lambda_{l+1}-\lambda_l)t}s(\theta_{l+1})c(\theta_l)c_{l+1,+}c_{l,+}^*-e^{i(\lambda_{l+1}-\lambda_l)t}c(\theta_{l+1})s(\theta_l)c_{l+1,-}c_{l,-}^* \\
        &\quad + e^{-i(\lambda_{l+1}+\lambda_l)t} s(\theta_{l+1}) s(\theta_l) c_{l+1,+}c_{l,-}^* - e^{i(\lambda_{l+1}+\lambda_l)t} c(\theta_{l+1}) c(\theta_l) c_{l+1,-}c_{l,+}^*.
        \end{aligned}
\end{equation}

For calculating the work done and degree of total quadrature spread on the resonator we also need
\begin{equation}\label{eq:nExpectValExact}
    \begin{aligned}
        \langle\hat{n}(t)\rangle&=\sum_{l=1}^{\infty} \left(l-s^2(\theta_l)\right)|c_{l,+}|^2+\left(l-c^2(\theta_l)\right)|c_{l,-}|^2 + s(\theta_l)c(\theta_l)\left(e^{-2i\lambda_l t}c^*_{l,-}c_{l,+}+e^{2i\lambda_l t}c^*_{l,+}c_{l,-}\right)
    \end{aligned}
\end{equation}
and
\begin{equation}\label{eq:aSquaredExpectValExact}
    \begin{aligned}
        \langle\hat{a}^2(t)\rangle&=c^*_{0,\uparrow}e^{\frac{i\Delta t}{2}}\sqrt{2}\left(c(\theta_2)c_{2,+}e^{-i\lambda_2t}+s(\theta_2)c_{2,-}e^{i\lambda_2t}\right)\\
        &\quad+\sum_{l=1}^{\infty} e^{-i(\lambda_{l+2}-\lambda_l)t}\left(\sqrt{\frac{(l+2)!}{l!}}c(\theta_l)c(\theta_{l+2})+\sqrt{\frac{(l+1)!}{(l-1)!}}s(\theta_l)s(\theta_{l+2})\right)c^*_{l,+}c_{l+2,+}\\
        &\qquad + e^{i(\lambda_{l+2}-\lambda_l)t}\left(\sqrt{\frac{(l+2)!}{l!}}s(\theta_l)s(\theta_{l+2})+\sqrt{\frac{(l+1)!}{(l-1)!}}c(\theta_l)c(\theta_{l+2})\right)c^*_{l,-}c_{l+2,-}\\
        &\qquad +e^{-i(\lambda_{l+2}+\lambda_l)t}\left(\sqrt{\frac{(l+2)!}{l!}}s(\theta_l)c(\theta_{l+2})-\sqrt{\frac{(l+1)!}{(l-1)!}}c(\theta_l)s(\theta_{l+2})\right)c^*_{l,-}c_{l+2,+}\\
        &\qquad + e^{i(\lambda_{l+2}+\lambda_l)t}\left(\sqrt{\frac{(l+2)!}{l!}}c(\theta_l)s(\theta_{l+2})-\sqrt{\frac{(l+1)!}{(l-1)!}}s(\theta_l)c(\theta_{l+2})\right)c^*_{l,+}c_{l+2,-}.
    \end{aligned}
\end{equation}

\cref{eq:aExpectValExact,eq:rhoQubitExact,eq:nExpectValExact,eq:aSquaredExpectValExact} are exact, but not very easy to interpret physically, as they still contain an infinite sum over the Fock index $l$. In App.~\ref{app:SaddlePoint}, we show how to apply the saddlepoint technique to transform the infinite sums into simple-to-interpret analytic functions of the physical variables $g,\Delta,N$, as well as the coefficients $c_\pm$ describing the initial qubit orientation.

\section{Saddlepoint technique}\label{app:SaddlePoint}
In this appendix, we show how to apply the saddlepoint technique, coupled with perturbation theory in $1/\sqrt{N}$ (where $N$ is the initial number of photons in the resonator), to evaluate the sums over Fock index $l$ in App.\ref{app:ObservablesJCmodelExactExpressions}. The method we develop is very general, and will work for any exactly solvable model with Poisson- or Gaussian-distributed occupation numbers in Fock space. Importantly, it can be made accurate to any desired power of $1/\sqrt{N}$, whereas semi-classical approaches (or at least those known to us) fail beyond the leading order. 

The key to this technique is the observation that the initial coherent state $\ket{\alpha_0}$ in the resonator causes the summand in all the observables formulas \cref{eq:aExpectValExact,eq:rhoQubitExact,eq:nExpectValExact,eq:aSquaredExpectValExact} to also broadly follow the Poisson distribution. The Poisson distribution is described by the $e^{-N}N^{l}/l!$ factor, which arises from the $c_{l,\pm}$ coefficients (see definition in \cref{eq:JCModelFullSystemWavefunctionCoeffs}), and has mean and standard deviation $N$. In addition to the Poisson factor, the summands contain two other types of terms: time-dependent complex exponentials $e^{-i(\alpha\lambda_{l+\Delta m}-\beta\lambda_{l+\Delta n})t}$, carrying information about the time evolution, and the remaining static terms, which are neither the Poisson factor nor the time-dependent exponential. Labeling the summand $f_{\alpha,\beta,\Delta m,\Delta n}(l)$, we thus have
\begin{equation}
    f_{\alpha,\beta,\Delta m,\Delta n}(l)=\underbrace{\frac{e^{-N}N^l}{l!}e^{-i(\alpha\lambda_{l+\Delta m}-\beta\lambda_{l+\Delta n})t}}_{\text{use to determine the saddle point}} s_{\alpha,\beta,\Delta m,\Delta n}(l),
\end{equation}
where $\alpha,\beta=\pm 1$ are the spin indices, and $\Delta m,\Delta n\in\mathbb{Z}$ denote integer deviations in the Fock index $l$. The static terms vary much more slowly with $l$ than the Poisson factor, and therefore do not greatly influence the location of the saddle point. On the other hand, the phase of the complex exponential factors can vary as fast as $\sqrt{\Delta^2+4g^2l} t$, so they do need to be included in the evaluation of the saddle point, otherwise the resulting expressions go wrong for $t>0$.

The idea of the saddlepoint technique is to write everything in terms of Gaussian integrals, which necessarily requires converting the discrete Riemann sums in \cref{eq:aExpectValExact,eq:rhoQubitExact,eq:nExpectValExact,eq:aSquaredExpectValExact} to continuous integrals, $\sum_{l=1}^{\infty}\sum_{\alpha,\beta=\pm}f_{\alpha,\beta,\Delta m,\Delta n}(l)\to \int_{1}^{\infty}\sum_{\alpha,\beta=\pm}\dd{l}f_{\alpha,\beta,\Delta m,\Delta n}(l)$. Although this may seem like a potential source of errors, the induced error is, in fact, exponentially small in $N$, as we show next. Applying the Euler-Maclaurin formula \cite{MITeulerMaclaurin}, the error between the Riemann sum and continuous integral is
\begin{equation}\label{eq:eulerMaclaurinError}
    \text{error}=\sum_{l=a}^{b}f(l)-\int_{a}^{b}\dd{l}f(l)=\frac{f(b)-f(a)}{2}+\sum_{k=1}^{\infty}\frac{B_{2k}}{(2k)!}\left[f^{(2k-1)}(b)-f^{(2k-1)}(a)\right],
\end{equation}
where we dropped the $\alpha,\beta,\Delta m,\Delta n$ indices on $f_{\alpha,\beta,\Delta m,\Delta n}(l)$ for notational simplicity. The lower and upper limits are $a=1$ and $b=\infty$. $B_{2k}=B_{2k}(1)$ are the even Bernoulli numbers, where the Bernoulli polynomials $B_k(x)$ are defined recursively by $B_k'(x)=kB_{k-1}(x)$, $\int_0^1\dd{x}B_k(x)=0$, with $B_0(x)=1$. For $N\gtrsim 1$, the summand scales approximately as $f_{\alpha,\beta,\Delta m,\Delta n}(l)\approx e^{-(l-N)^2/2N}$ with $l$, where we neglect small corrections from the non-Poisson factor contributions. From this, we immediately conclude that the $b$-limit terms vanish. For the $a$-limit, the leading order derivative $f_{\alpha,\beta,\Delta m,\Delta n}^{(k)}(1)\sim e^{-N/2}$ is independent of $k$. Then, using $\coth(\frac{1}{2})=2\sum_{k=0}^{\infty}\frac{B_{2k}}{(2k)!}$ \footnote{To show this, use the generating function for Bernoulli polynomials, $\frac{t e^{xt}}{e^t-1}=\sum_{n=0}^{\infty} \frac{B_n(x)}{n!}t^n$, (can be proved by differentiating both sides w.r.t. $x$), which yields $\frac{t}{e^t-1}=\sum_{n=0}^{\infty} \frac{B_n}{n!}t^n$ at $x=0$. Then, use $t\coth(t)=t+\frac{2t}{e^{2t}-1}=\sum_{m=0}^{\infty}\frac{B_{2m}}{(2m)!}(2t)^{2m}$, and set $2t=1$}, we conclude that the overall error term scales like $e^{-N/2}$. Ultimately, the highest accuracy we want to reach is $\mathcal{O}(1/N^2)$, in comparison to which the Euler-Maclaurin error is vanishingly small. Note that for the ensuing integrals, we will set $a=-\infty$ rather than $a=1$, as this also only introduces an error of order $e^{-N/2}$, i.e., insignificant in comparison to $\mathcal{O}(1/N^2)$.

Having convinced ourselves that switching from the Riemann sum to a continuous integral introduces only exponentially small in $N$ errors, we now treat $f_{\alpha,\beta,\Delta m,\Delta n}(l)$ as a continuous function of $l$, and identify the location of the saddle point. To this end, we define $g_{\alpha,\beta,\Delta m,\Delta n}(l)$, the logarithm of the product of the Poisson and time-dependent complex exponential factors,
\begin{equation}\label{eq:expgFuncDef}
    e^{g_{\alpha,\beta,\Delta m,\Delta n}(l)}=\frac{e^{-N}N^l}{l!/\sqrt{2\pi}}e^{-i(\alpha\lambda_{l+\Delta m}-\beta\lambda_{l+\Delta n})t},
\end{equation}
where we introduced a factor $1/\sqrt{2\pi}$ for reasons that will soon become clear. The saddlepoint value $L_{\alpha,\beta,\Delta m,\Delta n}$ maximizes $g_{\alpha,\beta,\Delta m,\Delta n}(l)$,
\begin{equation}\label{eq:saddlePointCondition}
    g^{(1)}_{\alpha,\beta,\Delta m,\Delta n}(L_{\alpha,\beta,\Delta m,\Delta n})=0,
\end{equation}
where we used the shorthand $g^{(k)}_{\alpha,\beta,\Delta m,\Delta n}(l)=\frac{\partial^k}{\partial l^k}\left(g_{\alpha,\beta,\Delta m,\Delta n}(l)\right)$ to denote the $k^{\text{th}}$ derivative with respect to $l$.

Next, we set up the perturbation theory in $1/\sqrt{N}$. The highest accuracy we require is $\mathcal{O}(1/N^2)$ for the $\alpha=\beta$ terms and $\mathcal{O}(1/N)$ for the $\alpha=-\beta$ terms, for reasons that will soon become clear. Using Stirling's approximation \cite{mermin1984stirling}, we replace $l!$ with
\begin{equation}\label{eq:StirlingApproximation}
l!=\sqrt{2\pi l}\left(\frac{l}{e}\right)^l\exp\left(\frac{1}{12l}+\mathcal{O}(l^{-3})\right),
\end{equation}
correct up to and including $\mathcal{O}(l^{-2})$, the highest degree of accuracy required for our subsequent calculations. Substituting \cref{eq:StirlingApproximation} into \cref{eq:gFuncDef}, the $\sqrt{2\pi}$-factors conveniently cancel, and after taking the $\log$ of both sides, we obtain
\begin{equation}\label{eq:gFuncDef}
    g_{\alpha,\beta,\Delta m,\Delta n}(l)=l-N+l\log(N)-l\log(l)-\frac{1}{2}\log(l)-\frac{1}{12l}-it\left(\alpha\lambda_{l+\Delta m}-\beta\lambda_{l+\Delta n}\right)+\mathcal{O}(l^{-3}).
\end{equation}
We also expand the saddle point $L_{\alpha,\beta,\Delta m,\Delta n}$ perturbatively in powers of $1/\sqrt{N}$,
\begin{equation}\label{eq:TaylorExpansionSaddlePoint}
    L_{\alpha,\beta,\Delta m,\Delta n}=N\sum_{k=0}^{\infty}N^{-k/2}L^{(k/2)}_{\alpha,\beta,\Delta m,\Delta n},
\end{equation}
where $L^{(k/2)}_{\alpha,\beta,\Delta m,\Delta n}$ are coefficients to be determined using the saddle point condition \cref{eq:saddlePointCondition}.
We define the shifted and rescaled coordinate 
\begin{equation}\label{eq:lShiftedRescaled}
    l\to L_{\alpha,\beta,\Delta m,\Delta n}+\sqrt{N}l,
\end{equation}
where the rescaling by $\sqrt{N}$ captures the natural width $\sigma\sim\sqrt{N}$ of the Gaussian; thus, only $|l|\lesssim 1$ contributes significantly to the Gaussian integral in this shifted and rescaled coordinate system. Substituting \cref{eq:lShiftedRescaled} into $g_{\alpha,\beta,\Delta m,\Delta n}(l)$ and Taylor expanding in $l$ gives
\begin{equation}\label{eq:gTaylorExpansionAtSaddlePoint}
    g_{\alpha,\beta,\Delta m,\Delta n}(L_{\alpha,\beta,\Delta m,\Delta n}+\sqrt{N}l)=\sum_{k=0}^{\infty}\tilde{g}^{(k)}_{\alpha,\beta,\Delta m,\Delta n}\frac{l^k}{k!},
\end{equation}
where $\tilde{g}^{(k)}_{\alpha,\beta,\Delta m,\Delta n}=N^{k/2}g^{(k)}_{\alpha,\beta,\Delta m,\Delta n}(L_{\alpha,\beta,\Delta m,\Delta n})$. To satisfy the saddle point condition $\tilde{g}^{(1)}_{\alpha,\beta,\Delta m,\Delta n}=0$ to order $N^{-k/2}$, we need to evaluate $L_{\alpha,\beta,\Delta m,\Delta n}$ up to and including order $N^{-(k+1)/2}$, due to the $\sqrt{N}$ rescaling factor introduced in \cref{eq:lShiftedRescaled}. We will now do this using perturbation theory. Introducing the perturbation theory bookkeeping parameter $\epsilon$, we send
\begin{equation}
    N\to N/\epsilon, \quad \Delta\to\Delta/\sqrt{\epsilon},
\end{equation}
such that $\Omega_R\to\Omega_R/\sqrt{\epsilon}$. Substituting \cref{eq:TaylorExpansionSaddlePoint} into \cref{eq:saddlePointCondition}, and solving order by order for the coefficients of $\sqrt{\epsilon}$, we find
\begin{equation}\label{eq:saddlePointFormula}
\begin{aligned}
    L_{\alpha,\alpha,\Delta m,\Delta n}&=\frac{N}{\epsilon}\Bigg(1-\textcolor{brown}{\epsilon\frac{1}{2N}}+\textcolor{gray}{\epsilon^{3/2}2i\alpha(\Delta m-\Delta n)\frac{\omega_s^2t}{\OmegaJC}}-\textcolor{orange}{\epsilon^2\frac{1}{24N^2}} \\
    &\qquad -\textcolor{olive}{\epsilon^{5/2}\frac{6i\alpha \omega_s^3 t}{\OmegaJC^2}\left(\Delta m(\Delta m-1)-\Delta n(\Delta n-1)\right)}+\mathcal{O}(N^{-3})\Bigg),\\
   L_{\alpha,-\alpha,\Delta m,\Delta n}&=\frac{N}{\epsilon}\Bigg(1-\textcolor{purple}{\sqrt{\epsilon}2i\alpha \omega_s t}-\textcolor{brown}{\epsilon\left(\frac{1}{2N}+\frac{2\omega_s^2\Delta^2t^2}{\OmegaJC^2}\right)}\\
   &\qquad +\textcolor{gray}{\epsilon^{3/2}\frac{2i\alpha\omega_s^2 t}{\OmegaJC}\left((\Delta m+\Delta n - 1) + \frac{2}{3}\frac{\omega_s t^2}{\OmegaJC^3}\left(\Delta^4-10g^2N\Delta^2+4g^4N^2\right)\right)}+\mathcal{O}(N^{-2})\Bigg),
\end{aligned}
\end{equation}
where we used $\omega_s=g^2/\sqrt{\Delta^2+4g^2N}$ and $\OmegaJC=\sqrt{\Delta^2+4g^2N}$. Note that the above calculation, along with the rest of the perturbation theory, is implemented on Mathematica in \texttt{paperSaddlePointJCmodelPublicationFinal.nb}, provided in the Supplementary Material. We also color-coded the various orders of $\sqrt{\epsilon}$ to facilitate their interpretation. 

Next, we truncate the formally infinite Taylor expansion in $k$ in \cref{eq:gTaylorExpansionAtSaddlePoint}, as we only wish to keep accuracy up to and including $\mathcal{O}(1/N^2)$ for $\alpha=\beta$ and $\mathcal{O}(1/N)$ for $\alpha=-\beta$. Using the fact that the leading order term of $g^{(k)}_{\alpha,\beta,\Delta m,\Delta n}(L_{\alpha,\beta,\Delta m,\Delta n})$ scales like $1/N^{k-1}$, we conclude that we may truncate at $k=6$ for $\alpha=\beta$ and at $k=4$ for $\alpha=-\beta$. Starting with the $\alpha=\beta$ terms, $\tilde{g}^{(k)}_{\alpha,\alpha,\Delta m,\Delta n}(L_{\alpha,\alpha,\Delta m,\Delta n})$ for $k=0,2,3,4,5,6$ (the $k=1$ case vanishes by construction) read
\begin{equation}\label{eq:gSlowTermsAtSP}
\begin{aligned}
    \tilde{g}^{(0)}_{\alpha,\alpha,\Delta m,\Delta n}&=-\frac{1}{2}\log(N/\epsilon)\textcolor{purple}{-\sqrt{\epsilon}i\alpha(\Delta m-\Delta n)\omega_s t}+\textcolor{brown}{\epsilon\frac{1}{24N}} \\
    &\qquad + \textcolor{gray}{\epsilon^{3/2}\frac{i\alpha\omega_s^2t}{\OmegaJC}\left(\Delta m(\Delta m-1)-\Delta n(\Delta n-1)\right)} - \textcolor{orange}{\epsilon^2\frac{2\omega_s^4Nt^2}{\OmegaJC^2}(\Delta m-\Delta n)^2},\\
    -\tilde{g}^{(2)}_{\alpha,\alpha,\Delta m,\Delta n}=\sigma^{-2}_{\alpha,\alpha,\Delta m,\Delta n}&=1+\textcolor{gray}{\epsilon^{3/2}\frac{2i\alpha\omega_s^2 t}{\OmegaJC^3}(2g^2N-\Delta^2)(\Delta m-\Delta n)}-\textcolor{orange}{\epsilon^2\frac{1}{24N^2}},\\
    \tilde{g}^{(3)}_{\alpha,\alpha,\Delta m,\Delta n}&=\textcolor{purple}{\sqrt{\epsilon}\frac{1}{\sqrt{N}}}-\textcolor{orange}{\epsilon^2\frac{4i\alpha\omega_s^2 t}{\sqrt{N}\OmegaJC^5}(\Delta^4+8\Delta^2g^2N-14g^4N^2)(\Delta m-\Delta n)},\\
    \tilde{g}^{(4)}_{\alpha,\alpha}&=-\textcolor{brown}{\epsilon\frac{2}{N}}+\mathcal{O}(N^{-5/2}),\\
    \tilde{g}^{(5)}_{\alpha,\alpha}&=\textcolor{gray}{\epsilon^{3/2}\frac{6}{N^{3/2}}}+\mathcal{O}(N^{-5/2}),\\
    \tilde{g}^{(6)}_{\alpha,\alpha}&=-\textcolor{orange}{\epsilon^{2}\frac{24}{N^2}},
\end{aligned}
\end{equation}
where we also defined the width $\sigma_{\alpha,\alpha,\Delta m,\Delta n}$ of the Gaussian in the third line. For $\alpha=-\beta$, the relevant terms are
\begin{equation}\label{eq:gFastTermsAtSP}
    \begin{aligned}
        \tilde{g}^{(0)}_{\alpha,-\alpha,\Delta m,\Delta n}&=-\frac{1}{2}\log(N/\epsilon)-\textcolor{teal}{\frac{1}{\sqrt{\epsilon}}i\alpha\OmegaJC t}-2N\omega_s^2t^2 - \textcolor{purple}{\sqrt{\epsilon}i\alpha\omega_s t\left(\Delta m+\Delta n-1+\frac{4}{3}\frac{\omega_s^2 t^2 N}{\OmegaJC^2}(2g^2N-\Delta^2)\right)}\\
        &\qquad +\textcolor{brown}{\epsilon\left(\frac{1}{24N}+\frac{4N\omega_s^3t^2}{\OmegaJC}(\Delta m+\Delta n-1)+\frac{2N\omega_s^4t^4\left(\Delta^4-16\Delta^2g^2N+16g^4N^2\right)}{3\OmegaJC^4}\right)},\\
        \sigma^2_{\alpha,-\alpha}&=1+\textcolor{purple}{\sqrt{\epsilon}\frac{2i\alpha\omega_s t}{\OmegaJC^2}(\Delta^2+2g^2N)}-\textcolor{brown}{\epsilon\frac{2\omega_s^2t^2}{\OmegaJC^4}(\Delta^4+12\Delta^2g^2N+8g^4N^2)},\\
        \tilde{g}^{(3)}_{\alpha,-\alpha}&=\textcolor{purple}{\sqrt{\epsilon}\frac{1}{\sqrt{N}}} + \textcolor{brown}{\epsilon \frac{4i \alpha \omega_s t}{\sqrt{N} \OmegaJC^4}(\Delta^4+ 8\Delta^2 g^2N + 10 g^4N^2)},\\
        \tilde{g}^{(4)}_{\alpha,-\alpha}&=-\textcolor{brown}{\epsilon\frac{2}{N}},
    \end{aligned}
\end{equation}
note that we drop the $\Delta m,\Delta n$ indices when the functions don't depend on them. 

Now we are finally ready to evaluate the Gaussian integrals using the saddlepoint technique. Expressing the Riemann sum as an integral, and shifting and rescaling $l$ according to \cref{eq:lShiftedRescaled}, we have
\begin{equation}\label{eq:riemannToIntegral}
    \mathcal{I}_{\alpha,\beta,\Delta m,\Delta n}=\sum_{l=0}^{\infty} e^{g_{\alpha,\beta,\Delta m,\Delta n}(l)}s_{\alpha,\beta,\Delta m,\Delta n}(l)\to \sqrt{\frac{N}{2\pi}}\int_{-\infty}^{\infty} \dd{l} \exp\left(\sum_{k'=0}^{\infty}\tilde{g}^{(k')}_{\alpha,\beta,\Delta m,\Delta n}\frac{l^{k'}}{k'!}\right) \sum_{k=0}^{\infty}\tilde{s}^{(k)}_{\alpha,\beta,\Delta m,\Delta n}\frac{l^{k}}{k!},
\end{equation}
where $\tilde{s}^{(k)}_{\alpha,\beta,\Delta m,\Delta n}=N^{k/2}s^{(k)}_{\alpha,\beta,\Delta m,\Delta n}(L_{\alpha,\beta,\Delta m,\Delta n})$ are the static terms and their derivatives with respect to $l$, evaluated at the saddle point and rescaled by the appropriate power of $\sqrt{N}$. It is useful to express the exponential as follows,
\begin{equation}\label{eq:exponentialFuncTaylorExpanded}
    \exp\left(\sum_{k'=0}^{\infty}\tilde{g}^{(k')}_{\alpha,\beta,\Delta m,\Delta n}\frac{l^{k'}}{k'!}\right)=e^{\tilde{g}^{(0)}_{\alpha,\beta,\Delta m,\Delta n}}e^{-l^2/(2\sigma^2_{\alpha,\beta,\Delta m,\Delta n})} \sum_{j=0}^{\infty} \frac{1}{j!}\left(\sum_{k'=3}^{\infty}\tilde{g}^{(k')}_{\alpha,\beta,\Delta m,\Delta n}\frac{l^{k'}}{k'!}\right)^j.
\end{equation}
In the above, the first exponential factor contains all the $l$-independent constant terms, while the second is the Gaussian factor. The last term is a Taylor expansion of the remaining terms of order $l^3$ and above. Substituting \cref{eq:exponentialFuncTaylorExpanded} into \cref{eq:riemannToIntegral}, we have
\begin{equation}\label{eq:riemannToIntegral1}
    \mathcal{I}_{\alpha,\beta,\Delta m,\Delta n}=e^{\tilde{g}^{(0)}_{\alpha,\beta,\Delta m,\Delta n}}\sqrt{\frac{N}{2\pi}}\int_{-\infty}^{\infty} \dd{l} e^{-l^2/(2\sigma^2_{\alpha,\beta,\Delta m,\Delta n})}\sum_{j,k=0}^{\infty}\frac{1}{j!}\left(\sum_{k'=3}^{\infty}\tilde{g}^{(k')}_{\alpha,\beta,\Delta m,\Delta n}\frac{l^{k'}}{k'!}\right)^j\tilde{s}^{(k)}_{\alpha,\beta,\Delta m,\Delta n}\frac{l^{k}}{k!}.
\end{equation}
The final step is to evaluate the Gaussian integrals in \cref{eq:riemannToIntegral}. These evaluate to
\begin{equation}\label{eq:gaussianIntegralMoments}
    \frac{1}{\sqrt{2\pi}}\int_{-\infty}^{\infty}\dd{l} l^{2k} e^{-l^2/(2\sigma^2_{\alpha,\beta,\Delta m,\Delta n})}=\frac{(2k)!}{2^k k!}\sigma^{2k+1}_{\alpha,\beta,\Delta m,\Delta n}
\end{equation}
for even powers of $l$, and vanish for odd powers of $l$. Crucially, we may use the fact that the $\tilde{g}^{(k')}_{\alpha,\beta,\Delta m,\Delta n}$ and $\tilde{s}^{(k)}_{\alpha,\beta,\Delta m,\Delta n}$ terms decay with $N$ as $N^{-(k'-2)/2}$ and $N^{-k/2}$, respectively, to truncate the sums over $j,k'$ and $k$ in \cref{eq:riemannToIntegral1}. We introduce the notation $j_{k'}$ to indicate what power(s) of the $\tilde{g}^{(k')}_{\alpha,\beta,\Delta m,\Delta n}$ we want to keep in \cref{eq:riemannToIntegral1}, with $j_{k'}=0$ by default, unless otherwise specified. For instance, $\{j_3=2,j_4=1\}$ means $(\tilde{g}^{(3)}_{\alpha,\beta,\Delta m,\Delta n})^2\cdot \tilde{g}^{(4)}_{\alpha,\beta,\Delta m,\Delta n}$. Note that the sum of all the $j_{k'}$ in a given configuration must match the power $j$ in \cref{eq:riemannToIntegral1}, $\sum_{\{j_{k'}\}}j_{k'}=j$. Using the fact that integrals with odd powers of $l$ vanish, we find that
\begin{equation}\label{eq:gaussianIntegralContributions}
    o(\{j_{k'}\},k)=\frac{1}{2}\left(k+\sum_{\{j_{k'}\}}k'\cdot j_{k'}\right)
\end{equation}
must be an integer, $o(\{j_{k'}\},k)\in\mathbb{Z}$, with $1/N^{o(\{j_{k'}\},k)-j}$ being the order of the resulting contribution. Using \cref{eq:riemannToIntegral1,eq:gaussianIntegralMoments}, the term corresponding to a given combination $\{\{j_{k'}\},k\}$ in \cref{table:SaddlePointContributionsOrders012} is given by
\begin{equation}\label{eq:generalTermSaddlePointApproximation}
    \sigma_{\alpha,\beta,\Delta m,\Delta n}\sqrt{N}e^{g^{(0)}_{\alpha,\beta,\Delta m,\Delta n}}\frac{\sigma^{2o}_{\alpha,\beta,\Delta m,\Delta n}(2o)!}{2^{o}o!}\frac{1}{k!}\tilde{s}^{(k)}_{\alpha,\beta,\Delta m,\Delta n}\Pi_{\{j_{k'}\}}\frac{(\tilde{g}^{(k')}_{\alpha,\beta,\Delta m,\Delta n})^{j_{k'}}}{j_{k'}!(k'!)^{j_{k'}}}.
\end{equation}
In the above, we included a combinatorial factor $\frac{j!}{\Pi_{\{j_{k'}\}}j_{k'}!}$ to correctly account for all possible combinations where more than one $j_{k'}$ is finite. Since we only require accuracy up to and including $1/N^2$, we need only consider the cases $o(\{j_{k'}\},k)-j=\{0,1,2\}$. We list all the possible combinations of $k$ and $\{j_{k'}\}$ satisfying $o(\{j_{k'}\},k)-j=\{0,1,2\}$ in \cref{table:SaddlePointContributionsOrders012},
\begin{equation}\label{table:SaddlePointContributionsOrders012}
\begin{tabular}{|l | l | l|} 
\hline
 $o-j=0$ & $o-j=1$ & $o-j=2$ \\ [0.5ex] 
 \hline
$k=0$ & $k=2$           & $k=4$ \\ 
      &  $j_3=1,k=1$    & $j_3=1,k=3$ \\
      &  $j_4=1,k=0$    & $j_4=1,k=2$  \\
      &  $j_3=2,k=0$    & $j_5=1,k=1$  \\
      &                 & $j_6=1,k=0$ \\ 
      &                 & $j_3=2,k=2$ \\
      &                 & $j_4=2,k=0$ \\ 
      &                 & $j_3=3,k=1$ \\
      &                 & $j_3=4,k=0$ \\
      &                 & $j_5=1,j_3=1,k=0$\\
      &                 & $j_4=1,j_3=2,k=0$\\
      &                 & $j_4=1,j_3=1,k=1$.\\
\hline
\end{tabular}
\end{equation}
The terms in \cref{table:SaddlePointContributionsOrders012} are to leading order $\mathcal{O}(N^0)$, $\mathcal{O}(N^{-1})$, and $\mathcal{O}(N^{-2})$, as one moves from the leftmost to the rightmost column.

Using \cref{eq:gaussianIntegralContributions,table:SaddlePointContributionsOrders012}, \cref{eq:riemannToIntegral1} evaluates to
\begin{equation}
\begin{aligned}\label{eq:sumOperatorExpectValNumberStatesSPtruncated}
   \mathcal{I}_{\alpha,\beta,\Delta m,\Delta n}&
    =\sigma_{\alpha,\beta,\Delta m,\Delta n}\sqrt{N}e^{g^{(0)}_{\alpha,\beta,\Delta m,\Delta n}}\Bigg[\nonumber \\
       o-j=0: &  \quad \tilde{s}^{(0)}_{\alpha,\beta,\Delta m,\Delta n} \nonumber \\
       \nonumber \\
    \textcolor{gray}{---}& \textcolor{gray}{--------------------------------------} \nonumber\\
       o-j=1:& +\frac{\sigma^2_{\alpha,\beta,\Delta m,\Delta n}}{2}\bigg[\tilde{s}^{(2)}_{\alpha,\beta,\Delta m,\Delta n} \nonumber \\
       &\quad +\sigma^2_{\alpha,\beta,\Delta m,\Delta n}\left(\tilde{s}^{(1)}_{\alpha,\beta,\Delta m,\Delta n}\tilde{g}^{(3)}_{\alpha,\beta,\Delta m,\Delta n}+\frac{1}{4}\tilde{g}^{(4)}_{\alpha,\beta,\Delta m,\Delta n}\tilde{s}^{(0)}_{\alpha,\beta,\Delta m,\Delta n}\right) \nonumber \\
      & \quad +\frac{5}{12}\sigma^4_{\alpha,\beta,\Delta m,\Delta n}(\tilde{g}^{(3)}_{\alpha,\beta,\Delta m,\Delta n})^2\tilde{s}^{(0)}_{\alpha,\beta,\Delta m,\Delta n}\bigg]\nonumber\\
      \nonumber \\
      \textcolor{gray}{---}& \textcolor{gray}{--------------------------------------} \nonumber\\
    o-j=2:& + \frac{\sigma^4_{\alpha,\beta,\Delta m,\Delta n}}{8}\bigg[ \frac{385}{144}\sigma^8_{\alpha,\beta,\Delta m,\Delta n}\tilde{s}^{(0)}_{\alpha,\beta,\Delta m,\Delta n}(\tilde{g}^{(3)}_{\alpha,\beta,\Delta m,\Delta n})^4 \nonumber \\
    \nonumber \\
    & \quad + \frac{35}{2}\sigma^6_{\alpha,\beta,\Delta m,\Delta n}\left(\frac{1}{3}\tilde{s}^{(1)}_{\alpha,\beta,\Delta m,\Delta n} (\tilde{g}^{(3)}_{\alpha,\beta,\Delta m,\Delta n})^3+\frac{1}{4}\tilde{s}^{(0)}_{\alpha,\beta,\Delta m,\Delta n} \tilde{g}^{(4)}_{\alpha,\beta,\Delta m,\Delta n} ( \tilde{g}^{(3)}_{\alpha,\beta,\Delta m,\Delta n})^2\right) \nonumber \\
    \nonumber \\
    & \quad + \frac{35}{6}\sigma^4_{\alpha,\beta,\Delta m,\Delta n}\Big(\tilde{s}^{(2)}_{\alpha,\beta,\Delta m,\Delta n}(\tilde{g}^{(3)}_{\alpha,\beta,\Delta m,\Delta n})^2+\tilde{s}^{(1)}_{\alpha,\beta,\Delta m,\Delta n}\tilde{g}^{(4)}_{\alpha,\beta,\Delta m,\Delta n}\tilde{g}^{(3)}_{\alpha,\beta,\Delta m,\Delta n} \nonumber \\
    &\qquad+ \frac{7}{35}\tilde{s}^{(0)}_{\alpha,\beta,\Delta m,\Delta n} \tilde{g}^{(5)}_{\alpha,\beta,\Delta m,\Delta n}\tilde{g}^{(3)}_{\alpha,\beta,\Delta m,\Delta n} + \frac{1}{8}\tilde{s}^{(0)}_{\alpha,\beta,\Delta m,\Delta n} (\tilde{g}^{(4)}_{\alpha,\beta,\Delta m,\Delta n})^2\Big)\nonumber \\
    \nonumber \\
    & \quad +\frac{1}{6} \sigma^2_{\alpha,\beta,\Delta m,\Delta n}\Big(20\tilde{s}^{(3)}_{\alpha,\beta,\Delta m,\Delta n} \tilde{g}^{(3)}_{\alpha,\beta,\Delta m,\Delta n} +15 \tilde{s}^{(2)}_{\alpha,\beta,\Delta m,\Delta n} \tilde{g}^{(4)}_{\alpha,\beta,\Delta m,\Delta n} \nonumber \\
    &\qquad +6\tilde{s}^{(1)}_{\alpha,\beta,\Delta m,\Delta n} \tilde{g}^{(5)}_{\alpha,\beta,\Delta m,\Delta n} + \tilde{s}^{(0)}_{\alpha,\beta,\Delta m,\Delta n} \tilde{g}^{(6)}_{\alpha,\beta,\Delta m,\Delta n}\Big) \nonumber \\
    \nonumber \\
    & \quad + \tilde{s}^{(4)}_{\alpha,\beta,\Delta m,\Delta n}
    \bigg] \nonumber \\
    \nonumber \\
    &+ \mathcal{O}\left(N^{-3}\right)\Bigg],
\end{aligned}
\end{equation}
where the $o-j=0,1,2$ labels designate terms coming from the first, second and third columns in \cref{table:SaddlePointContributionsOrders012}, respectively. See also \texttt{paperSaddlePointJCmodelPublicationFinal.nb}, for an implementation of \cref{eq:sumOperatorExpectValNumberStatesSPtruncated}.  

\subsection{Applying \texorpdfstring{\cref{eq:sumOperatorExpectValNumberStatesSPtruncated}}{C21}: a worked example}
To finish this appendix, we present one worked example where we use \cref{eq:sumOperatorExpectValNumberStatesSPtruncated} to calculate $\mathcal{I}_{\alpha,\beta,\Delta m,\Delta n}$ for the $[\rho_\text{q}]_{\uparrow\uparrow}$ component of the reduced qubit density matrix. By summing over the relevant $\{\alpha,\beta,\Delta m,\Delta n\}$, we then obtain $[\rho_{\text{q}}]_{\uparrow\uparrow}$. We will calculate the $\alpha=-\beta$ (``fast-decaying'') terms to order $\mathcal{O}(1/N)$, and the $\alpha=\beta$ (those which survive at $t\gtrsim t_r$) terms to order $\mathcal{O}(1/N^2)$. See also the \texttt{saddlePointUpToSecondOrder} function in Section 3 of \texttt{paperSaddlePointJCmodelPublicationFinal.nb} for an implementation of the method in \textit{Mathematica}, up to and including $\mathcal{O}(1/N^2)$ terms. From \cref{eq:rhoQubitExact}, the $s^{(0)}_{\alpha,\beta,\Delta m,\Delta n}(l)$ components for $[\rho_\text{q}]_{\uparrow\uparrow}$ read
\begin{equation}\label{eq:sCoeffsQubitDMupup}
    \begin{aligned}
    s^{(0)}_{+,+,0,0}(l)&=c^2(\theta_l)|\tilde{c}_{l,+}|^2, & s^{(0)}_{+,-,0,0}(l)&=s(\theta_l)c(\theta_l)\tilde{c}^*_{l,-}\tilde{c}_{l,+}, \\
    s^{(0)}_{-,-,0,0}(l)&=s^2(\theta_l)|\tilde{c}_{l,-}|^2, &
    s^{(0)}_{-,+,0,0}(l)&=s(\theta_l)c(\theta_l)\tilde{c}^*_{l,+}\tilde{c}_{l,-}.
    \end{aligned}
\end{equation}
where $\tilde{c}_{l,\pm}=c_{l,\pm}/\left(\frac{e^{-N/2}N^{l/2}}{\sqrt{l!}}\right)$, and the $c_{l,\pm}$ coefficients are defined in \cref{eq:JCModelFullSystemWavefunctionCoeffs}. We grouped the terms in two columns: $\alpha=\beta$ on the left and $\alpha=-\beta$ on the right. We replace $c_\pm$ by $r$ and $\phi_{\pm}$ using the parametrization given in \cref{eq:cPlusMinusParametrization},
\begin{equation}
\begin{aligned}
    \tilde{c}_{l,+}&=e^{il\phiJC}\left[\sqrt{\frac{1}{2}(1+r)}e^{i\phi_+}\left(c(\theta_l)c(\thetaJC)+\sqrt{\frac{l}{N}}s(\theta_l)s(\thetaJC)\right)+\sqrt{\frac{1}{2}(1-r)}e^{i\phi_-}\left(c(\theta_l)s(\thetaJC)-\sqrt{\frac{l}{N}}s(\theta_l)c(\thetaJC)\right)\right],\\
    \tilde{c}_{l,-}&=e^{il\phiJC}\left[\sqrt{\frac{1}{2}(1+r)}e^{i\phi_+}\left(s(\theta_l)c(\thetaJC)-\sqrt{\frac{l}{N}}c(\theta_l)s(\thetaJC)\right)+\sqrt{\frac{1}{2}(1-r)}e^{i\phi_-}\left(s(\theta_l)s(\thetaJC)+\sqrt{\frac{l}{N}}c(\theta_l)c(\thetaJC)\right)\right].
    \end{aligned}
\end{equation}
Using \cref{eq:sCoeffsQubitDMupup,eq:saddlePointFormula}, we calculate the coefficients $\tilde{s}^{(k)}_{\alpha,\beta,\Delta m,\Delta n}=N^{k/2}\tilde{s}^{(k)}_{\alpha,\beta,\Delta m,\Delta n}(L_{\alpha,\beta,\Delta m,\Delta n})$ for $k=0,1,2,3,4$, i.e., all the values of $k$ that appear in \cref{eq:sumOperatorExpectValNumberStatesSPtruncated}. We Taylor expand in $\epsilon$, keeping all terms up to and including $\mathcal{O}(\epsilon^2)$ for $\alpha=\beta$ and $\mathcal{O}(\epsilon)$ for $\alpha=-\beta$. As an example, the  $\tilde{s}^{(0)}_{+,+,0,0}$ component reads
\begin{equation}\label{eq:sTilderhouuKequals0}
    \begin{aligned}
        \tilde{s}^{(0)}_{+,+,0,0}&=\frac{1}{4}(1+r)\left(1+\frac{\Delta}{\OmegaJC}\right)+\textcolor{brown}{\frac{\epsilon}{N}\frac{g\sqrt{N}}{8\OmegaJC}\sqrt{1+\frac{\Delta}{\OmegaJC}}\left(-(1+r)\left(1-\frac{\Delta}{\OmegaJC}\right)^{3/2}+\sqrt{1-r^2}\cos(\Delta\Phi)\left(1+\frac{\Delta}{\OmegaJC}\right)^{3/2}\right)} \\
        & \textcolor{orange}{+\frac{\epsilon ^2}{N^2}\frac{g \sqrt{N}}{48\OmegaJC^5} \Bigg(g \sqrt{N} \left(2 g^2N(11 \Delta +2 (\Delta  r-\OmegaJC)-5 r\OmegaJC)+\Delta ^2 (\Delta +5 (\OmegaJC-\Delta  r)-r \OmegaJC)\right)}\\
        &\qquad \textcolor{orange}{+\sqrt{1-r^2} \cos(\Delta\Phi) \left(\Delta^3 (\Delta +\OmegaJC)+8 g^4N^2+2 \Delta  g^2N(9\Delta+8\OmegaJC)\right)\Bigg)}.
    \end{aligned}
\end{equation}
The last step is to sum \cref{eq:sumOperatorExpectValNumberStatesSPtruncated} over the four relevant combinations of $\{\alpha,\beta,\Delta m,\Delta n\}$. For $[\rho_{\text{q}}]_{\uparrow\uparrow}$, these are $\alpha,\beta=\pm 1$ and $\Delta m=\Delta n=0$, see \cref{eq:rhoQubitExact}. Due to the $-2N\omega_s^2t^2$ factor in \cref{eq:gFastTermsAtSP}, the terms with $\alpha=-\beta$ decay much faster than those with $\alpha=\beta$, disappearing at readout time $t_r\sim\gamma_f^{-1}=(\sqrt{2N}\omega_s)^{-1}$. We keep track of the orders of $1/\sqrt{N}$ in the resulting operator expectation value using the following notation,
\begin{equation}\label{eq:SzComponentJCmodelUpToSecondOrder}
    \langle\hat{O}\rangle=\sum_{k=0}^{4}\left(\frac{\epsilon}{N}\right)^{\frac{k}{2}}\underbrace{\langle\hat{O}\rangle|_{\alpha=\beta,\frac{k}{2}}}_{\text{surviving}} + \sum_{k=0}^{2}\left(\frac{\epsilon}{N}\right)^{\frac{k}{2}}\underbrace{\langle\hat{O}\rangle|_{\alpha=-\beta,\frac{k}{2}}}_{\text{fast-decaying}} + \mathcal{O}(N^{-3/2}),
\end{equation}
going up to order $\mathcal{O}(1/N^2)$ for the $\alpha=\beta$ terms and $\mathcal{O}(1/N)$ for the $\alpha=-\beta$ terms.
For $\langle 
\hat{S}^z\rangle =2[\rho_{\text{q}}]_{\uparrow\uparrow}-1$, we find the result given in \cref{eq:SzSaddlePointGeneral}.

\section{Saddlepoint expressions \texorpdfstring{for $\langle\hat{\vb{S}}(t)\rangle$, $\langle\hat{a}(t)\rangle$, $\langle\hat{a}^2(t)\rangle$ and $\langle\hat{n}(t)\rangle$ in the JC model}{}}\label{app:SaddlePointExpressions}

In this appendix, we collect the saddlepoint expressions for the expectation values of relevant observables in the JC model, which we work out using the method presented in App.~\ref{app:SaddlePoint}. We calculate the Bloch vector for the qubit, $\langle\hat{\vb{S}}(t)\rangle$, as well as the annihilation, annihilation squared and number operators for the resonator $\langle\hat{a}(t)\rangle$, $\langle\hat{a}^2(t)\rangle$ and $\langle\hat{n}(t)\rangle$. With these quantities, we can also calculate the purity of the qubit $\Tr[\hat{\rho}_{\text{q}}^2]=\frac{1}{2}(1+|\langle\hat{\vb{S}}\rangle|^2)$ and the work done and degree of quadrature spread in the resonator state.

Starting with the $\langle\hat{S}^z\rangle$ component of the qubit Bloch vector, we have
\begin{equation}\label{eq:SzSaddlePointGeneral}
    \begin{aligned}
    &\textcolor{gray}{\textit{*surviving*}} \\
        \\
    \langle\hat{S}^z\rangle|_{\alpha=\beta,0}&=\frac{\Delta}{\OmegaJC}\frac{1}{2}(1+r)-\frac{\Delta}{\OmegaJC}\frac{1}{2}(1-r), \\
    \textcolor{purple}{\langle\hat{S}^z\rangle|_{\alpha=\beta,\frac{1}{2}}}&=\textcolor{purple}{0}, \\
    \textcolor{brown}{\langle\hat{S}^z\rangle|_{\alpha=\beta,1}}&=\textcolor{brown}{\frac{2g^2N\Delta^2}{\OmegaJC^4}\left(1-r\frac{\Delta}{\OmegaJC}+\frac{2g\sqrt{N}}{\OmegaJC}\sqrt{1-r^2}\cos(\Delta\Phi)\right)}, \\
    \textcolor{gray}{\langle\hat{S}^z\rangle|_{\alpha=\beta,\frac{3}{2}}}&=\textcolor{gray}{0},\\
    \textcolor{orange}{\langle\hat{S}^z\rangle|_{\alpha=\beta,2}}&=\textcolor{orange}{\frac{\epsilon^2}{N^2}\frac{4\omega_s^2 N^2\Delta^2}{\OmegaJC^6}(8g^2N-\Delta^2)\left(\frac{2g\sqrt{N}}{\OmegaJC}\sqrt{1-r^2}\cos(\Delta\Phi)+\left(1-r\frac{\Delta}{\OmegaJC}\right)\right)}, \\
    \\
    &\textcolor{gray}{\textit{*fast-decaying*}} \\
    \\
    \langle\hat{S}^z\rangle|_{\alpha=-\beta,0}&=e^{-\gamma_f^2t^2}\frac{2g\sqrt{N}}{\OmegaJC}\frac{1}{2}\sqrt{1-r^2}\cos(\OmegaJC t-\Delta\Phi)\left(-\frac{\OmegaJC}{\Delta}\left(1-\frac{\Delta}{\OmegaJC}\right)+\frac{\OmegaJC}{\Delta}\left(1+\frac{\Delta}{\OmegaJC}\right)\right), \\
    \textcolor{purple}{\langle\hat{S}^z\rangle|_{\alpha=-\beta,\frac{1}{2}}}&=\textcolor{purple}{e^{-\gamma_f^2t^2}\frac{4gN}{\OmegaJC}\omega_s t\Bigg(\sin(\OmegaJC t)\frac{g\sqrt{N}}{\OmegaJC}\left(1-r\frac{\Delta}{\OmegaJC}\right)} \\
    &\qquad \textcolor{purple}{+\sqrt{1-r^2}\sin(\OmegaJC t-\Delta\Phi)\left(\frac{2}{3}\frac{N\omega_s^2t^2}{\OmegaJC^2}(\Delta^2-2g^2N)-\frac{1}{\OmegaJC^2}(\Delta^2+g^2N)\right)\Bigg) }, \\
    \textcolor{brown}{\langle\hat{S}^z\rangle|_{\alpha=-\beta,1}}&=\textcolor{brown}{-e^{\gamma_f^2t^2}\frac{2g^2N}{\OmegaJC^2}\Bigg(\left(1-r\frac{\Delta}{\OmegaJC}\right)\Big(\frac{\Delta^2}{\OmegaJC^2}-\frac{2g^4N}{\OmegaJC^4}t^2\left(3\Delta^2-2g^2N\right) -\frac{8}{3}\frac{g^8N^2}{\OmegaJC^6}t^4(2g^2N-\Delta^2)\Big)\cos(\OmegaJC t)} \\
     &\qquad \textcolor{brown}{+ \frac{2g\sqrt{N}}{\OmegaJC^3}\sqrt{1-r^2}\bigg( \cos(\OmegaJC t-\Delta\Phi)\Big(\Delta^2 + \frac{\omega_s}{\OmegaJC}t^2(\Delta^4-6\Delta^2 g^2N + 3g^4N^2)} \\
    &\qquad\quad \textcolor{brown}{- \frac{1}{3} \frac{\omega_s^3 N}{\OmegaJC}t^4(5\Delta^4-20\Delta^2g^2N + 8g^4N^2) + \frac{4}{9}\frac{\omega_s^5 N^2}{\OmegaJC}t^6(\Delta^2-2g^2N)^2\Big)\bigg)} \\
    &\qquad \textcolor{brown}{ + g^2N\left(1-4\omega_s^2Nt^2 \right)\sin(\OmegaJC t)\sin(\Delta\Phi) \Bigg)}.
    \end{aligned}
\end{equation}
A sanity check for \cref{eq:SzComponentJCmodelUpToSecondOrder} (and all subsequent expressions) is to set $t=0$. At this time, the qubit and cavity are unentangled, and we should recover the Rabi result, \cref{eq:RabiModelBlochVectorTimeEvolution}. Indeed, we can check that the $1/\sqrt{N}$, $1/N$ corrections from the surviving and fast-decaying terms cancel each other out at $t=0$. This, in fact, extends to all higher orders of $N^{-k/2}$, $k\geq 3$, but we only calculated the fast-decaying terms up to $\mathcal{O}(1/N)$ in \cref{eq:SzComponentJCmodelUpToSecondOrder}. 

For the $\langle\hat{S}^+\rangle=\langle\hat{S}^x\rangle+i\langle\hat{S}^y\rangle$ component of the Bloch vector, we have
\begin{equation}\label{eq:SPlusSaddlePointGeneral}
    \begin{aligned}
    &\textcolor{gray}{\textit{*surviving*}} \\
        \\
  \langle\hat{S}^+\rangle|_{\alpha=\beta,0}&=e^{i \phiJC}\frac{2g\sqrt{N}}{\OmegaJC}\left(e^{-i\omega_s t}\frac{1}{2}(1+r)+e^{i(\pi+\omega_s t)}\frac{1}{2}(1-r)\right), \\
  \textcolor{purple}{\langle\hat{S}^+\rangle|_{\alpha=\beta,\frac{1}{2}}}&=\textcolor{purple}{0}, \\
\textcolor{brown}{\langle\hat{S}^+\rangle|_{\alpha=\beta,1}}&=-\textcolor{brown}{e^{i \phiJC}\frac{\omega_s N}{\OmegaJC}\Bigg(\frac{2g\sqrt{N}}{\OmegaJC}\Big(\cos(\omega_s t)\left(\frac{r}{\OmegaJC^2}(3\Delta^2+4g^2N)-\frac{2\Delta}{\OmegaJC}\right)-i\sin(\omega_s t)\left(\frac{2}{\OmegaJC^2}(\Delta^2+g^2N)-r\frac{\Delta}{\OmegaJC}\right)\Big)},  \\
    &\textcolor{brown}{+\sqrt{1-r^2}\Big(\cos(\omega_s t)\left(\frac{\Delta}{\OmegaJC^3}(\Delta^2-4g^2N)\cos(\Delta\Phi)-i\sin(\Delta\Phi)\right) -\frac{\Delta}{\OmegaJC}\sin(\omega_s t)\left(i\frac{\Delta}{\OmegaJC} \cos(\Delta\Phi) + \sin(\Delta\Phi)\right)\Big) \Bigg)}, \\
   \textcolor{gray}{\langle\hat{S}^+\rangle|_{\alpha=\beta,\frac{3}{2}}}&=\textcolor{gray}{-e^{i\phiJC}N^{\frac{3}{2}}2i\Bigg(+2\sqrt{1-r^2}\frac{\omega_s^3 N t}{\OmegaJC^2}\cos(\Delta\Phi)\left(\cos(\omega_s t)-i\frac{\Delta}{\OmegaJC}\sin(\omega_s t)\right)} \\
    &\qquad\quad \textcolor{gray}{+\frac{g\sqrt{N}\omega_s^2 t}{\OmegaJC^2}\Big(\left(\frac{2}{\OmegaJC^2}(g^2N-\Delta^2)+r\frac{\Delta}{\OmegaJC}\right)\cos(\omega_s t)-i\left(\frac{2}{\OmegaJC^2}(g^2N-\Delta^2)r+\frac{\Delta}{\OmegaJC}\right)\sin(\omega_s t)\Big)\Bigg)},\\
\textcolor{orange}{\langle\hat{S}^+\rangle|_{\alpha=\beta,2}}&=\textcolor{orange}{-e^{i\phiJC}\frac{\omega_s^2 N^2}{\OmegaJC^2}\Bigg(
\sqrt{1-r^2}\Bigg(\sin (\omega_s t)\Big(\frac{3\Delta}{\OmegaJC^3} (\Delta ^2-6 g^2 N)(\sin (\Delta \phi )+i \frac{\Delta}{\OmegaJC}\cos(\Delta \phi))\Big)}
\\
&\textcolor{orange}{+\cos (\omega_s t)\Big(\frac{\Delta}{\OmegaJC} \cos(\Delta \phi )\frac{1}{\OmegaJC^4}(-3 \Delta^4-112 g^4N^2+56\Delta^2g^2N)+i\frac{1}{\OmegaJC^2}\sin(\Delta\Phi)(3 \Delta^2-4 g^2 N)\Big)\Bigg)}\\
&\textcolor{orange}{\frac{2g\sqrt{N}}{\OmegaJC}\Bigg(\cos (\omega_s t)\Big(2 \omega_s^2 N r t^2 +16 \frac{\omega_s^2 N^2 r}{\OmegaJC^2}+8 \frac{\Delta \omega_s N}{\OmegaJC^2}\left(3\frac{\Delta}{\OmegaJC} r-4\right)+\frac{\Delta^3}{\OmegaJC}\left(16 -19 \frac{\Delta}{\OmegaJC} r \right)\Big)}\\
&\textcolor{orange}{-i \sin (\omega_s t)\Big(2\omega_s^2N t^2+14 \frac{\omega_s^2}{\OmegaJC^2} N^2+2 \frac{\Delta \omega_s}{\OmegaJC^2} N\left(16 \frac{\Delta}{\OmegaJC} -9 r\right)+3\frac{\Delta^3}{\OmegaJC^3} \left(r-2\frac{\Delta}{\OmegaJC} \right)\Big)\Bigg)\Bigg)},\\
\\
&\textcolor{gray}{\textit{*fast-decaying*}} \\
    \\
    \langle\hat{S}^+\rangle|_{\alpha=-\beta,0}&=e^{i\phiJC}e^{-\gamma_f^2t^2}\frac{1}{2}\sqrt{1-r^2}\left(e^{i(-\omega_s t-(\OmegaJC t-\Delta\Phi))}\left(1-\frac{\Delta}{\OmegaJC}\right) +e^{i(\pi+\omega_s t+(\OmegaJC t-\Delta\Phi))}\left(1+\frac{\Delta}{\OmegaJC}\right)\right) , \\
    \textcolor{purple}{\langle\hat{S}^+\rangle|_{\alpha=-\beta,\frac{1}{2}}}&=\textcolor{purple}{ i e^{i\phiJC} e^{-\gamma_f^2t^2}\frac{2g\sqrt{N}}{\OmegaJC}\sqrt{N}\omega_s t\Bigg( \left(1-r\frac{\Delta}{\OmegaJC}\right)\left(\cos((\OmegaJC+\omega_s)t)+i\frac{\Delta}{\OmegaJC}\sin((\OmegaJC+\omega_s)t)\right)}  \\
    &\quad \textcolor{purple}{+\frac{g\sqrt{N}}{\OmegaJC}\sqrt{1-r^2}\bigg(\left(1+\frac{2}{3}\frac{\omega_s t^2}{\OmegaJC}(\Delta^2-2g^2N)\right)\cos((\OmegaJC+\omega_s)t-\Delta\Phi)} \\
    &\qquad \quad \textcolor{purple}{ + i\frac{\Delta}{\OmegaJC}\left(3+\frac{2}{3}\frac{\omega_s t^2}{\OmegaJC}(\Delta^2-2g^2N)\right)\sin((\OmegaJC+\omega_s)t-\Delta\Phi)\bigg)\Bigg)}, \\
\end{aligned}
\end{equation}

\begin{equation*}
\begin{aligned}
    \textcolor{brown}{\langle\hat{S}^+\rangle|_{\alpha=-\beta,1}}&=\textcolor{brown}{e^{i\phiJC}e^{-\gamma_f^2 t^2}\frac{2\omega_sN}{\OmegaJC}
\Bigg( \frac{g \sqrt{N}}{\OmegaJC}\Bigg[}\\
&\textcolor{brown}{\qquad \cos ((\OmegaJC+\omega_s) t)
\Bigg(
-\frac{2 \Delta}{\OmegaJC}
+\left(1-\frac{\Delta r}{\OmegaJC}\right)
\Bigg[
\frac{\Delta \omega_s t^2}{\OmegaJC^2}(10 g^2 N-\Delta^2)
-\frac{4 \omega_s^3 \Delta N t^4}{3 \OmegaJC^2}(2 g^2 N-\Delta^2)
\Bigg]
+r \left(\frac{2 \Delta^2}{\OmegaJC^2}+1\right)
\Bigg)}
\\
&\textcolor{brown}{\qquad +\, i \sin ((\OmegaJC+\omega_s)t)
\Bigg(
\left(1-\frac{\Delta r}{\OmegaJC}\right)
\Bigg[
\frac{\omega_s t^2}{\OmegaJC}\left(-\frac{\Delta^4}{\OmegaJC^2}+16 N^2 \omega_s^2+2 N \OmegaJC \omega_s\right)
-\frac{4\omega_s^3t^4 N}{3 \OmegaJC}(2 g^2 N-\Delta^2)
\Bigg]
+\frac{\Delta r}{\OmegaJC}
\Bigg)\Bigg]}
\\
&\textcolor{brown}{\quad + \sqrt{1-r^2}
\Bigg[i \sin ((\OmegaJC+\omega_s)t-\Delta\Phi)
\Bigg(\frac{\Delta^2+3 g^2 N}{2 \OmegaJC^2}+
\frac{4 \omega_s^5 t^6 N^2}{9 \OmegaJC^3}(\Delta^2-2 g^2 N)^2
}
\\
&\textcolor{brown}{\qquad\quad
+\frac{\omega_s^2 t^2N}{\OmegaJC^2}(g^2 N-4\Delta^2)
-\frac{\omega_s^3t^4N}{3 \OmegaJC^3}(\Delta^4+24 g^4 N^2-20 \Delta^2 g^2 N)
\Bigg)}
\\
&\textcolor{brown}{\qquad -\frac{\omega_s N}{2 \OmegaJC} (4 N t^2 \omega_s^2-1)\sin((\OmegaJC+\omega_s)t + \Delta \phi)}
\\
&\textcolor{brown}{\qquad +\frac{\Delta}{\OmegaJC}\cos((\OmegaJC+\omega_s)t-\Delta \phi)
\Bigg(
\frac{4 \omega_s^5 t^6 N^2}{9 \OmegaJC^3}(\Delta^2-2 g^2 N)^2
-\frac{1}{2 \OmegaJC^2}(3 g^2 N-\Delta^2)
}
\\
&\textcolor{brown}{\qquad\quad
+\frac{\omega_s^2 t^2 N}{\OmegaJC^2}(17 g^2 N-6 \Delta^2)
-\frac{\omega_s^3 t^4 N}{3\OmegaJC^3}(\Delta^4+40 g^4 N^2-28 \Delta^2 g^2 N)
\Bigg)}
\\
&\textcolor{brown}{\qquad +\frac{\omega_s \Delta N }{2\OmegaJC^2}(4 N t^2 \omega_s^2-1)\cos((\OmegaJC+\omega_s)t+\Delta \phi) \Bigg]
\Bigg)}.
    \end{aligned}
\end{equation*}

Moving on to the resonator, we only need to calculate terms up to and including $\mathcal{O}(1/N)$ corrections, both for the surviving and fast-decaying terms. For the expectation value of the annihilation operator, we obtain
\begin{equation}\label{eq:annihilationSaddlePointGeneral}
\begin{aligned}
&\textcolor{gray}{\textit{*surviving*}} \\
\langle\hat{a}\rangle|_{\alpha=\beta,0} &= \sqrt{N}e^{i\phiJC} \left(e^{-i\omega_s t}\frac{1}{2}(1+r) + e^{i\omega_s t}\frac{1}{2}(1-r)\right), \\
\textcolor{purple}{\langle\hat{a}\rangle|_{\alpha=\beta,\frac{1}{2}} }&=\textcolor{purple}{0}, \\
\textcolor{brown}{\langle\hat{a}\rangle|_{\alpha=\beta,1} }&=\textcolor{brown}{\sqrt{N} e^{i\phiJC}\frac{g\sqrt{N}}{\OmegaJC}\bigg(\sin(\omega_s t)\frac{ig\sqrt{N}}{\OmegaJC}\left(\frac{1}{\OmegaJC^2}(\Delta^2+2g^2N)r-\frac{\Delta}{\OmegaJC}\right)}\\
&\textcolor{brown}{-\frac{1}{2}\sqrt{1-r^2}\left(\cos(\omega_s t)\left(\cos(\Delta\Phi)-i\frac{\Delta}{\OmegaJC}\sin(\Delta\Phi)\right)-i\sin(\omega_s t)\left(\frac{\Delta^3}{\OmegaJC^3}\cos(\Delta\Phi)-i\sin(\Delta\Phi)\right)\right)\bigg)}, \\
\textcolor{gray}{\langle\hat{a}\rangle|_{\alpha=\beta,\frac{3}{2}}}&=\textcolor{gray}{\sqrt{N}e^{i\phiJC}\frac{i\omega_s^2t}{2\OmegaJC^3}\bigg[\left(2(\Delta^2+g^2N)-\Delta\OmegaJC\right)(1+r)e^{-i\omega_s t}+\left(2(\Delta^2+g^2N)+\Delta\OmegaJC\right)(-1+r)e^{i\omega_s t}}\\
&+\textcolor{gray}{4g\sqrt{N}\sqrt{1-r^2}\cos(\Delta\Phi)\left(-\Delta\cos(\omega_s t)+i\OmegaJC\sin(\omega_s t)\right)\bigg]}
\\
\textcolor{orange}{\langle\hat{a}\rangle|_{\alpha=\beta,2}}&=\textcolor{orange}{\frac{\omega_s^2}{\OmegaJC^6}\bigg[-\OmegaJC^2\left(\OmegaJC^2-2r\Delta\OmegaJC+2g^4Nt^2\right)\cos(\omega_s t)}\\
&\textcolor{orange}{+i\sin(\omega_s t)\left(-2g^4N^2r-6g^2N\Delta(r\Delta+\OmegaJC)+\Delta^3(6\OmegaJC-7r\Delta)+2g^4N\OmegaJC^2 r t^2\right)\bigg]}\\
&\textcolor{orange}{\sqrt{1-r^2}\frac{\omega_s g\sqrt{N}}{2N\OmegaJC^8}\bigg[2\OmegaJC^4\left(\Delta^2\cos(\omega_s t)\cos(\Delta\Phi)-(\Delta^2+g^2N)\sin(\omega_s t)\sin(\Delta\Phi)\right)}\\
&\textcolor{orange}{+i\Delta\OmegaJC\left(g^2N(32g^2N-7\Delta^2)\sin(\Delta\Phi-\omega_s t) + \Delta^2(7g^2N-2\Delta^2)\sin(\Delta\Phi+\omega_s t)\right)\bigg]} \\
\\
&\textcolor{gray}{\textit{*fast-decaying*}} \\
\langle\hat{a}\rangle|_{\alpha=-\beta,0} &=0, \\
\textcolor{purple}{\langle\hat{a}\rangle|_{\alpha=-\beta,\frac{1}{2}} }&=\textcolor{purple}{0}, \\
\textcolor{brown}{\langle\hat{a}\rangle|_{\alpha=-\beta,1} }&=\textcolor{brown}{ \sqrt{N} e^{i\phiJC}e^{-\gamma_f^2 t^2} \frac{g\sqrt{N}}{2\OmegaJC}\sqrt{1-r^2}\left(\cos((\OmegaJC+\omega_s)t-\Delta\Phi) + i\frac{\Delta}{\OmegaJC}\sin((\OmegaJC+\omega_s)t-\Delta\Phi)\right)}.
\end{aligned}
\end{equation}
The annihilation squared operator has expectation value
\begin{equation}\label{eq:annihilationSquaredSaddlePointGeneral}
\begin{aligned}
&\textcolor{gray}{\textit{*surviving*}} \\
\langle\hat{a}^2\rangle|_{\alpha=\beta,0} &= N e^{2i\phiJC} \left(e^{-2i\omega_s t}\frac{1}{2}(1+r) + e^{2i\omega_s t}\frac{1}{2}(1-r)\right), \\
\textcolor{purple}{\langle\hat{a}^2\rangle|_{\alpha=\beta,\frac{1}{2}} }&=\textcolor{purple}{0}, \\
\textcolor{brown}{\langle\hat{a}^2\rangle|_{\alpha=\beta,1} }&=\textcolor{brown}{N e^{2i\phiJC}\frac{g\sqrt{N}}{\OmegaJC}\bigg(\sin(2\omega_s t)\left(\frac{ig\sqrt{N}}{\OmegaJC}\left(\frac{1}{\OmegaJC^2}(\Delta^2+2g^2N)r-\frac{\Delta}{\OmegaJC}\right)\right)}\\
&\textcolor{brown}{-\sqrt{1-r^2}\bigg(\cos(2\omega_s t)\left(\cos(\Delta\Phi)-i\frac{\Delta}{\OmegaJC}\sin(\Delta\Phi)\right)}\\
&\textcolor{brown}{-i\sin(2\omega_s t)\left(\frac{\Delta}{\OmegaJC^3}(\Delta^2+2g^2N)\cos(\Delta\Phi)-i\sin(\Delta\Phi)\right)\bigg)\bigg)}, \\
\textcolor{gray}{\langle\hat{a}^2\rangle|_{\alpha=\beta,\frac{3}{2}}}&=\textcolor{gray}{N e^{2i\phiJC}\frac{i\omega_s^2t}{\OmegaJC^3}\bigg[\left(3(\Delta^2+2g^2N)-\Delta\OmegaJC\right)(1+r)e^{-2i\omega_s t}}\\
&\textcolor{gray}{\quad +\left(3(\Delta^2+2g^2N)+\Delta\OmegaJC\right)(-1+r)e^{2i\omega_s t}}\\
&+\textcolor{gray}{4g\sqrt{N}\sqrt{1-r^2}\cos(\Delta\Phi)\left(-\Delta\cos(2\omega_s t)+i\OmegaJC\sin(2\omega_s t)\right)\bigg]}
\\
&+\textcolor{orange}{\frac{\omega_s^2}{\OmegaJC^2}\Bigg[4\left(r\frac{\Delta}{\OmegaJC}-\frac{2g^4Nt^2}{\OmegaJC^2}\right)\cos(2\omega_s t)+i\sin(2\omega_s t)\bigg(\frac{\Delta^4}{\OmegaJC^4}\left(9\frac{\Delta}{\OmegaJC}-13 r\right)}\\
&\textcolor{orange}{+\frac{42g^2N\Delta^2}{\OmegaJC^4}\left(\frac{\Delta}{\OmegaJC}-r\right) + \frac{2g^4N^2}{\OmegaJC^4}\left(12\frac{\Delta}{\OmegaJC}-25r\right)+r\frac{8g^4Nt^2}{\OmegaJC^2}\bigg)\Bigg]}
\\
&\textcolor{orange}{+\sqrt{1-r^2}\frac{2g\omega_s}{\sqrt{N}\OmegaJC^8}\Bigg[2\cos(2\omega_st)\cos(\Delta\Phi)\left(\Delta^6+32g^6N^3+2g^2N\Delta^2(5\Delta^2+16g^2N)\right)}\\
&\textcolor{orange}{-\OmegaJC^4(2\Delta^2+5g^2N)\sin(2\omega_s t)\sin(\Delta\Phi)-2i\Delta^3\OmegaJC^3\cos(2\omega_st)\sin(\Delta\Phi)}\\
&\textcolor{orange}{-2i\Delta\OmegaJC\left(\Delta^4+2\Delta^2g^2N+7g^4N^2\right)\sin(2\omega_s t)\cos(\Delta\Phi)\Bigg]} \\
&\textcolor{gray}{\textit{*fast-decaying*}} \\
\langle\hat{a}^2\rangle|_{\alpha=-\beta,0} &=0, \\
\textcolor{purple}{\langle\hat{a}^2\rangle|_{\alpha=-\beta,\frac{1}{2}} }&=\textcolor{purple}{0}, \\
\textcolor{brown}{\langle\hat{a}^2\rangle|_{\alpha=-\beta,1} }&=\textcolor{brown}{N e^{2i\phiJC}e^{-\gamma_f^2 t^2} \frac{g\sqrt{N}}{\OmegaJC}\sqrt{1-r^2}\bigg(\cos((\OmegaJC+2\omega_s)t-\Delta\Phi) }\\
&\textcolor{brown}{ \quad + i\frac{\Delta}{\OmegaJC}\sin((\OmegaJC+2\omega_s)t-\Delta\Phi)\bigg)}.
\end{aligned}
\end{equation}
Finally, the number of photons in the resonator is given by
\begin{equation}\label{eq:numberOperatorSaddlePointGeneral}
\begin{aligned}
&\textcolor{gray}{\textit{*surviving*}} \\
\langle\hat{n}\rangle|_{\alpha=\beta,0} &=N, \\
\textcolor{purple}{\langle\hat{n}\rangle|_{\alpha=\beta,\frac{1}{2}} }&=\textcolor{purple}{0}, \\
\textcolor{brown}{\langle\hat{n}\rangle|_{\alpha=\beta,1} }&=\textcolor{brown}{-\frac{g\sqrt{N}}{\OmegaJC}\sqrt{1-r^2}\cos(\Delta\Phi)},\\
\textcolor{gray}{\langle\hat{n}\rangle|_{\alpha=\beta,\frac{3}{2}} }&=\textcolor{gray}{0},\\
\textcolor{orange}{\langle\hat{n}\rangle|_{\alpha=\beta,2} }&=\textcolor{orange}{\frac{g^2\Delta^2}{\OmegaJC^4}\left(\left(1-r\frac{\Delta}{\OmegaJC}\right)+\sqrt{1-r^2}\frac{2g\sqrt{N}}{\OmegaJC}\cos(\Delta\Phi)\right)},\\
\\
&\textcolor{gray}{\textit{*fast-decaying*}} \\
\langle\hat{n}\rangle|_{\alpha=-\beta,0} &=0, \\
\textcolor{purple}{\langle\hat{n}\rangle|_{\alpha=-\beta,\frac{1}{2}} }&=\textcolor{purple}{0}, \\
\textcolor{brown}{\langle\hat{n}\rangle|_{\alpha=-\beta,1} }&=\textcolor{brown}{e^{-\gamma_f^2 t^2} \frac{g\sqrt{N}}{\OmegaJC}\sqrt{1-r^2}\cos(\OmegaJC t-\Delta\Phi)}.
\end{aligned}
\end{equation}

\section{Saddlepoint expressions \texorpdfstring{for observables in the JC model at the sweet-spot angles, $r=\pm 1$, at $t\gtrsim t_r$}{}}\label{app:SaddlePointExpressionsAtSweetSpot}

In this appendix, we collect expressions for JC model observables at the sweet-spot angles, for $t\gtrsim t_r$. Setting $r=\pm 1$ in \cref{eq:SzSaddlePointGeneral,eq:SPlusSaddlePointGeneral,eq:annihilationSaddlePointGeneral,eq:annihilationSquaredSaddlePointGeneral,eq:numberOperatorSaddlePointGeneral}, and keeping only the surviving $\alpha=\beta$ terms, we obtain
\begin{equation}\label{eq:qubitDMsweetspotReadoutTime}
\begin{aligned}
    \langle\hat{S}^+(t\gtrsim t_r)\rangle&=\pm e^{i(\phiJC \mp i \omega_s t)}\frac{2g\sqrt{N}}{\OmegaJC}\Bigg(1-\textcolor{brown}{\frac{\omega_s}{\OmegaJC}\frac{1}{2}\left(\frac{1}{\OmegaJC^2}(5\Delta^2+6g^2N)\mp\frac{3\Delta}{\OmegaJC}+e^{\pm 2 i \omega_s t}\frac{1}{2}\left(1\mp\frac{\Delta}{\OmegaJC}\right)^2\right)} \\
    &\textcolor{gray}{\pm  \frac{i\omega_s^2t}{\OmegaJC}\left(\frac{2}{\OmegaJC^2}(\Delta^2-g^2N)\mp \frac{\Delta}{\OmegaJC}\right) }\\
    &\textcolor{orange}{+\frac{\omega_s^2}{2\OmegaJC^2}\Bigg(\left(1+e^{\pm 2i\omega_s t}\right) \left(\frac{1}{\OmegaJC^4}(19\Delta^4-16 g^4 N^2-24\Delta^2g^2N)\mp\frac{16\Delta}{\OmegaJC}\frac{1}{\OmegaJC^2}\left(\Delta^2-2g^2N\right)-2N\omega_s^2t^2\right)}\\
    &\quad \textcolor{orange}{+\left(1-e^{\pm 2i\omega_s t}\right)\left(\mp\frac{ 3\Delta}{\OmegaJC}\frac{1}{\OmegaJC^2}\left(\Delta^2-6g^2N\right)+\frac{1}{\OmegaJC^4}\left(6\Delta^4-14 g^4N^2-32\Delta^2g^2N\right)-2N\omega_s^2t^2\right)\Bigg)}\\
    &+\mathcal{O}(N^{-5/2}),
    \\
    \langle\hat{S}^z(t\gtrsim t_r)\rangle_{\pm}&=\pm\frac{\Delta}{\OmegaJC}\left(1\pm\textcolor{brown}{\frac{2\omega_s\Delta}{\OmegaJC^2}\left(1\mp\frac{\Delta}{\OmegaJC}\right)}\mp\textcolor{orange}{\frac{4\omega_s^2\Delta}{\OmegaJC^5}(\Delta^2-8g^2N)\left(1\mp\frac{\Delta}{\OmegaJC}\right)}\right)+\mathcal{O}(N^{-3}),
    \end{aligned}
\end{equation}
for the spin components and
\begin{equation}\label{eq:resonatorSweetspotReadoutTime}
\begin{aligned}
    \langle\hat{a}(t\gtrsim t_r)\rangle_{\pm}&=\sqrt{N}e^{i(\phiJC\mp\omega_s t)}\Bigg(1+\textcolor{brown}{\left(e^{\pm 2 i\omega_s t}-1\right)\frac{\omega_s}{\OmegaJC}\frac{1}{4}\left(1\mp\frac{\Delta}{\OmegaJC}\right)^2}  \textcolor{gray}{\pm \frac{i\omega_s^2t}{\OmegaJC^3}\left(2(\Delta^2+g^2N)\mp\Delta\OmegaJC\right)}\\
    &+\textcolor{orange}{\frac{\omega_s^2}{2\OmegaJC^2}\bigg(-(1+e^{\pm 2i\omega_s t})\left(1\mp \frac{2\Delta}{\OmegaJC}+2\omega_s^2Nt^2\right)}\\
    &\textcolor{orange}{+(1-e^{\pm 2 i\omega_s t})\left(\frac{2\omega_s^2N^2}{\OmegaJC^2}+\frac{6\omega_s N\Delta^2}{\OmegaJC^3}+\frac{7\Delta^4}{\OmegaJC^4}-2\omega_s^2Nt^2 \mp \frac{6\Delta}{\OmegaJC^3}(\Delta^2-g^2N)\right)\bigg)}+\mathcal{O}(N^{-5/2})\Bigg), \\
    \langle\hat{a}^2(t\gtrsim t_r)\rangle_{\pm}&=Ne^{2i(\phiJC\mp\omega_s t)}\Bigg(1 +  \textcolor{brown}{\left(e^{\pm4 i \omega_s t}-1\right)\frac{\omega_s}{\OmegaJC}\frac{1}{4}\left(1\mp\frac{\Delta}{\OmegaJC}\right)^2} \textcolor{gray}{\pm\frac{2i\omega_s^2 t}{\OmegaJC^3}\left(3(\Delta^2+2g^2N)\mp\Delta\OmegaJC\right)}\\
    &+\textcolor{orange}{\frac{\omega_s^2}{2\OmegaJC^2}\Bigg[-4\left(2\omega_s^2Nt^2\mp\frac{\Delta}{\OmegaJC}\right)(1+e^{\pm 4i\omega_s t})+(1-e^{\pm i4\omega_s t})\bigg(\frac{\Delta^4}{\OmegaJC^4}\left(13\mp 9\frac{\Delta}{\OmegaJC}\right)}\\
    &\textcolor{orange}{+\frac{42\omega_s N\Delta^2}{\OmegaJC^3}\left(1\mp\frac{\Delta}{\OmegaJC}\right)+\frac{2\omega_s^2N^2}{\OmegaJC^2}\left(25\mp \frac{12\Delta}{\OmegaJC}\right)-8\omega_s^2Nt^2\Bigg]} +\mathcal{O}(N^{-5/2}) \Bigg),\\
    \langle\hat{n}(t\gtrsim t_r)\rangle_{\pm}&=N\left(1+\textcolor{orange}{\frac{\omega_s\Delta^2}{N\OmegaJC^3}\left(1\mp\frac{\Delta}{\OmegaJC}\right)}+\mathcal{O}(N^{-3}) \right).
\end{aligned}
\end{equation}
for the resonator observables. 

We can use $\langle\hat{S}^+(t\gtrsim t_r)\rangle,\langle\hat{S}^z(t\gtrsim t_r)\rangle_{\pm}$ in \cref{eq:qubitDMsweetspotReadoutTime} to calculate the purity and angles of the qubit close to the readout time. Using $\Tr[\hat{\rho}_{\text{q}}^2(t\gtrsim t_r)]=\frac{1}{2}\left(1+|\langle\hat{\vb{S}}(t\gtrsim t_r)\rangle|^2\right)$, we have
\begin{equation}\label{eq:purityJCpostReadoutTime}
\begin{aligned}
    \Tr[\hat{\rho}_{\text{q}}^2(t\gtrsim t_r)]&=1-\textcolor{brown}{\frac{\omega_s}{\OmegaJC}\left(1\mp \frac{\Delta}{\OmegaJC}\right)^2\frac{1}{\OmegaJC^2}\left(\Delta^2+3g^2N+g^2N\cos(2\omega_s t)\right)}\\
    &+\textcolor{orange}{\frac{2\omega_s^2}{\OmegaJC^2}\Bigg(\frac{2}{\OmegaJC^6}\left(2\Delta^6+11\Delta^4g^2N-15\Delta^2g^4N^2-10g^6N^3\right)\mp \frac{\Delta}{\OmegaJC}\frac{1}{\OmegaJC^4}\left(4\Delta^4+11\Delta^2g^2N-40 g^4N^2\right)} \\
    &\textcolor{orange}{-\frac{4\omega_s}{\OmegaJC}N^2\omega_s^2t^2+\frac{\omega_s N}{\OmegaJC}\left(\frac{1}{\OmegaJC^4}\left(17\Delta^4+22\Delta^2g^2N+4g^4N^2\right) \mp \frac{\Delta}{\OmegaJC}\frac{1}{\OmegaJC^2}\left(17\Delta^2-8g^2N\right)\right)\cos(2\omega_s t)\Bigg)}\\
    &+\mathcal{O}(N^{-3}).
\end{aligned}
\end{equation}
For the measurement calculation in Sec.~\ref{sec:Measurements}, we need $\Theta_{\pm}(t\gtrsim t_r)$, $\Phi_{\pm}(t\gtrsim t_r)$, the polar and azimuthal angles describing the motion of the qubit initialized in $\ket{\pm}$. These are obtained from \cref{eq:qubitDMsweetspotReadoutTime} to order $\mathcal{O}(1/N^2)$ as follows
\begin{equation}\label{eq:PhiQubitSlowTerms}
    \begin{aligned}
        \Phi_{\pm}(t\gtrsim t_r)&=\frac{1}{i}\log\left(\frac{\langle\hat{S}^+(t\gtrsim t_r)\rangle}{|\langle\hat{S}^+(t\gtrsim t_r)\rangle|}\right)\\
        &=\frac{1}{2}(1\mp 1)\pi+\phiJC\mp \textcolor{purple}{\omega_s t}\mp \textcolor{brown}{\frac{1}{4}\frac{\omega_s}{\OmegaJC}\left(1\mp \frac{\Delta}{\OmegaJC}\right)^2\sin(2\omega_s t)}\,\textcolor{gray}{\pm  \frac{\omega_s^2t}{\OmegaJC}\left(\frac{2}{\OmegaJC^2}(\Delta^2-g^2N)\mp \frac{\Delta}{\OmegaJC}\right) } \\
        &\textcolor{orange}{\mp\frac{\omega_s^2}{4\OmegaJC^2}\Bigg(\left(\frac{1}{\OmegaJC^4}\left(\Delta^4+4\Delta^2g^2N+2g^4N^2\right)\mp \frac{\Delta}{\OmegaJC}\frac{1}{\OmegaJC}\left(\Delta^2+2g^2N\right)\right)\sin(4\omega_s t)} \\
        &\textcolor{orange}{-2\left(\frac{1}{\OmegaJC^4}\left(9\Delta^4-6\Delta^2g^2N-8g^4N^2\right)\mp \frac{\Delta}{\OmegaJC}\frac{1}{\OmegaJC^2}\left(9\Delta^2-20g^2N\right)\right)\sin(2\omega_s t)\Bigg)} + \mathcal{O}(N^{-5/2}),\\
    \end{aligned}
\end{equation}
\begin{equation}\label{eq:ThetaQubitSlowTerms}
    \begin{aligned}
        \Theta_{\pm}(t\gtrsim t_r)&=\arctan\left(\frac{|\langle\hat{S}^+(t\gtrsim t_r)\rangle|}{\langle\hat{S}^z(t\gtrsim t_r)\rangle}\right)\\
        &=\frac{1}{2}(1\mp 1)\pi \pm \arctan\left(\frac{2g\sqrt{N}}{\Delta}\right) \textcolor{brown}{\mp \frac{g\sqrt{N}\Delta\omega_s}{\OmegaJC^3}\left(\frac{1}{\OmegaJC^2}(\Delta^2+6g^2N) \pm \frac{\Delta}{\OmegaJC}+\frac{1}{2}\left(1\mp \frac{\Delta}{\OmegaJC}\right)^2\cos(2\omega_s t)\right)} \\
        &\textcolor{orange}{\pm \frac{g\sqrt{N}\Delta\omega_s^2}{4\OmegaJC^4}\Bigg(-16N\omega_s^2t^2 \mp 3\frac{\Delta}{\OmegaJC}\frac{1}{\OmegaJC^4}\left(15\Delta^4+66\Delta^2g^2N+40g^4N^2\right)}\\
        &\quad \textcolor{orange}{+\frac{1}{\OmegaJC^6}\left(69\Delta^6 + 272\Delta^4 g^2N - 198\Delta^2g^4N^2 -776 g^6N^3\right)}\\
        &\quad \textcolor{orange}{+4\cos(2\omega_s t)\Big(\frac{1}{\OmegaJC^6}\left(9\Delta^6+32\Delta^4g^2N-34\Delta^2g^4N^2-56g^6N^3\right) \mp \frac{\Delta}{\OmegaJC}\frac{1}{\OmegaJC^4}\left(9\Delta^4+18\Delta^2g^2N-88g^4N^2\right)\Big)} \\
        &\quad \textcolor{orange}{-\cos(4\omega_s t)\frac{1}{\OmegaJC^2}(\Delta^2+12g^2N)\Big(\frac{1}{\OmegaJC^4}(\Delta^4+4\Delta^2g^2N+2g^4N^2) \mp \frac{\Delta}{\OmegaJC}\frac{1}{\OmegaJC^2}(\Delta^2+2g^2N)\Big) \Bigg)}+\mathcal{O}(N^{-3}).
    \end{aligned}
\end{equation}
In the immediate vicinity of the readout time $t\approx t_r$, we can Taylor expand $\cos(2\omega_s t)\sim 1-2\omega_s^2t^2$, $\sin(2\omega_s t)\sim 2\omega_s t$ in \cref{eq:ThetaQubitSlowTerms,eq:PhiQubitSlowTerms}. Keeping terms up to and including $\mathcal{O}(1/N^2)$ yields
\begin{equation}\label{eq:PhiThetaPMSaddlePointReadoutTime}
    \begin{aligned}
        \Phi_{\pm}(t\approx t_r)&=\frac{1}{2}(1\mp 1)\pi+\phiJC\mp \textcolor{purple}{\omega_s t}\mp \textcolor{gray}{\frac{4\omega_s^3tN}{\OmegaJC^2}}  + \mathcal{O}(N^{-5/2}),\\
        \Theta_{\pm}(t\approx t_r)&=\arctan\left(\frac{2g\sqrt{N}}{\Delta}\right)  \mp\textcolor{brown}{ \frac{2g\sqrt{N}\Delta\omega_s}{\OmegaJC^3}} -\textcolor{orange}{\frac{2g\sqrt{N}\Delta^2\omega_s^3t^2}{\OmegaJC^4}\left(1\mp\frac{\Delta}{\OmegaJC}\right)} \\
        &\textcolor{orange}{\pm\frac{2g\sqrt{N}\Delta\omega_s^2}{\OmegaJC^4}\left(\frac{1}{\OmegaJC^4}\left(13\Delta^4-4\Delta^2g^2N-32g^4N^2\right)\mp 2\frac{\Delta}{\OmegaJC}\frac{1}{\OmegaJC^2}(5\Delta^2-4g^2N)\right)}+ \mathcal{O}(N^{-3}).
    \end{aligned}
\end{equation}
Next, we turn our attention to the resonator state. At $t=0$, this starts out as a perfectly coherent state $\hat{D}^{\dagger}(\alpha_0)\ket{0}$, where $\hat{D}(\alpha)=e^{\alpha \hat{a}^{\dagger}-\alpha^*\hat{a}}$ is the photonic displacement operator. For $t>0$, as the resonator interacts with the qubit, it becomes squeezed and stretched \cite{walls1983}. We use the following definition for the total quadrature spread parameter
\begin{equation}\label{eq:squeezingParameter}
    \zeta(t)=\frac{1}{2}\log\left(\Delta x^2(t)+\Delta p^2 (t)\right),
\end{equation}
where $\hat{x}=\sqrt{\frac{m\omega}{2\hbar}}\hat{X}$, $\hat{p}=\frac{1}{\sqrt{2m\omega\hbar}}\hat{P}$ are dimensionless position and momentum coordinates, and $\hat{x}$ and $\hat{p}$ are defined in the usual way via $\hat{a}=\sqrt{\frac{m\omega}{2\hbar}}\left(\hat{x}+i\hat{p}/(m\omega)\right)$. The variances of $\hat{x}$ and $\hat{p}$ are given by
\begin{equation}\label{eq:resonatorStateWidths}
    \begin{aligned}
    \Delta x^2&=\frac{1}{2}+(\langle\hat{n}\rangle-|\langle\hat{a}\rangle|^2)+\Re\left(\langle\hat{a}^2\rangle-\langle\hat{a}\rangle^2\right), \\
    \Delta p^2&=\frac{1}{2}+(\langle\hat{n}\rangle-|\langle\hat{a}\rangle|^2)-\Re\left(\langle\hat{a}^2\rangle-\langle\hat{a}\rangle^2\right).
    \end{aligned}
\end{equation}
As a quick sanity check of \cref{eq:resonatorStateWidths}, at $t=0$ we have $\langle\hat{a}\rangle=\alpha_0$, $\langle\hat{a}^2\rangle=\alpha_0^2$, $\langle\hat{n}\rangle=|\alpha_0|^2$, so that $\Delta x^2=\Delta p^2=\frac{1}{2}$, the minimum allowed by the Heisenberg uncertainty principle.
Substituting \cref{eq:resonatorSweetspotReadoutTime} into \cref{eq:resonatorStateWidths} and taking the sum and difference of $\Delta x^2_{\pm}$, $\Delta p^2_{\pm}$ gives
\begin{equation}\label{eq:sumDeltaXsquaredDeltaPsquared}
   \begin{aligned}
   \Delta x^2_{\pm}(t\gtrsim t_r)&+\Delta p^2_{\pm}(t\gtrsim t_r)=1+\frac{2N\omega_s}{\OmegaJC}\sin^2(\omega_s t)\left(1\mp\frac{\Delta}{\OmegaJC}\right)^2+\textcolor{brown}{\frac{2\omega_s}{\OmegaJC}\bigg[\frac{\Delta^2}{\OmegaJC^2}\left(1\mp\frac{\Delta}{\OmegaJC}\right)} \\
   &\textcolor{brown}{+\frac{\omega_s N}{\OmegaJC}\bigg(2\cos^2(\omega_s t)\left(1\mp\frac{2\Delta}{\OmegaJC}+2\omega_s^2Nt^2\right)-\frac{1}{4}\sin^2(\omega_s t)\left(1\mp\frac{\Delta}{\OmegaJC}\right)^4}\\
   &\textcolor{brown}{-2\sin^2(\omega_s t)\left(\frac{2\omega_s^2N^2}{\OmegaJC^2}+\frac{6\omega_s\Delta^2N}{\OmegaJC^3}+\frac{7\Delta^4}{\OmegaJC^4}\mp\frac{6\Delta}{\OmegaJC^3}(\Delta^2-g^2N)-2\omega_s^2Nt^2\right) \bigg)\bigg]}\\
   &+\mathcal{O}(N^{-2})
   \end{aligned}
\end{equation}
and
\begin{equation}\label{eq:differenceDeltaXsquaredDeltaPsquared}
   \begin{aligned}
   \Delta x^2_{\pm}(t\gtrsim t_r)&-\Delta p^2_{\pm}(t\gtrsim t_r)=2\Bigg[-\cos(2\phiJC)\frac{N\omega_s}{\OmegaJC}\sin^2(\omega_s t)\left(1\mp\frac{\Delta}{\OmegaJC}\right)^2\mp\textcolor{purple}{\frac{2N\omega_s^2t}{\OmegaJC}\sin(2\phiJC\mp 2\omega_s t)}\\
   &+\textcolor{brown}{\frac{2N\omega_s^2}{\OmegaJC^2}\bigg[-2\cos(2\phiJC)\cos(2\omega_s t)\left(2N\omega_s^2t^2\mp\frac{\Delta}{\OmegaJC}\right)} \\
   &\textcolor{brown}{+\cos(2\phiJC\mp\omega_s t)\cos(\omega_s t)\left(1+2N\omega_s^2t^2\mp\frac{2\Delta}{\OmegaJC}\right)+\frac{1}{8} \cos (2\phiJC) \left(\left(1\mp\frac{\Delta}{\OmegaJC}\right)^4\sin^2(\omega_s t)\right)}\\
   &\textcolor{brown}{\pm\frac{1}{2}\sin(2\phiJC)\sin(2\omega_s t)\bigg(\frac{2 \omega_s^2 N^2 \left(25\mp\frac{12 \Delta }{\OmegaJC}\right)}{\OmegaJC^2}-8N\omega_s^2t^2+\frac{42\Delta^2\omega_s N}{\OmegaJC^3}\left(1\mp\frac{\Delta}{\OmegaJC}\right)+\frac{\Delta^4\left(13\mp\frac{9\Delta}{\OmegaJC}\right)}{\OmegaJC^4}\bigg)}\\
   &\textcolor{brown}{\mp\sin(\omega_s t)\sin(2\phiJC\mp \omega_s t) \left(\frac{7\Delta ^4}{\OmegaJC^4}+\frac{2\omega_s^2N^2}{\OmegaJC^2}-2N\omega_s^2t^2+\frac{6\Delta^2\omega_s N}{\OmegaJC^3}\mp\frac{6\Delta\left(\Delta^2-g^2N\right)}{\OmegaJC^3}\right)\bigg]}\Bigg]\\
   &+\mathcal{O}(N^{-2}).
   \end{aligned}
\end{equation}

\section{Saddlepoint expressions for observables in the two-drive model\texorpdfstring{,  \cref{eq:JCplusRabiModel}}{}}\label{app:SaddlePointExpressionsSweetspot}
Using $\hat{D}(s)\hat{a}\hat{D}^{\dagger}(s)=\hat{a}-s$, $\hat{D}(s)\hat{a}^2\hat{D}^{\dagger}(s)=(\hat{a}-s)^2$ and $\hat{D}(s)\hat{n}\hat{D}^{\dagger}(s)=(\hat{a}^{\dagger}-s^*)(\hat{a}-s)$, we have
\begin{equation}\label{eq:resonatorObservablesTwoDrivesSP}
    \begin{aligned}
        \langle\hat{a}\rangle_{\text{JC+R}}&=\langle\hat{a}\rangle_{\sqrt{N}\to\sqrt{N}+|s|e^{i\varphi}}-s,\\
        \langle\hat{a}^2\rangle_{\text{JC+R}}&=\langle\hat{a}^2\rangle_{\sqrt{N}\to\sqrt{N}+|s|e^{i\varphi}}-2s\langle\hat{a}\rangle_{\sqrt{N}\to\sqrt{N}+se^{i\varphi}} +s^2, \\
        \langle\hat{n}\rangle_{\text{JC+R}}&=\langle\hat{n}\rangle_{\sqrt{N}\to\sqrt{N}+|s|e^{i\varphi}}-s\langle\hat{a}\rangle^*_{\sqrt{N}\to\sqrt{N}+|s|e^{i\varphi}}-s^*\langle\hat{a}\rangle_{\sqrt{N}\to\sqrt{N}+|s|e^{i\varphi}}+|s|^2,
    \end{aligned}
\end{equation}
where $s=\Omega_d/g$, $\varphi=\arg(s/\alpha_0)$ and $\langle\hat{a}\rangle$, $\langle\hat{a}^2\rangle$ and $\langle\hat{n}\rangle$ on the right sides of the equations where defined in \cref{eq:resonatorSweetspotReadoutTime}. For the qubit spin components, the expressions are even more simple, 
\begin{equation}\label{eq:qubitObservablesTwoDrivesSP}
    \begin{aligned}
        \langle\hat{S}^+\rangle_{\text{JC+R}}&=\langle\hat{S}^+\rangle_{\sqrt{N}\to\sqrt{N}+|s|e^{i\varphi}},\\
        \langle\hat{S}^z\rangle_{\text{JC+R}}&=\langle\hat{S}^z\rangle_{\sqrt{N}\to\sqrt{N}+|s|e^{i\varphi}}.
    \end{aligned}
\end{equation}

\section{Measurement, Fidelity and QNDness in the dispersive model\texorpdfstring{, \cref{eq:dispersiveHamiltonian}}{}}\label{app:measurementDispersiveModel}

\subsection{Evaluation of overlaps}
In this subsection, we evaluate the $\bra{\alpha_{\uparrow/\downarrow}}\hat{P}_{\gtrless}\ket{\alpha_{\uparrow/\downarrow}}$ half-plane overlaps in \cref{eq:postMeasurementDensityMatrixDispersive}. We demonstrate the procedure for $\bra{\alpha_{\uparrow}}\hat{P}_{<}\ket{\alpha_{\uparrow}}$; the other overlaps are computed analogously. Inserting two resolutions of the identity $\mathcal{I}=\int\dd{p}\ket{p}\bra{p}$ before and after the projection operator $\hat{P}_{\gtrless}$, and using $\bra{p}\ket{\alpha}=\pi^{-\frac{1}{4}}e^{-\Re[\alpha]^2/2}\exp\left(-\frac{1}{2}(p+i\alpha)^2\right)$, we have
\begin{equation*}
\begin{aligned}
    \bra{\alpha_{\uparrow}}\hat{P}_{<}\ket{\alpha_{\uparrow}}&=\int_{-\infty}^0\dd{p}\frac{1}{\sqrt{\pi}}e^{-\left(p+\sqrt{N}\sin(\chi t)\right)^2}=\int_{-\infty}^{\sqrt{N}\sin(\chi t)}\dd{p'}\frac{e^{-p'^2}}{\sqrt{\pi}}=\frac{1}{2}(1+f),
\end{aligned}
\end{equation*}
where $f=\erf(A(t))$, $A(t)=\sqrt{N}\sin(\chi t)$, and $\erf(z)=\frac{2}{\sqrt{\pi}}\int_0^z\dd{t}e^{-t^2}$. We collect the expressions for the six possible overlaps below,
\begin{equation}\label{eq:halfPlaneOverlapsDispersiveLimit}
    \begin{aligned}
        \bra{\alpha_{\uparrow}}\hat{P}_{<}\ket{\alpha_{\uparrow}}&=\bra{\alpha_{\downarrow}}\hat{P}_{>}\ket{\alpha_{\downarrow}}=\frac{1}{2}(1+f),\\
        \bra{\alpha_{\uparrow}}\hat{P}_{>}\ket{\alpha_{\uparrow}}&=\bra{\alpha_{\downarrow}}\hat{P}_{<}\ket{\alpha_{\downarrow}}=\frac{1}{2}(1-f),\\
        \bra{\alpha_{\downarrow}}\hat{P}_{\gtrless}\ket{\alpha_{\uparrow}}&=\frac{1}{2}\exp(-A^2-iB),
    \end{aligned}
\end{equation}
where $B(t)=N\sin(2\chi t)$.

\subsection{Values of \texorpdfstring{$r$ which minimize $\mathcal{F}_{\text{disp.}}$ and $\text{QNDness}_{\text{disp.}}$}{r which minimize F disp and QNDness}}

To find the worst values of $r$ for the fidelity, we differentiate \cref{eq:fidelityDispersive} with respect to $r$. The derivative of $\mathcal{F}$ reads
\begin{equation}\label{eq:fidelityDerivative}
\begin{aligned}
    \frac{\partial\mathcal{F}_{\text{disp.}}}{\partial r}&=\frac{1}{4}\left(\frac{1+f+2f\,r}{\sqrt{(1+r)(1+f\,r)}}-\frac{1+f-2f\,r}{\sqrt{(1-r)(1-f\,r)}}\right).
\end{aligned}
\end{equation}
\cref{eq:fidelityDerivative} has only one root, at $r=0$. For $r>0$, $\frac{\partial\mathcal{F}}{\partial r}$ is always negative, whereas for $r<0$, it is always positive. To show this, we use the fact that both terms inside the brackets are always positive, and then compare their sizes by squaring them and taking the difference, which is proportional to
\begin{equation}\label{eq:diffFidelityDerivativeTerms}
    (1+f+2f\, r)^2(1-r)(1-f\, r)- (1+f-2f\, r)^2(1+r)(1+f\, r)=-2r(1-f)^2(1+f).
\end{equation}
Since $f\leq 1$, the sign of the RHS in \cref{eq:diffFidelityDerivativeTerms} is determined purely by $\sign(r)$. In addition, we see that the magnitude increases monotonically with $|r|$, i.e., the gradient $\frac{\partial\mathcal{F}}{\partial r}$ has the biggest magnitude at $r=\pm 1$. The worst fidelity therefore occurs at $r=\pm 1$ and is given by
\begin{equation}\label{eq:fidelityDispersiveWorst}
    \min(\mathcal{F}_{\text{disp.}})=\frac{1}{\sqrt{2}}\sqrt{1+f}.
\end{equation}

\section{Measurement, Fidelity and QNDness in the nondispersive two-drive model\texorpdfstring{, \cref{eq:JCplusRabiModel}}{}}\label{app:measurementNondispersiveModel}

\subsection{Exact expressions for resonator wavefunctions \texorpdfstring{$\ket{R_{\pm,\perp/\parallel}(t)}$}{R}}

To isolate $\ket{R_{\pm}(t)}$, we set $c_{\pm}=1$, $c_{\mp}=0$ in $\ket{\Psi(t)}$. The resonator wavefunctions $\ket{R_{\pm,\parallel/\perp}(t)}$ can then be obtained by projecting the qubit basis vectors onto $\ket{R_{\pm}(t)}$, $\ket{R_{\pm,\parallel/\perp}(t)}=\bra{\pm_{\parallel/\perp}}\ket{R_{\pm}}$. Substituting in the definitions \cref{eq:timeDependenceNonDispersiveBasisStates,eq:psiTwoDrives} for $\ket{\Psi(t)}$ and $\bra{\pm_\parallel},\bra{\pm_\perp}$, respectively, gives
\begin{equation}\label{eq:resonatorPlusMinusParallelPerpKets}
\begin{aligned}
    &\ket{R_{+,\parallel}(t)}=\hat{D}^{\dagger}(s)e^{-\Neff/2}\Bigg[c(\thetaJCR)c(\Theta_+)e^{-\frac{i\Delta t}{2}}\ket{0}\\
    &+\sum_{m=1}^{\infty}\frac{\Neff^{\frac{m}{2}}}{\sqrt{m!}}\Big[e^{-i\lambda_{m}t}\left(c(\theta_m)c(\thetaJCR)+\sqrt{\frac{m}{\Neff}}s(\theta_m)s(\thetaJCR)\right)\left(s(\theta_m)s(\Theta_+)e^{-i\Phi_+}\ket{m-1}+c(\theta_m)c(\Theta_+)\ket{m}\right)\\
    &\quad +e^{i\lambda_m t}\left(-s(\theta_m)c(\thetaJCR)+\sqrt{\frac{m}{\Neff}}c(\theta_m)s(\thetaJCR)\right)\left(c(\theta_m)s(\Theta_+)e^{-i\Phi_+}\ket{m-1}-s(\theta_m)c(\Theta_+(t))\ket{m}\right)\Big]\Bigg],\\
    \\
    &\ket{R_{+,\perp}(t)}=\hat{D}^{\dagger}(s)e^{-\Neff/2}\Bigg[c(\thetaJCR)s(\Theta_+)e^{-\frac{i\Delta t}{2}}\ket{0}\\
    &+\sum_{m=1}^{\infty}\frac{\Neff^{\frac{m}{2}}}{\sqrt{m!}}\Big[e^{-i\lambda_{m}t}\left(c(\theta_m)c(\thetaJCR)+\sqrt{\frac{m}{\Neff}}s(\theta_m)s(\thetaJCR)\right)\left(-s(\theta_m)c(\Theta_+)e^{-i\Phi_+}\ket{m-1}+c(\theta_m)s(\Theta_+)\ket{m}\right)\\
    &\quad +e^{i\lambda_m t}\left(-s(\theta_m)c(\thetaJCR)+\sqrt{\frac{m}{\Neff}}c(\theta_m)s(\thetaJCR)\right)\left(-c(\theta_m)c(\Theta_+)e^{-i\Phi_+(t)}\ket{m-1}-s(\theta_m)s(\Theta_+)\ket{m}\right)\Big]\Bigg],\\
    \\
    &\ket{R_{-,\parallel}(t)}=\hat{D}^{\dagger}(s)e^{-\Neff/2}\Bigg[s(\thetaJCR)c(\Theta_-)e^{-\frac{i\Delta t}{2}}\ket{0}\\
    &+\sum_{m=1}^{\infty}\frac{\Neff^{\frac{m}{2}}}{\sqrt{m!}}\Big[e^{-i\lambda_{m}t}\left(c(\theta_m)s(\thetaJCR)-\sqrt{\frac{m}{N}}s(\theta_m)c(\thetaJCR)\right)\left(s(\theta_m)s(\Theta_-)e^{-i\Phi_-}\ket{m-1}+c(\theta_m)c(\Theta_-)\ket{m}\right)\\
    &\quad +e^{i\lambda_m t}\left(-s(\theta_m)s(\thetaJCR)-\sqrt{\frac{m}{N}}c(\theta_m)c(\thetaJCR)\right)\left(c(\theta_m)s(\Theta_-)e^{-i\Phi_-}\ket{m-1}-s(\theta_m)c(\Theta_-(t))\ket{m}\right)\Big]\Bigg],\\
    \\
    &\ket{R_{-,\perp}(t)}=\hat{D}^{\dagger}(s)e^{-\Neff/2}\Bigg[s(\thetaJCR)s(\Theta_-)e^{-\frac{i\Delta t}{2}}\ket{0}\\
    &+\sum_{m=1}^{\infty}\frac{\Neff^{\frac{m}{2}}}{\sqrt{m!}}\Big[e^{-i\lambda_{m}t}\left(c(\theta_m)s(\thetaJCR)-\sqrt{\frac{m}{N}}s(\theta_m)c(\thetaJCR)\right)\left(-s(\theta_m)c(\Theta_-)e^{-i\Phi_-}\ket{m-1}+c(\theta_m)s(\Theta_-)\ket{m}\right)\\
    &\quad +e^{i\lambda_m t}\left(-s(\theta_m)s(\thetaJCR)-\sqrt{\frac{m}{\Neff}}c(\theta_m)c(\thetaJCR)\right)\left(-c(\theta_m)c(\Theta_-)e^{-i\Phi_-(t)}\ket{m-1}-s(\theta_m)s(\Theta_-)\ket{m}\right)\Big]\Bigg].
\end{aligned}
\end{equation}
where $c(\theta)=\cos(\theta/2)$, $s(\theta)=\sin(\theta/2)$, $\tan(\thetaJCR)=2g\sqrt{\Neff}/\Delta$.

\subsection{Components of post-measurement density matrices \texorpdfstring{$\hat{\rho}_{\text{q},\gtrless}$}{}}

We express $\hat{\rho}_{\text{q},\gtrless}$ in the $\{\ket{\mp_{\parallel}(t)},\ket{\mp_\perp(t)}\}$ bases, respectively. We denote the components by $[\rho_{\text{q},\gtrless}]_{s s'}$, $s,s'=\{\parallel,\perp\}$. We need only calculate three of these, as the off-diagonal components are complex conjugates of each other. Note that we really do need to calculate \emph{both} $[\rho_{\text{q},\gtrless}]_{\parallel\parallel}$ and $[\rho_{\text{q},\gtrless}]_{\perp\perp}$, as the post-measurement density matrices are not necessarily normalized, $\Tr[\hat{\rho}_{\text{q},\gtrless}]\neq 1$. Each term in $[\rho_{\text{q},\gtrless}]_{ss'}$ is a quadratic combination of the $c_\pm$ coefficients, which describe the initial state of the qubit: $|c_{\pm}|^2,c_\pm c_\mp^*$. Defining $\vb{c}_{\pm}=(c_{\pm},c_{\mp})^T$, we can write $\rho_{\gtrless,ss'}$ as a sandwich of $\vb{c}_{\pm}^{\dagger}$, a $2\times 2$ matrix $\mathcal{R}_{\gtrless,\sigma\sigma'}$, and $\vb{c}_{\pm}$, 
\begin{equation}\label{eq:rhoPostMeasurementComponents}
    [\rho_{\text{q},\gtrless}]_{ss'}=\vb{c}^{\dagger}_{\mp}\mathcal{R}_{\gtrless,ss'}\vb{c}_{\mp}.
\end{equation}
The $t,t'=\{\pm,\mp\}$ components of $[\mathcal{R}_{\gtrless,ss'}]_{tt'}$ are given by
\begin{equation}\label{eq:overlapHalfPlaneRmatComponents}
    \begin{aligned}
        [\mathcal{R}_{\gtrless,ss'}]_{\mp\mp}&=\bra{R_{\mp,s'}}\hat{P}_{\gtrless}\ket{R_{\mp,s}}, \\
        [\mathcal{R}_{\gtrless,ss'}]_{\pm\pm}&=\sum_{t,t'=\{\parallel,\perp\}}\bra{\mp_s}\ket{\pm_t}\bra{\pm_{t'}}\ket{\mp_{s'}}\bra{R_{\pm,t'}}\hat{P}_{\gtrless}\ket{R_{\pm,t}},\\
        [\mathcal{R}_{\gtrless,ss'}]_{\pm\mp}&=\sum_{t=\{\parallel,\perp\}}\bra{\pm_t}\ket{\mp_{s'}}\bra{R_{\pm,t}}\hat{P}_{\gtrless}\ket{R_{\mp,s}},\\
        [\mathcal{R}_{\gtrless,ss'}]_{\mp\pm}&=\sum_{t=\{\parallel,\perp\}} \bra{\mp_s}\ket{\pm_t} \bra{R_{\mp,s'}}\hat{P}_{\gtrless}\ket{R_{\pm,t}}.
    \end{aligned}
\end{equation}
\cref{eq:overlapHalfPlaneRmatComponents} contains three different inner products of qubit states and ten different half-plane overlaps of resonator states.

Using the definition for $\ket{\pm_{\parallel}(t)}$ in \cref{eq:timeDependenceNonDispersiveBasisStates}, and recalling that for $\ket{\pm_{\perp}(t)}$, we send $\Theta\to\pi-\Theta$, $\Phi\to\pi+\Phi$, the three inner products of the qubit states evaluate to
\begin{equation}\label{eq:overlapHalfPlaneRmatComponentsBreakdown}
    \begin{aligned}
        \bra{\mp_{\parallel}}\ket{\pm_{\parallel}}&=\cos(\frac{\Theta_{\mp}}{2})\cos(\frac{\Theta_{\pm}}{2})+e^{i(\Phi_{\pm}-\Phi_{\mp})}\sin(\frac{\Theta_{\mp}}{2})\sin(\frac{\Theta_{\pm}}{2}),\\
        \bra{\mp_{\perp}}\ket{\pm_{\perp}}&=\sin(\frac{\Theta_{\mp}}{2})\sin(\frac{\Theta_{\pm}}{2})+e^{i(\Phi_{\pm}-\Phi_{\mp})}\cos(\frac{\Theta_{\mp}}{2})\cos(\frac{\Theta_{\pm}}{2}),\\
        \bra{\mp_{\perp}}\ket{\pm_{\parallel}}&=\sin(\frac{\Theta_{\mp}}{2})\cos(\frac{\Theta_{\pm}}{2})-e^{i(\Phi_{\pm}-\Phi_{\mp})}\cos(\frac{\Theta_{\mp}}{2})\sin(\frac{\Theta_{\pm}}{2}),
    \end{aligned}
\end{equation}
where $\Theta_\pm(t)=\arctan(\sqrt{\langle \hat{S}^x(t)\rangle^2_\pm+\langle \hat{S}^y(t)\rangle^2_\pm}/\langle \hat{S}^z(t)\rangle_\pm)$, $\Phi_\pm(t)=\arctan(\langle \hat{S}^y(t)\rangle_\pm/\langle \hat{S}^x(t)\rangle_\pm)$ are determined using the time-dependent Bloch vector $\langle\vb{S}(t)\rangle_{\pm}$ for a system where the qubit is initialized in $\ket{\pm}$. 

The half-plane overlaps of the resonator states have the form $\bra{R_{\sigma',s'}}\hat{P}_{\gtrless}\ket{R_{\sigma,s}}$, where $\sigma,\sigma'=\{+,-\}$, $s,s'=\{\parallel,\perp\}$. There are four half-plane overlaps with $\sigma=\sigma'$, $s=s'$, two with $\sigma=\sigma'$, $s\neq s'$, and two with $\sigma=-\sigma'$, $s=s'$. 
To evaluate them explicitly, we insert two resolutions of the identity, $\mathcal{I}=\sum_m \ket{m}\bra{m}$, before and after $\hat{P}_{\gtrless}$, turning the overlap into
\begin{equation}\label{eq:halfPlaneOverlapsWithKernel}
    \bra{R_{\sigma',s'}}\hat{P}_{\gtrless}\ket{R_{\sigma,s}}=\sum_{n,m}K_{\gtrless}(n,m)R^*_{\sigma',s'}(n) R_{\sigma,s}(m).
\end{equation}
In \cref{eq:halfPlaneOverlapsWithKernel}, $R_{\sigma,s}(m)=\bra{m}\ket{R_{\sigma,s}(t)}$ are the Fock-basis coefficients of the resonator wavefunctions $\ket{R_{\sigma,s}(t)}$, defined in \cref{eq:resonatorPlusMinusParallelPerpKets}. Using that $s\in\mathbb{R}$, $\hat{D}(s)\ket{p}=e^{-2isp}\ket{p}$; i.e., $\hat{D}^{\dagger}$ just produces an $m$-independent phase $e^{-2isp}$. This phase will be canceled by the corresponding $e^{2isp}$ factor in $R^*_{\sigma,s}(m)$, so we can just ignore the $\hat{D}(s)$, $\hat{D}^{\dagger}(s)$ operators. $K_{\gtrless}(n,m)=\int_{p\gtrless 0}\dd{p}\psi_n^*(p)\psi_m(p)$ is the half-line integration kernel, where are the SHO eigenstates in the (dimensionless) momentum representation. $K_{\gtrless}(n,m)$ evaluates to
\begin{equation}\label{eq:HalfPlaneIntegrationKernel}
    K_{\gtrless}(n,m)=\int_{p\gtrless 0}\dd{p}\psi_n^*(p)\psi_m(p)=\begin{cases}
    \frac{1}{2}, & n=m, \\
    \pm i\sqrt{\frac{2}{\pi}} \frac{\sqrt{(2l)!(2j+1)}}{2^{j+l+1}l!j!}\frac{1}{2(j-l)+1}, & n=2j+1, m=2l,\\
    \mp i\sqrt{\frac{2}{\pi}} \frac{\sqrt{(2l)!(2j+1)}}{2^{j+l+1}l!j!}\frac{1}{2(j-l)+1}, & n=2l, m=2j+1,\\
    0, & \text{otherwise},
    \end{cases}
\end{equation}
see App.~(\ref{subsubapp:HalfPlaneIntegrationKernelProof}) for a proof of this result by induction.

This concludes the derivation of the exact expressions for the components of the post-measurement matrices $\hat{\rho}_{\text{q,}\gtrless}$. See ``Measurements'' section in \texttt{paperSaddlePointJCmodelPublicationFinal.nb} for an implementation. In App.~\ref{app:PostMeasurementDensityMatricesSPapproach}, we show how the saddlepoint approach can be applied to \cref{eq:halfPlaneOverlapsWithKernel} to obtain approximate expressions correct up to and including $\mathcal{O}(N^{-2})$. This allows us to compute the leading order $\mathcal{O}(N^{-1})$ error in fidelity and QNDness in the nondispersive model.

\subsubsection{Proof by induction\texorpdfstring{ of \cref{eq:HalfPlaneIntegrationKernel}}{}}\label{subsubapp:HalfPlaneIntegrationKernelProof}
In \cref{eq:HalfPlaneIntegrationKernel}, the $n=m$ $m,n\in\text{odd}$, and $m,n\in\text{even}$ cases trivially follow from the orthogonality property $\int_{-\infty}^{\infty} \dd{p}\psi^*_n(p)\psi_m(p)=\delta_{mn}$ and even/odd parity of $\psi_m(p)=\pm \psi_m(-p)$. Thus, we only need to prove the cases where $m\in\text{odd}$, $n\in\text{even}$ or vice versa. The explicit formula for $\psi_m(p)$ is
\begin{equation}\label{eq:SHOeigenstateMomentumRep}
    \psi_m(p)=\frac{(-i)^m}{\sqrt{2^n n!}\pi^{1/4}}e^{-\frac{p^2}{2}}H_n(p),
\end{equation}
where $H_n(z)$ are Hermite polynomials defined by Rodrigues' formula
\begin{equation}\label{eq:hermitePolynomial}
    H_n(z)=(-1)^ne^{z^2}\frac{d^n}{dz^n}(e^{-z^2}).
\end{equation}
Taylor expanding \cref{eq:hermitePolynomial} and setting $z=0$ gives 
\begin{equation}\label{eq:hermitePolynomialAtZero}
    H_n(0)=\begin{cases}
        0, & n \text{ odd}, \\
        \frac{(-1)^{\frac{n}{2}}n!}{\left(\frac{n}{2}\right)!}, & n \text{ even.}
    \end{cases}
\end{equation}
Inserting \cref{eq:SHOeigenstateMomentumRep} into the definition of $K_{\gtrless}(m,n)$ in \cref{eq:HalfPlaneIntegrationKernel}, we find that it is sufficient to show that
\begin{equation}\label{eq:Jkernel}
\begin{aligned}
        J_{\gtrless}(2j+1,2l)=J_{\gtrless}(2l,2j+1)=\int_{p\gtrless 0}\dd{p} e^{-p^2} H_{2j+1}(p) H_{2l}(p)=\pm \frac{(2j+1)!(2l)!}{j!l!}\frac{(-1)^{j+l}}{2(j-l)+1},
\end{aligned}
\end{equation}
to prove \cref{eq:HalfPlaneIntegrationKernel}. We will do this by induction. The base case is $J_{n,0}=J_{0,n}$, where $n=|2(j-l)+1|\in\text{odd}$. $J_{n,0}$ can be integrated directly, and evaluates to
\begin{equation}\label{eq:JkernelBaseCase}
    J_{\gtrless}(n,0)=\int_{p\gtrless 0}\dd{p}e^{-p^2}H_n(p)\underbrace{H_0(p)}_{=1}=\int_{p\gtrless 0}\dd{p}(-1)^n\frac{d^n}{dp^n}(e^{-z^2})=\mp(-1)^n \frac{d^{n-1}}{dp^{n-1}}(e^{-p^2})\Big|_0=\pm (-1)^{\frac{n-1}{2}}\frac{(n-1)!}{(\frac{n-1}{2})!}.
\end{equation}
Setting $l=0$ in \cref{eq:Jkernel} and $n=2j+1$ in \cref{eq:JkernelBaseCase}, we see that \cref{eq:Jkernel} satisfies the base case. Next, we write out the induction step. Integrating \cref{eq:Jkernel} by parts and using $\frac{d}{dz}\left(H_{n-1}(z)e^{-z^2}\right)=-H_{n}e^{-z^2}$ and
$\frac{dH_n}{dz}=2nH_{n-1}$ gives
\begin{equation}\label{eq:recursionRelationJkernel}
    J_{\gtrless}(n,m)=\int_{p\gtrless 0}\dd{p} e^{-p^2} H_n(p) H_m(p)=\pm H_{\max(m,n)}(0)H_{\min(n,m)-1}(0) + 2\max(n,m)J_{\gtrless}(n-1,m-1).
\end{equation}
It is sufficient to check that the recurrence relation \cref{eq:recursionRelationJkernel} satisfies \cref{eq:Jkernel} in just one of the two cases $n<m$ or $n>m$; the other follows analogously. Assuming $n>m$ and $n=2j+1$, $m=2l$ to begin with, $H_{n}(0)$ and $H_{m-1}(0)$ both vanish because $n,m-1\in\text{odd}$, giving 
\begin{equation}
\begin{aligned}
        \text{RHS}&=2(2j+1)J_{\gtrless}(2j,2(l-1)+1)= 2(2j+1)\cdot\left(\pm\frac{(2l-1)!(2j)!}{(l-1)!j!}\frac{(-1)^{l-1+j}}{2(l-1-j)+1}\right)=J_{\gtrless}(2j+1,2l).
\end{aligned}
\end{equation}
For the other case $n>m$ and $n=2l$, $m=2j+1$, we have
\begin{equation}
\begin{aligned}
    \text{RHS}&=\pm H_{2l}(0) H_{2j}(0) + 2 (2l) J_{2l-1, 2j}\\
    &=\pm \frac{(-1)^{j+l}(2j)!(2l)!}{j!l!}+2(2l)\cdot\left(\pm\frac{(2l-1)!(2j)!}{(l-1)!j!}\frac{(-1)^{l-1+j}}{2(l-1-j)+1}\right)\\
    &=\pm \frac{(-1)^{j+l}(2j)!(2l)!}{j!l!}\left[1+\frac{2l}{2(j-l)+1}\right]=\frac{(2j+1)!(2l)!}{j!l!}\frac{(-1)^{j+l}}{2(j-l)+1}\\
    &=J_{\gtrless}(2l,2j+1).
\end{aligned}
\end{equation}
This completes the proof of the inductive step. Since the base case and the inductive step are both true, \cref{eq:JkernelBaseCase}, and consequently also \cref{eq:HalfPlaneIntegrationKernel}, are true for all integer $j,l\geq 0$.

\subsection{Evaluation of \texorpdfstring{$\bra{R_{\sigma',s'}}\hat{P}_{\gtrless}\ket{R_{\sigma,s}}$ in \cref{eq:halfPlaneOverlapsWithKernel}}{overlap} using the saddlepoint approach}\label{app:PostMeasurementDensityMatricesSPapproach}

In this subsection, we use the saddlepoint approach to approximate the sums over $m,n$ in \cref{eq:halfPlaneOverlapsWithKernel}, obtaining expressions for the fidelity and QNDness in terms of $g,\Delta,\Neff$ and the initial qubit state parameters $r,\Delta\Phi$.

Inserting the definition of the kernel \cref{eq:HalfPlaneIntegrationKernel}, the sum can be separated into two kinds of contributions,
\begin{equation}\label{eq:halfPlaneOverlapsWithKernelSeparation}
    \bra{R_{\sigma',s'}}\hat{P}_{\gtrless}\ket{R_{\sigma,s}}=\bra{R_{\sigma',s'}}\hat{P}_{\gtrless}\ket{R_{\sigma,s}}_{m=n}+\bra{R_{\sigma',s'}}\hat{P}_{\gtrless}\ket{R_{\sigma,s}}_{m\neq n},
\end{equation}
where
\begin{equation}\label{eq:halfPlaneOverlapsTwoKinds}
\begin{aligned}
        \bra{R_{\sigma',s'}}\hat{P}_{\gtrless}\ket{R_{\sigma,s}}_{m=n}&=\frac{1}{2}\sum_{m}R^*_{\sigma',s'}(m) R_{\sigma,s}(m)=\frac{1}{2}\bra{R_{\sigma',s'}}\ket{R_{\sigma,s}},\\
        \bra{R_{\sigma',s'}}\hat{P}_{\gtrless}\ket{R_{\sigma,s}}_{m\neq n}&=\sum_{j,l}K_{\gtrless}(2j+1,2l)R^*_{\sigma',s'}(2j+1) R_{\sigma,s}(2l)+\sum_{j,l}K_{\gtrless}(2l,2j+1)R^*_{\sigma',s'}(2l) R_{\sigma,s}(2j+1).
\end{aligned}
\end{equation}
$\bra{R_{\sigma',s'}}\hat{P}_{\gtrless}\ket{R_{\sigma,s}}_{m=n}$ contains a single sum over $m$, allowing us to directly apply the saddlepoint approach developed in App.~\ref{app:SaddlePoint}. On the other hand, $\bra{R_{\sigma',s'}}\hat{P}_{\gtrless}\ket{R_{\sigma,s}}_{m\neq n}$ contains a double sum over $m,n$, which means that the same approach cannot be applied directly. Remarkably, a modified version of the saddlepoint approach exists, which is also able to evaluate the more difficult double sums over $m,n$; see App.~\ref{app:SaddlepointPostMeasurementMnotEqualNterms} for technical details of this procedure. For a list of the saddlepoint expressions of the eight relevant overlaps at readout time $t\approx t_r$, skip directly to App.~\ref{app:halfPlaneOverlapsSaddlePointReadoutTime}.

\subsubsection{Saddlepoint approach for calculating \texorpdfstring{$\bra{R_{\sigma,s}}\hat{P}_{\gtrless}\ket{R_{\sigma',s'}}_{m\neq n}$}{overlap}}\label{app:SaddlepointPostMeasurementMnotEqualNterms}

Inserting the definition of $K_{\gtrless}(2j+1,2l)$ into \cref{eq:HalfPlaneIntegrationKernel}, we see that $\bra{R_{\sigma,s}}\hat{P}_{>}\ket{R_{\sigma',s'}}_{m\neq n}=-\bra{R_{\sigma,s}}\hat{P}_{<}\ket{R_{\sigma',s'}}_{m\neq n}$. Therefore, it is sufficient to calculate just one of them; we will choose $\bra{R_{\sigma,s}}\hat{P}_{>}\ket{R_{\sigma',s'}}_{m\neq n}$. Replacing the discrete sum over $j,l$ by a continuous integral, $\sum_{j,l}\to\iint_{-\infty}^{\infty}\dd{l}\dd{j}$, inserting \cref{eq:HalfPlaneIntegrationKernel}, and sending $j\to j-1/2$ gives
\begin{equation}\label{eq:MunequalNoverlap1}
\begin{aligned}
        \bra{R_{\sigma',s'}}\hat{P}_{\gtrless}\ket{R_{\sigma,s}}_{m\neq n}&=\sum_{\substack{\Delta m,\Delta n=0,1}}\sum_{\alpha,\beta=\pm 1}\mathcal{I}_{oe}-\mathcal{I}_{eo},
\end{aligned}
\end{equation}
where
\begin{equation}\label{eq:IevenoddDefs1}
\begin{aligned}
        \mathcal{I}_{oe}&=\frac{1}{4}\frac{i}{\pi^{3/2}}\iint_{-\infty}^{\infty}\dd{j}\dd{l}\frac{e^{g_{J+L,\alpha,\beta,\Delta m,\Delta n}(j,l)}}{j-l}R^*_{\sigma',s',\beta,\Delta n}(2j)R_{\sigma,s,\alpha,\Delta m}(2l),\\
    \mathcal{I}_{eo}&=\frac{1}{4}\frac{i}{\pi^{3/2}}\iint_{-\infty}^{\infty}\dd{j}\dd{l}\frac{e^{g_{J+L,-\beta,-\alpha,\Delta n,\Delta m}(j,l)}}{j-l}R^*_{\sigma',s',\beta,\Delta n}(2l)R_{\sigma,s,\alpha,\Delta m}(2j),
\end{aligned}
\end{equation}
where
\begin{equation}
    g_{J+L,\alpha,\beta,\Delta m,\Delta n}(j,l)=g_{J,-\beta,\Delta n}(j)+g_{L,\alpha,\Delta m}(l)
\end{equation}
and $g_{L,\alpha,\Delta m}(l)$, $g_{J,\alpha,\Delta m}(j)$ are defined as
\begin{equation}\label{eq:gFuncsSPunequalTerms1}
    \begin{aligned}
        e^{g_{L,\alpha,\Delta m}(l)}&=\frac{e^{-N/2}(N/2)^l}{l!/\sqrt{2\pi}}e^{-i\alpha\lambda_{2l+\Delta m}t},& 
        e^{g_{J,\alpha,\Delta m}(j)}&=\frac{e^{-N/2}(N/2)^j}{(j-1/2)!/\sqrt{2\pi}}e^{-i\alpha\lambda_{2j+\Delta m}t}.
    \end{aligned}
\end{equation}
$R_{\sigma,s,\alpha,\Delta m}(2l)=\frac{1}{2\pi}\int_{-\infty}^{\infty}\dd{t} e^{i\alpha\lambda_{2l+\Delta m} t}R_{\sigma,s}(2l)$ is the Fourier transform in time of $R_{\sigma,s}(2l)$, yielding the prefactor of the $e^{-i\alpha\lambda_{2l+\Delta m} t}$ term in $R_{\sigma,s}(2l)$. As before, we approximate $l!$ and $(j-1/2)!$ using Stirling's approximation, \cref{eq:StirlingApproximation}. Note that we used $N$ instead of $\Neff$ in the definition \cref{eq:gFuncsSPunequalTerms1} in order to save space, as $N$ will reappear multiple times over the course of this derivation. Ultimately, however, everything we do also holds for $N\to \Neff$.

Consider the integrands in \cref{eq:IevenoddDefs1}. Most of the weight is concentrated within a disk-shaped region of radius $\sim\sqrt{N/2}$ centered at $l,j\sim N/2$, due to the $(N/2)^l/l!$ and $(N/2)^j/(j-1/2)!$ factors in \cref{eq:gFuncsSPunequalTerms1}. It is therefore natural to switch to the center of mass and difference coordinates $x,y$, defined as follows
\begin{equation}\label{eq:XYcoordinates}
\begin{aligned}
&\mathcal{I}_{oe}: &    j&=\frac{1}{2}\left(X_{\alpha,\beta,\Delta m,\Delta n}+\sqrt{N}(x+y)\right), & l&=\frac{1}{2}\left(X_{\alpha,\beta,\Delta m,\Delta n}+\sqrt{N}(x-y)\right),\\
&\mathcal{I}_{eo}: &    j&=\frac{1}{2}\left(X_{-\beta,-\alpha,\Delta n,\Delta m}+\sqrt{N}(x+y)\right), & l&=\frac{1}{2}\left(X_{-\beta,-\alpha,\Delta n,\Delta m}+\sqrt{N}(x-y)\right),
\end{aligned}
\end{equation}
which rescales the integration element to $\dd{j}\dd{l}=\frac{N}{2} \dd{x}\dd{y}$.
In these new coordinates, the integrands in \cref{eq:IevenoddDefs1} are largest when $x,y\sim 0$ and decay exponentially to zero when $x,y\gtrsim 1$. We choose $X_{\alpha,\beta,\Delta m,\Delta n}$ such that $x=y=0$ is a local maximum in the $x$-direction,
\begin{equation}\label{eq:localMaximumXdirection}
    \frac{\partial}{\partial x}(g_{J+L,\alpha,\beta,\Delta m,\Delta n})|_{x,y=0}=0.
\end{equation}
We proceed exactly as in App.~\ref{app:SaddlePoint}, \cref{eq:TaylorExpansionSaddlePoint}, Taylor expanding $X_{\alpha,\beta,\Delta m,\Delta n}$ in powers of $1/\sqrt{N}$, $X_{\alpha,\beta,\Delta m,\Delta n}=N\sum_{k=0}^{\infty}N^{-k/2}X^{(k/2)}_{\alpha,\beta,\Delta m,\Delta n}$. Solving for the coefficients $X^{(k/2)}_{\alpha,\beta,\Delta m,\Delta n}$, we find that they are identical to the coefficients $L^{(k/2)}_{\alpha,\beta,\Delta m,\Delta n}$ in \cref{eq:saddlePointFormula}. To apply the saddlepoint method, we Taylor expand $g_{J+L,\alpha,\beta,\Delta m,\Delta n}$ in $x$ and $y$ around the local maximum in the $x$-direction $x,y=0$, giving
\begin{equation}\label{eq:gOEsaddlePointExpansion}
\begin{aligned}
    g_{J+L,\alpha,\beta,\Delta m,\Delta n}&=g^{(0,0)}_{J+L,\alpha,\beta,\Delta m,\Delta n}|_{x,y=0} - \frac{x^2+y^2}{2\sigma_{\alpha,\beta,\Delta m,\Delta n}^2} -iz_{\alpha,\beta,\Delta m,\Delta n}y  \\
    &+ \sum_{k_x+k_y=3}^{\infty}\tilde{g}^{(k_x,k_y)}_{J+L,\alpha,\beta,\Delta m,\Delta n}\frac{x^{k_x}y^{k_y}}{k_x! k_y!},
\end{aligned}
\end{equation}
where
\begin{equation}\label{eq:gOEDefAndExpansionCoeffs}
\begin{aligned}
        \tilde{g}^{(k_x,k_y)}_{J+L,\alpha,\beta,\Delta m,\Delta n}&=\frac{\partial^{k_x+k_y}g_{J+L,\alpha,\beta,\Delta m,\Delta n}}{\partial y^{k_y}\partial x^{k_x}}\bigg|_{x,y=0}, &
        z_{\alpha,\beta,\Delta m,\Delta n}&=i\left(\tilde{g}^{(0,1)}_{J+L,\alpha,\beta,\Delta m,\Delta n} + \tilde{g}^{(1,1)}_{oe} x\right).
\end{aligned}
\end{equation}
Note that there is no $\tilde{g}_{J+L,\alpha,\beta,\Delta m,\Delta n}^{(1,0)}x$ term in \cref{eq:gOEsaddlePointExpansion} by construction. Using the fact that each derivative in $x$ or $y$ reduces the order by $\sqrt{N}\cdot 1/N$, the leading order term of $\tilde{g}^{(k_x,k_y)}_{oe}$ is $\mathcal{O}(N^{1-(k_x+k_y)/2})$. Hence, $\tilde{g}^{(k_x,k_y)}_{oe}$ becomes less and less relevant as $k_x,k_y$ increases, allowing us to safely truncate the sum over $k_x,k_y$. To obtain accuracy up to and including $\mathcal{O}(N^{-2})$, it is sufficient to truncate the sum at $k_x+k_y=6$, just as we did in App.~\ref{app:SaddlePoint}. Next, we substitute the perturbative solution $X_{\alpha,\beta,\Delta m,\Delta n}$ into $\tilde{g}^{(k_x,k_y)}_{J+L,\alpha,\beta,\Delta m,\Delta n}$ and evaluate the coefficients up to $k_x+k_y=6$ using the definition in \cref{eq:gOEDefAndExpansionCoeffs}. We find that $\tilde{g}_{J+L,\alpha,\beta,\Delta m,\Delta n}^{(0,0)}=\sqrt{2}\tilde{g}^{(0)}_{\alpha,\beta,\Delta m,\Delta n}$, defined in the first few lines of \cref{eq:gSlowTermsAtSP,eq:gFastTermsAtSP}. For $k_x>0$, the $\tilde{g}_{J+L,\alpha,\beta,\Delta m,\Delta n}^{(k_x,0)}$ coefficients are identical to the $\tilde{g}^{(k)}_{\alpha,\beta,\Delta m,\Delta n}$ coefficients in \cref{eq:gSlowTermsAtSP,eq:gFastTermsAtSP}. This is not generally true for derivatives $\tilde{g}^{(0,k)}_{J+L,\alpha,\beta,\Delta m,\Delta n}$ in the $y$-direction. However, it \emph{is} true for $k=2$; hence, the $x^2$ and $y^2$ terms in \cref{eq:gOEsaddlePointExpansion} share the same prefactor $1/(2\sigma_{\alpha,\beta,\Delta m,\Delta n}^2)$. To leading order, the prefactor of the linear in $y$ term reads
\begin{equation}\label{eq:zTerm}
\begin{aligned}
    z_{\alpha,\beta,\Delta m,\Delta n}= -(\alpha+\beta)\sqrt{N}\omega_s t + \mathcal{O}(N^{-1/2}).
\end{aligned}
\end{equation}

We are now ready to evaluate the integrals in \cref{eq:IevenoddDefs1}. Substituting \cref{eq:XYcoordinates} along with the rescaling $\dd{j}\dd{l}=\frac{N}{2}\dd{x}\dd{y}$ into $\mathcal{I}_{oe}$, $\mathcal{I}_{eo}$ gives
\begin{equation}\label{eq:integralsInXandMuCoords}
    \begin{aligned}
    \mathcal{I}_{oe}&=\frac{1}{4}\frac{i e^{\tilde{g}^{(0,0)}_{oe}}}{\pi^{3/2}}\frac{\sqrt{N}}{2}\int_{-\infty}^{\infty}\dd{x}e^{-x^2/(2\sigma^2_{oe})}\int_{-\infty}^{\infty}\dd{y}e^{iz_{oe}y}\left(\frac{1}{y} e^{-y^2/(2\sigma^2_{oe})} \right) Q_{oe}(x,y),\\
    \mathcal{I}_{eo}&=\frac{1}{4}\frac{i e^{\tilde{g}^{(0,0)}_{eo}}}{\pi^{3/2}}\frac{\sqrt{N}}{2}\int_{-\infty}^{\infty}\dd{x}e^{-x^2/(2\sigma^2_{eo})}\int_{-\infty}^{\infty}\dd{y}e^{iz_{eo}y}\left(\frac{1}{y} e^{-y^2/(2\sigma^2_{eo})} \right) Q_{eo}(x,y),
    \end{aligned}
\end{equation}
where we introduced $\tilde{g}^{(k_x,k_y)}_{oe}=\tilde{g}^{(k_x,k_y)}_{\alpha,\beta,\Delta m,\Delta n}$, $z_{oe}=z_{\alpha,\beta,\Delta m,\Delta n}$, $\sigma_{oe}=\sigma_{\alpha,\beta,\Delta m,\Delta n}$, and for $eo$ we send $\alpha\leftrightarrow-\beta$, $\Delta m\leftrightarrow\Delta n$ to make the expressions more concise. $Q_{oe}(x,y)$ and $Q_{eo}(x,y)$ are polynomial functions of $x,y$ which take the form
\begin{equation}\label{eq:Qpolynomial}
    \begin{aligned}
        Q_{oe}(x,y)&=\sum_{j=0}^{\infty}\left(\sum_{k_x+k_y=3}^{\infty}\tilde{g}^{(k_x,k_y)}_{oe}\frac{x^{k_x}y^{k_y}}{k_x! k_y!}\right)^j\sum_{k_x',k_y'}x^{k_x'}y^{k_y'}\frac{\partial^{k_x'+k_y'}}{\partial k_x'\partial k_y'}\left(R^*_{\sigma',s',\beta,\Delta n}(2j(x,y))R_{\sigma,s,\alpha,\Delta m}(2l(x,y))\right)|_{x,y=0},\\
         Q_{eo}(x,y)&=\sum_{j=0}^{\infty}\left(\sum_{k_x+k_y=3}^{\infty}\tilde{g}^{(k_x,k_y)}_{eo}\frac{x^{k_x}y^{k_y}}{k_x! k_y!}\right)^j\sum_{k_x',k_y'}x^{k_x'}y^{k_y'}\frac{\partial^{k_x'+k_y'}}{\partial k_x'\partial k_y'}\left(R^*_{\sigma',s',\beta,\Delta n}(2l(x,y))R_{\sigma,s,\alpha,\Delta m}(2j(x,y))\right)|_{x,y=0},
    \end{aligned}
\end{equation}
where we Taylor expanded the resonator amplitudes $R_{\sigma,s,\alpha,\Delta m}$ around the saddle point $x,y=0$. The choice of the order of integration in \cref{eq:integralsInXandMuCoords} --- first over $y$, then over $x$ --- is not arbitrary. We will now show that, at $t\gtrsim t_r$, the only surviving contribution of the inner integral over $y$ comes from terms of the form $e^{-y^2/(2\sigma^2)}/y$, which greatly simplifies the evaluation of the outer integral over $x$. To show why this is the case, we use the standard Fourier transform of $y^n e^{-y^2/(2\sigma^2)}$,
\begin{equation}\label{eq:standardFourierTransformGaussianPolynomial}
    \int_{-\infty}^{\infty}\dd{y} e^{-izy} \,\,  y^{n} e^{-y^2/(2\sigma^2)}=\begin{cases}
        \sqrt{2\pi\sigma^2}\left(i\frac{\partial}{\partial z}\right)^n [e^{-\sigma^2z^2/2}],& n\geq 0,\\
        -i\pi \erf(\frac{\sigma z}{\sqrt{2}}), & n=-1.
    \end{cases}
\end{equation}
where $\erf(z)=\frac{2}{\sqrt{\pi}}\int_0^z\dd{t}e^{-t^2}$ is the error function \footnote{to prove the identity in \cref{eq:standardFourierTransformGaussianPolynomial}, differentiate both sides with respect to $z$ and use that the Fourier transform of a Gaussian is another Gaussian, $\int_{-\infty}^{\infty}\dd{y}e^{-izy}e^{-y^2/(2\sigma^2)}=\sqrt{2\pi\sigma^2}e^{-\sigma^2z^2/2}$}. Note that we did not include, and do not care about, the values $n<-1$. This is because the polynomials $Q_{oe/eo}(j(x,y),l(x,y))$ in \cref{eq:integralsInXandMuCoords} only contribute $y^m$ with $m\geq 0$, from the definition in \cref{eq:Qpolynomial}. Therefore, the smallest possible power of $y$ is $-1$, arising from the $1/y$ factor in the inner integral over $y$ in \cref{eq:integralsInXandMuCoords}. Consider first the $n\gtrsim 0$ case in \cref{eq:standardFourierTransformGaussianPolynomial}. Setting $\sigma=\sigma_{\alpha,\beta,\Delta m,\Delta n}$, $z=z_{\alpha,\beta,\Delta m,\Delta n}$, and using the definitions in \cref{eq:gSlowTermsAtSP,eq:gFastTermsAtSP,eq:zTerm}, we find that the argument of the exponential function on the RHS of \cref{eq:standardFourierTransformGaussianPolynomial} is $-2N\omega_s^2 t^2\delta_{\alpha,\beta}+\mathcal{O}(N^{-1/2})$. Meanwhile, the leading order decay term in the $e^{\tilde{g}^{0,0}}_{oe/eo}$ prefactor in \cref{eq:integralsInXandMuCoords} is $-2N\omega_s^2 t^2\delta_{\alpha,-\beta}$, where we used the definitions of $\tilde{g}^{(0)}_{\alpha,\beta,\Delta m,\Delta n}$ in \cref{eq:gSlowTermsAtSP,eq:gFastTermsAtSP}. Thus, for $t\gtrsim t_r$, the $\alpha=-\beta$ contributions decay for all choices of $n$, while the $\alpha=\beta$ contributions decay for $n\geq 0$. This leaves only one surviving contribution: $n=-1$ and $\alpha=\beta$. In the limit $t\gtrsim t_r$, we can also simplify the error function, $\lim_{t\gtrsim t_r}\erf(\frac{\sigma z_{\alpha,\alpha,\Delta m,\Delta n}}{\sqrt{2}})=-\alpha$, where used that the sign of $z_{\alpha,\alpha,\Delta m,\Delta n}$ depends on the sign of $\alpha$, see \cref{eq:zTerm}. Thus, in the $t\gtrsim t_r$ limit, the $\mathcal{I}_{oe/eo}$ in \cref{eq:integralsInXandMuCoords} evaluate to
\begin{equation}\label{eq:integralsInXandMuCoordsReadoutTime}
    \begin{aligned}
    \mathcal{I}_{oe}(t\gtrsim t_r)&=-\frac{\alpha}{4} e^{\tilde{g}^{(0,0)}_{oe}} \sqrt{\frac{N}{2}} \frac{1}{\sqrt{2\pi}}\int_{-\infty}^{\infty}\dd{x}e^{-x^2/(2\sigma^2_{oe})} Q_{oe}(x,0),\\
    \mathcal{I}_{eo}(t\gtrsim t_r)&=\frac{\alpha}{4} e^{\tilde{g}^{(0,0)}_{eo}} \sqrt{\frac{N}{2}} \frac{1}{\sqrt{2\pi}} \int_{-\infty}^{\infty}\dd{x}e^{-x^2/(2\sigma^2_{eo})} Q_{eo}(x,0).
    \end{aligned}
\end{equation}
Substituting \cref{eq:integralsInXandMuCoordsReadoutTime} into \cref{eq:MunequalNoverlap1} and using $\tilde{g}^{0,0}_{-\alpha,-\alpha,\Delta n,\Delta m}=\tilde{g}^{0,0}_{\alpha,\alpha,\Delta m,\Delta n}$, $Q_{oe}(x,0)=Q_{oe}(x,0)$, we obtain
\begin{equation}\label{eq:MunequalNoverlap1_2}
\begin{aligned}
        \bra{R_{\sigma',s'}}\hat{P}_{\gtrless}\ket{R_{\sigma,s}}_{m\neq n,t\gtrsim t_r}&=-\frac{1}{2}\sum_{\substack{\Delta m,\Delta n=0,1}}\sum_{\alpha=\pm 1}\alpha e^{\tilde{g}^{(0,0)}_{oe}} \sqrt{\frac{N}{2}} \frac{1}{\sqrt{2\pi}}\int_{-\infty}^{\infty}\dd{x}e^{-x^2/(2\sigma^2_{oe})} Q_{oe}(x,0).
\end{aligned}
\end{equation}
The form of \cref{eq:integralsInXandMuCoordsReadoutTime} becomes almost identical to \cref{eq:riemannToIntegral1} if we let $x\to l, N/2 \to N$, the only difference being the additional $-\alpha$ factor resulting from the error function integral in \cref{eq:standardFourierTransformGaussianPolynomial}. We already evaluated the Taylor expansion coefficients $\frac{\partial^k}{\partial x^k}\left(R^*_{\sigma',s',\beta,\Delta n}(2l(x,y))R_{\sigma,s,\alpha,\Delta m}(2j(x,y))\right)|_{x,y=0}$ for the saddlepoint calculation of the $\bra{R_{\sigma',s'}}\hat{P}_{\gtrless}\ket{R_{\sigma,s}}_{m=n}$ overlaps in \cref{eq:halfPlaneOverlapsTwoKinds}. Reusing these, we sum over the $\alpha=\beta=\pm 1$, $\Delta m,\Delta n=0,1$ indices in $\mathcal{I}_{oe}$, $\mathcal{I}_{eo}$ and obtain \cref{eq:resonatorHalfPlaneOverlapsMneqN}.

\subsubsection{Saddlepoint expressions for \texorpdfstring{$\bra{R_{\sigma',s'}}\hat{P}_{\gtrless}\ket{R_{\sigma,s}}_{t\gtrsim t_r}$}{overlaps II}}\label{app:halfPlaneOverlapsSaddlePointReadoutTime}

There are ten different overlaps $\bra{R_{\sigma',s'}}\hat{P}_{\gtrless}\ket{R_{\sigma,s}}$: four with $\sigma=\sigma'$, $s=s'$, two with $\sigma=\sigma'$, $s\neq s'$, two with $\sigma=-\sigma'$, $s=s'$ and two with $\sigma=-\sigma'$, $s\neq s'$. All other overlaps in \cref{eq:overlapHalfPlaneRmatComponents} can be obtained by taking the complex conjugates of the ten overlaps we have just listed. 

For the $\bra{R_{\sigma',s'}}\ket{R_{\sigma,s}}_{m=n,t\gtrsim t_r}$ contributions to the overlaps, we directly apply the method from App.~\ref{app:SaddlePoint}, keeping only the slow $\alpha=\beta$ terms. Inserting the saddlepoint expressions for $\Theta_\pm(t\gtrsim t_r)$, $\Phi_\pm(t\gtrsim t_r)$ in \cref{eq:PhiThetaPMSaddlePointReadoutTime}, and expanding in powers of $1/\sqrt{N}$ up to and including $\mathcal{O}(N^{-2})$, gives
\begin{equation}\label{eq:resonatorHalfPlaneOverlapsMM}
    \begin{aligned}
        \bra{R_{\pm,\parallel}}\hat{P}_{\gtrless}\ket{R_{\pm,\parallel}}_{m=n,t\gtrsim t_r}&=\frac{1}{2}\Bigg[1-\textcolor{brown}{\frac{1}{2}\frac{\omega_s}{\OmegaJC}\left(1\mp\frac{\Delta}{\OmegaJC}\right)^2}  \\
        &\textcolor{orange}{+\frac{ \omega_s^2}{\OmegaJC^{3}}\left(1\mp\frac{\Delta}{\OmegaJC}\right)\left(\frac{1}{4\OmegaJC^{2}}\left(\mp\Delta(9\Delta^2-56g^2N)+\OmegaJC(7\Delta^2-8g^2N)\right)\mp 2\omega_s^2N\Delta t^2\right)} + \mathcal{O}(N^{-5/2}) \Bigg], \\
        \bra{R_{\pm,\perp}}\hat{P}_{\gtrless}\ket{R_{\pm,\perp}}_{m=n,t\gtrsim t_r}&=\frac{1}{2}\Bigg[\textcolor{brown}{\frac{1}{2}\frac{\omega_s}{\OmegaJC}\left(1\mp \frac{\Delta}{\OmegaJC}\right)^2}  \\
        &\textcolor{orange}{-\frac{ \omega_s^2}{\OmegaJC^{3}}\left(1\mp\frac{\Delta}{\OmegaJC}\right)\left(\frac{1}{4\OmegaJC^{2}}\left(\mp\Delta(9\Delta^2-56g^2N)+\OmegaJC(7\Delta^2-8g^2N)\right)\mp 2\omega_s^2N\Delta t^2\right)}+ \mathcal{O}(N^{-5/2})\Bigg], \\
        \bra{R_{\pm,\perp}}\hat{P}_{\gtrless}\ket{R_{\pm,\parallel}}_{m=n,t\gtrsim t_r}&=\frac{1}{2}\Bigg[\textcolor{orange}{\frac{4\omega_s^2 g\sqrt{N}\Delta^2}{\OmegaJC^7}(8g^2N-\Delta^2)\left(1\mp\frac{\Delta}{\OmegaJC}\right)} + \mathcal{O}(N^{-5/2})\Bigg]\\
        \bra{R_{-,\parallel}}\hat{P}_{\gtrless}\ket{R_{+,\parallel}}_{m=n,t\gtrsim t_r}&=\frac{1}{2}\Bigg[\textcolor{brown}{\frac{\omega_s(\Delta^2+2g^2N)}{\OmegaJC^4}} + \textcolor{gray}{\frac{2i\omega_s^2\Delta^2 t}{\OmegaJC^3}} \\
        &+\textcolor{orange}{\frac{\omega_s^2}{\OmegaJC^2}\left(\frac{1}{\OmegaJC^4}\left(-3\Delta^4+12\Delta^2 g^2N + 8g^4N^2\right) - \frac{2\omega_s\Delta^2t^2(\Delta^2+3g^2N)}{\OmegaJC^3}\right)}+ \mathcal{O}(N^{-5/2})\Bigg],\\
        \bra{R_{-,\perp}}\hat{P}_{\gtrless}\ket{R_{+,\perp}}_{m=n,t\gtrsim t_r}&=\frac{1}{2}\Bigg[\textcolor{brown}{-\frac{2N\omega_s^2}{\OmegaJC^2}}+\textcolor{orange}{\frac{2\omega_s^3N}{\OmegaJC^5}\left(10\Delta^2-4g^2N + \Delta^2g^2t^2\right)}\Bigg],\\
        \bra{R_{\mp,\perp}}\hat{P}_{\gtrless}\ket{R_{\pm,\parallel}}_{m=n,t\gtrsim t_r}&=\mp\frac{1}{2}\frac{\omega_s g\sqrt{N} \Delta}{\OmegaJC^3}\Bigg[\textcolor{brown}{-1} +\textcolor{gray}{\frac{2i\omega_s \Delta t}{\OmegaJC}} +\textcolor{orange}{\frac{2 \omega_s}{\OmegaJC}\left[\frac{2}{\OmegaJC^2}(\Delta^2-7g^2N)+\omega_st^2\left(2\omega_s N\mp\Delta\right)\right]} + \mathcal{O}(N^{-5/2}) \Bigg],
    \end{aligned}
\end{equation}
Note that $\bra{R_{\pm,\parallel}}\hat{P}_{\gtrless}\ket{R_{\pm,\parallel}}_{m=n,t\gtrsim t_r}+\bra{R_{\pm,\perp}}\hat{P}_{\gtrless}\ket{R_{\pm,\perp}}_{m=n,t\gtrsim t_r}=1/2$, a property that is enforced to all orders of $1/\sqrt{N}$. To see why, set $c_\pm=1,c_{\mp}=0$ in $\ket{\Psi(t)}$, using the definition in \cref{eq:timeEvolvedWavefunctionNonDispersiveLimit}. Then, $\bra{\Psi(t)}\ket{\Psi(t)}=1$ implies $\bra{\pm,R_{\pm}}\ket{\pm,R_{\pm}}=1$, which, after inserting the definition of $\ket{\pm,R_{\pm}}$ in \cref{eq:ResonatorPMstate} and expanding, produces the identity $\bra{R_{\pm,\parallel}}\ket{R_{\pm,\parallel}}+\bra{R_{\pm,\perp}}\ket{R_{\pm,\perp}}=1$. The half-plane $m=n$ overlaps $\bra{R_{\pm,\parallel}}\hat{P}_{\gtrless}\ket{R_{\pm,\parallel}}_{m=n}$ and $\bra{R_{\pm,\perp}}\hat{P}_{\gtrless}\ket{R_{\pm,\perp}}_{m=n}$ are exactly half of the full-plane overlaps, from which the identity follows to all orders $1/N$.

For the $\bra{R_{\sigma,s}}\ket{\sigma',s'}_{m\neq n,t\gtrsim t_r}$ contributions to the overlaps, we use the modified saddlepoint formula \cref{eq:MunequalNoverlap1}. After simplifying some hefty algebra (using FullSimplify in Mathematica with plenty of hand-holding), we obtain
\begin{equation}\label{eq:resonatorHalfPlaneOverlapsMneqN}
    \begin{aligned}
        \bra{R_{\pm,\parallel}}\hat{P}_{>}\ket{R_{\pm,\parallel}}_{m\neq n,t\gtrsim t_r}&=\mp\frac{1}{2}\Bigg[ 1-\textcolor{brown}{\frac{1}{2}\frac{\omega_s}{\OmegaJC}\left(1\mp\frac{\Delta}{\OmegaJC}\right)^2}  \\
        &\quad \textcolor{orange}{-\frac{\omega_s^2}{\OmegaJC^2}\left[\frac{\omega_s N}{\OmegaJC^3}\left(14g^2N+13 \Delta^2 \mp 18\Delta \OmegaJC\right) +\left(\mp\frac{3\Delta^3}{\OmegaJC^3}+ 2N\omega_s^2 t^2\right)\left(1\mp \frac{\Delta}{\OmegaJC}\right)\right]} + \mathcal{O}(N^{-5/2})\Bigg], \\
        \bra{R_{\pm,\perp}}\hat{P}_{>}\ket{R_{\pm,\perp}}_{m\neq n,t\gtrsim t_r}&=\mp\frac{1}{2}\Bigg[ \textcolor{orange}{\frac{\omega_s^3N}{\OmegaJC^3}\left(-\frac{\Delta^2}{\OmegaJC^2}+2g^2t^2\left(1\mp\frac{\Delta}{\OmegaJC}\right)\right)} + \mathcal{O}(N^{-5/2})\Bigg], \\
        \bra{R_{\pm,\perp}}\hat{P}_{>}\ket{R_{\pm,\parallel}}_{m\neq n,t\gtrsim t_r}&=\pm\frac{\omega_s^2 g\sqrt{N}}{2\OmegaJC^3}\Bigg[\textcolor{gray}{\mp i\OmegaJC t\left(1\mp \frac{\Delta}{\OmegaJC}\right)^2}\\
        &\quad \textcolor{orange}{\mp\frac{\Delta}{\OmegaJC}\left(-\frac{4\omega_s N}{\OmegaJC}\left(\mp\frac{\Delta}{\OmegaJC}-2-\frac{4g^2N}{\OmegaJC^2}\right) \mp \frac{11\Delta^3}{\OmegaJC^3}\left(1\mp \frac{\Delta}{\OmegaJC}\right) + \OmegaJC\omega_s t^2\left(1\mp \frac{\Delta}{\OmegaJC}\right)^2\right)} \\
        &+ \mathcal{O}(N^{-5/2}) \Bigg], \\
        \bra{R_{-,\parallel}}\hat{P}_{>}\ket{R_{+,\parallel}}_{m\neq n,t\gtrsim t_r}&=\frac{1}{2}\frac{\omega_s\Delta}{\OmegaJC^2}\Bigg[-\textcolor{brown}{1} - \textcolor{gray}{2i\omega_s t\frac{\Delta^2}{\OmegaJC^2}} + \textcolor{orange}{\frac{2\omega_s}{\OmegaJC}\left(\frac{3}{2\OmegaJC^2}(\Delta^2-6g^2N) + \frac{\omega_s t^2(\Delta^2-g^2N)}{\OmegaJC}\right)}+ \mathcal{O}(N^{-5/2}) \Bigg],\\
        \bra{R_{-,\perp}}\hat{P}_{>}\ket{R_{+,\perp}}_{m\neq n,t\gtrsim t_r}&=\frac{1}{2} \frac{2\omega_s^3 \Delta N t}{\OmegaJC^3}\Bigg[-\textcolor{gray}{2i}+ \textcolor{orange}{\omega_s t} + \mathcal{O}(N^{-5/2}) \Bigg],\\
        \bra{R_{\mp,\perp}}\hat{P}_{>}\ket{R_{\pm,\parallel}}_{m\neq n,t\gtrsim t_r}&=\mp\frac{\omega_s g\sqrt{N}}{2\OmegaJC^2}\Bigg[\textcolor{brown}{\pm\frac{\Delta}{\OmegaJC}}  \mp \textcolor{gray}{\frac{2i\omega_s t(\Delta^2+2g^2N)}{\OmegaJC^2}} \\
        &\quad \textcolor{orange}{\mp\frac{\Delta\omega_s}{\OmegaJC^2}\left[\mp\frac{\Delta}{\OmegaJC}-\frac{3}{\OmegaJC^2}(6g^2N-\Delta^2)\mp 2\omega_s\Delta t^2\right]} + \mathcal{O}(N^{-5/2})\Bigg],
    \end{aligned}
\end{equation}
with $\bra{R_{\sigma',s'}}\hat{P}_{<}\ket{R_{\sigma,s}}_{m\neq n,t\gtrsim t_r}=-\bra{R_{\sigma',s'}}\hat{P}_{>}\ket{R_{\sigma,s}}_{m\neq n,t\gtrsim t_r}$, note the extra minus sign in comparison to $\bra{R_{\sigma',s'}}\hat{P}_{<}\ket{R_{\sigma,s}}_{m=n,t\gtrsim t_r}=\bra{R_{\sigma',s'}}\hat{P}_{>}\ket{R_{\sigma,s}}_{m=n,t\gtrsim t_r}$. To obtain $\bra{R_{\sigma',s'}}\hat{P}_{\gtrless}\ket{R_{\sigma,s}}_{t\gtrsim t_r}$, we therefore simply take the sum or difference of the $m=n$ and $m\neq n$ contributions in \cref{eq:resonatorHalfPlaneOverlapsMM,eq:resonatorHalfPlaneOverlapsMneqN},
\begin{equation}\label{eq:totalOverlapPosNegMom}
    \begin{aligned}
    \bra{R_{\sigma',s'}}\hat{P}_{\gtrless}\ket{R_{\sigma,s}}_{t\gtrsim t_r}&=\bra{R_{\sigma',s'}}\hat{P}_{\gtrless}\ket{R_{\sigma,s}}_{m=n,t\gtrsim t_r}\pm\bra{R_{\sigma',s'}}\hat{P}_{>}\ket{R_{\sigma,s}}_{m\neq n,t\gtrsim t_r}.
    \end{aligned}
\end{equation}
If the readout scheme were ideal, we would have $\bra{R_{\sigma',s'}}\hat{P}_{\gtrless}\ket{R_{\sigma,s}}_{t\gtrsim t_r}=\delta_{\sigma,\sigma'}\delta_{s',\parallel}\delta_{s,\parallel}$. This would make the post-measurement density matrices defined in \cref{eq:rhoPostMeasurementComponents} pure and equal to $\rho_{\gtrless}=\frac{1}{2}(1\mp r)\frac{1}{2}(1+\sigma^z)$. I.e., if we measure $p<0$ in the resonator, the qubit is exactly in $\ket{+_{\parallel}(t)}$, with probability $\frac{1}{2}(1+r)$, whilst if we measure $p>0$, it is exactly in $\ket{-_{\parallel}(t)}$, with probability $\frac{1}{2}(1-r)$. The $\mathcal{O}(N^{-1})$ and $\mathcal{O}(N^{-2})$ terms in \cref{eq:resonatorHalfPlaneOverlapsMM,eq:resonatorHalfPlaneOverlapsMneqN} therefore describe the imperfections in our readout scheme. By keeping track of them, we will be able to evaluate the leading order errors in the nondispersive readout scheme.

Finally, we also need the overlaps of the qubit states in \cref{eq:overlapHalfPlaneRmatComponentsBreakdown} at $t\gtrsim t_r$. Substituting in \cref{eq:PhiThetaPMSaddlePointReadoutTime} and Taylor expanding up to and including $\mathcal{O}(N^{-2})$ gives
\begin{equation}\label{eq:spinOverlapsReadoutTime}
    \begin{aligned}
        \bra{\mp_{\parallel}}\ket{\pm_{\parallel}}&=\textcolor{purple}{\pm\frac{2i\omega_s g\sqrt{N}t}{\OmegaJC}} + \textcolor{brown}{\frac{2g\sqrt{N}\omega_s^2t^2}{\OmegaJC}}  \textcolor{gray}{\mp\frac{4ig\sqrt{N}\omega_s t}{\OmegaJC}\left(\frac{\omega_s}{\OmegaJC^3}(\Delta^2-2g^2N)+\frac{\omega_s^2t^2}{3}\right)}\\
        &\qquad +\textcolor{orange}{\frac{2\omega_s^2 g\sqrt{N}}{\OmegaJC^3}\left[\frac{2\Delta^2}{\OmegaJC^4}(5\Delta^2-4g^2N)+\frac{2\omega_st^2}{\OmegaJC}(4g^2N-\Delta^2)-\frac{g^4t^4}{3}\right]} +\mathcal{O}(N^{-5/2}),\\
        \bra{\mp_{\perp}}\ket{\pm_{\perp}}&=\textcolor{purple}{\pm\frac{2i\omega_s g\sqrt{N}t}{\OmegaJC}} + \textcolor{brown}{\frac{2g\sqrt{N}\omega_s^2t^2}{\OmegaJC}}  \textcolor{gray}{\mp\frac{4ig\sqrt{N}\omega_s t}{\OmegaJC}\left(\frac{\omega_s}{\OmegaJC^3}(\Delta^2-2g^2N)+\frac{\omega_s^2t^2}{3}\right)}\\
        &\qquad +\textcolor{orange}{\frac{2\omega_s^2 g\sqrt{N}}{\OmegaJC^3}\left[-\frac{2\Delta^2}{\OmegaJC^4}(5\Delta^2-4g^2N)+\frac{4\omega_st^2}{\OmegaJC}(2g^2N-\Delta^2)-\frac{g^4t^4}{3}\right]} +\mathcal{O}(N^{-5/2}),\\
        \bra{\mp_{\perp}}\ket{\pm_{\parallel}}&=1\mp\textcolor{purple}{i\omega_s t\left(1\mp\frac{\Delta}{\OmegaJC}\right)}-\textcolor{brown}{\omega_s^2t^2\left(1\mp\frac{\Delta}{\OmegaJC}\right)} \\
        &\qquad \textcolor{gray}{\pm i\omega_s t\left[\frac{\omega_s}{\OmegaJC^3}\left(\Delta^2-4g^2N\mp\frac{\Delta}{\OmegaJC}(\Delta^2-8g^2N)\right)+\frac{2\omega_s^2t^2}{3}\left(1\mp\frac{\Delta}{\OmegaJC}\right)\right]}\\
        &\qquad +\textcolor{orange}{\frac{\omega_s^3t^2}{\OmegaJC}\left[\frac{2}{\OmegaJC^3}\left(\mp\Delta(\Delta^2-6g^2N)+\OmegaJC(\Delta^2-4g^2N)\right)+\frac{g^2t^2}{3}\left(1\mp\frac{\Delta}{\OmegaJC}\right)\right]} +\mathcal{O}(N^{-5/2}).
    \end{aligned}
\end{equation}

\subsection{Saddlepoint formulas for fidelity and QNDness in the nondispersive two-drive model\texorpdfstring{, \cref{eq:JCplusRabiModel}}{}}

To obtain the saddlepoint formulas for the post-measurement density matrices $\hat{\rho}_\gtrless$, we substitute \cref{eq:resonatorHalfPlaneOverlapsMM,eq:resonatorHalfPlaneOverlapsMneqN,eq:totalOverlapPosNegMom,eq:spinOverlapsReadoutTime} into \cref{eq:overlapHalfPlaneRmatComponents} and evaluate the density matrix components using \cref{eq:rhoPostMeasurementComponents}. After some simplifications, we obtain the following post-measurement density components 
\begin{equation}\label{eq:PostMeasurementDensityMatrixComponentsSaddlepoint}
    \begin{aligned}
    [\rho_{\text{q,}\gtrless}]_{\parallel\parallel}(t_r)&=\frac{1}{2}(1\mp r) - \textcolor{brown}{\frac{\omega_s}{4\OmegaJC^3}\left[2g^2N(1\mp 3r) \pm\Delta\left(\pm\Delta(1\mp 3r)+(3\mp r)\OmegaJC\right) \pm 4g\sqrt{N}\Delta\sqrt{1-r^2}\cos(\Delta\Phi)\right]}\\
    &+\textcolor{orange}{\frac{\omega_s^2}{4\OmegaJC^2}\Bigg[\frac{2\omega_s^2N^2}{\OmegaJC^2}(-7\pm 15 r) + \frac{8\omega_s^3N^2t^2}{\OmegaJC}(-1\pm r) + \frac{4\omega_s\Delta^2N}{\OmegaJC^3}(-3\pm 8 r) + (3\mp 11 r)\frac{\Delta^4}{\OmegaJC^4}}\\
    &\qquad \textcolor{orange}{\pm\frac{\Delta}{\OmegaJC}\left(\frac{2\omega_s N}{\OmegaJC}(-25\pm 9 r) + (11\mp 3 r)\frac{\Delta^2}{\OmegaJC^2}\right)}\\
    &\qquad \textcolor{orange}{\pm\frac{\Delta}{\OmegaJC}\frac{2g\sqrt{N}}{\OmegaJC}\sqrt{1-r^2}\cos(\Delta\Phi)\left(\pm\frac{\Delta}{\OmegaJC}+\frac{1}{\OmegaJC^2}(7\Delta^2-46g^2N)\right)\Bigg]}+ \mathcal{O}(N^{-5/2}),\\
    [\rho_{\text{q,}\gtrless}]_{\perp\perp}(t_r)&=\textcolor{brown}{\mp r\frac{\omega_s}{8\OmegaJC}\left(1\mp\frac{\Delta}{\OmegaJC}\right)^2}\\
    &-\textcolor{orange}{\frac{\omega_s^2}{4\OmegaJC^2}\Bigg[\pm 2\cos(\Delta\Phi)\sqrt{1-r^2}\frac{g\sqrt{N}}{\OmegaJC}\frac{\Delta}{\OmegaJC}\left(\pm\frac{\Delta}{\OmegaJC} + \frac{1}{\OmegaJC^2}(\Delta^2-10g^2N)\right)}\\
    &\qquad \textcolor{orange}{+\frac{8\omega_s^3 N^2t^2}{\OmegaJC}(-1\pm r) + \frac{\Delta^4}{\OmegaJC^4}\left(\pm\frac{\Delta}{\OmegaJC}(5\mp 3r)+(3\mp 5 r)\right) + \frac{2\omega_s^2N^2}{\OmegaJC^2}\left(\pm\frac{4\Delta}{\OmegaJC}(-7\pm 9 r) - 7 \pm r\right)}\\
    &\qquad \textcolor{orange}{+\frac{2\omega_s N}{\OmegaJC}\frac{\Delta^2}{\OmegaJC^2}\left(\pm \frac{3\Delta}{\OmegaJC}(1\pm r) + 2(-3\pm r)\right)\Bigg]} + \mathcal{O}(N^{-5/2}),\\
    [\rho_{\text{q,}\gtrless}]_{\perp\parallel}(t_r)&=\textcolor{brown}{\pm\frac{\omega_s}{4\OmegaJC}\sqrt{1-r^2}\left(4i\sin(\Delta\Phi)\frac{\omega_s N}{\OmegaJC} - e^{\pm i \Delta\Phi}\frac{\Delta}{\OmegaJC}\left(1\mp \frac{\Delta}{\OmegaJC}\right)\right)}\\
    &\qquad \textcolor{gray}{\pm\frac{i\omega_s^2 g\sqrt{N}t}{2\OmegaJC^2}\sqrt{1\mp r}\left(\frac{1}{2}\sqrt{1\mp r}\left(1\pm\frac{\Delta}{\OmegaJC}\right)^2 - \frac{2g\sqrt{N}}{\OmegaJC}\sqrt{1\pm r}\left(1\mp\frac{\Delta}{\OmegaJC}\right)\cos(\Delta\Phi)\right)}\\
    &\quad+\textcolor{orange}{\frac{\omega_s^2}{4\OmegaJC^2}\Bigg[ \pm\frac{\Delta}{\OmegaJC}\bigg[\pm\frac{\Delta}{\OmegaJC}\sqrt{1-r^2}\left(\frac{4e^{\mp i \Delta\Phi}\omega_s N}{\OmegaJC}\frac{1}{\OmegaJC^2}(18g^2N+5\Delta^2) + \frac{e^{\pm i \Delta\Phi}}{\OmegaJC^4}(56g^4N^2-3\Delta^4)\right)}\\
    &\qquad\textcolor{orange}{-\frac{2g\sqrt{N}}{\OmegaJC}\frac{1}{\OmegaJC^4}\left(48g^4N^2 + 8g^2N\Delta^2(1\pm 4 r)+ \Delta^4(11\mp 4r)\right)}\\
    &\qquad \textcolor{orange}{\pm \frac{\Delta}{\OmegaJC}\frac{2g\sqrt{N}}{\OmegaJC}\left(-\frac{4\omega_s N}{\OmegaJC}(-8\pm r)+ \frac{\Delta^2}{\OmegaJC^2}(-4\pm 11 r)\right)}\\
    &\qquad \textcolor{orange}{- 3e^{\pm i\Delta\Phi}\sqrt{1-r^2}\frac{1}{\OmegaJC^4}(24g^4N^2+2g^2N\Delta^2-\Delta^4)\bigg]} \\
    &\qquad \textcolor{orange}{\pm 64 i\frac{\omega_s^3N^3}{\OmegaJC^3}\sqrt{1-r^2}\sin(\Delta\Phi)\pm\frac{\Delta}{\OmegaJC}\omega_s g\sqrt{N}t^2(-1\pm r)\left(1\pm\frac{\Delta}{\OmegaJC}\right)^2\Bigg]}+\mathcal{O}(N^{-5/2}),
    \end{aligned}
\end{equation}
with $[\rho_{\text{q,}\gtrless}]_{\parallel\perp}=[\rho_{\text{q,}\gtrless}]^*_{\perp\parallel}$. Recall that the $\rho_{\text{q,}\gtrless}$ are not normalized. Indeed, $\Tr[\rho_{\text{q},\gtrless}]$ gives the probability $P_{\gtrless}$ of measuring $p\gtrless 0$ in the resonator state. At readout time $t=t_r$, the probabilities $P_{\gtrless}(t_r)$ are given by
\begin{equation}\label{eq:probabilityPosNegMomentumReadoutTimeSaddlepoint}
\begin{aligned}
    P_{\gtrless}(t_r)=\Tr[\hat{\rho}_\gtrless(t_r)]&=\frac{1}{2}\left(1\mp r\mp 2\delta P\right).
\end{aligned}
\end{equation}
In an ideal readout scheme, $\delta P=0$, and $P_{\gtrless}(t_r)=|c_{\mp}|^2=\frac{1}{2}(1\mp r)$ perfectly replicates the state in which the qubit is initialized. For the nondispersive JC model, $\delta P$ is finite and proportional to $1/N$ to leading order (brown terms),
\begin{equation}\label{eq:deltaP}
    \begin{aligned}
       \delta P&= \frac{1}{2}\bigg[\textcolor{brown}{\frac{\omega_s}{\OmegaJC}\left(\frac{\Delta}{\OmegaJC}-r\frac{1}{2}\left(1+\frac{\Delta^2}{\OmegaJC^2}\right)+2\sqrt{1-r^2}\frac{g\sqrt{N}}{\OmegaJC}\frac{\Delta}{\OmegaJC}\cos(\Delta\Phi)\right)}\\
    &+\textcolor{orange}{\frac{\omega_s^2}{\OmegaJC^2}\left(\frac{3\Delta}{\OmegaJC}\frac{1}{\OmegaJC^2}(6g^2N-\Delta^2)\left(1+\frac{2g\sqrt{N}}{\OmegaJC}\sqrt{1-r^2}\cos(\Delta\Phi)\right)+\frac{r}{\OmegaJC^4}(3\Delta^4-14\Delta^2g^2N-14g^4N^2)\right)}\bigg],
    \end{aligned}
\end{equation}
where we also included the $\mathcal{O}(N^{-2})$ terms (orange), since our calculation gave us access to them. Note that there is special value $r^*$ given by 
\begin{equation}\label{eq:rStar}
    r^*=\frac{\Delta\OmegaJC}{\Delta^4+8\Delta^2g^2N+4g^4N^2}\left[\Delta^2+2g^2N\left(1+\sign(\Delta) \frac{2\sqrt{\Delta^2+g^2N}}{\OmegaJC}\right)\right],
\end{equation}
where the $\mathcal{O}(N^{-1})$ terms (brown terms in \cref{eq:deltaP}) vanish. At $r=r^*$, the fidelity approaches $1$ to within $\mathcal{O}(N^{-2})$. Hence,  on a logarithmic scale $1-\mathcal{F}(t_r)$ shows a pronounced negative peak at $r^*$, see dashed lines in \cref{fig:fidelityQNDnessVaryRandDeltaPhi}(b). Inserting \cref{eq:probabilityPosNegMomentumReadoutTimeSaddlepoint} into the definition of the fidelity, \cref{eq:fidelityDef}, gives
\begin{equation}\label{eq:fidelityErrorNonDispersive}
    1-\mathcal{F}(t_{\text{off}}\gtrsim t_r)=\begin{cases}
         \textcolor{brown}{\frac{1}{8}\frac{\omega_s}{\OmegaJC}\left(1\mp\frac{\Delta}{\OmegaJC}\right)^2}, &  r=\pm 1\\
        \frac{\delta P^2}{2}\frac{1}{1-r^2}, & r\neq \pm 1
    \end{cases},
\end{equation}
where the cases $r\neq \pm 1$ are correct as long as $\delta P\ll (1\pm r)$. Repeating the procedure for the QNDness in \cref{eq:QNDnessDef} gives
\begin{equation}\label{eq:qndnessErrorNonDispersive}
\begin{aligned}
    1-\text{QNDness}(t_{\text{off}}\gtrsim t_r)&=\textcolor{brown}{\frac{\omega_s}{2\OmegaJC}\left(1+\frac{\Delta^2}{\OmegaJC^2}-2r\frac{\Delta}{\OmegaJC}\right)} \textcolor{orange}{+\frac{\omega_s^2}{2\OmegaJC^2}\Bigg[\frac{8\omega_s^3N^2t_{\text{off}}^2}{\OmegaJC} + \frac{14\omega_s^2N^2}{\OmegaJC^2} + \frac{12N\omega_s\Delta^2}{\OmegaJC^3} - \frac{3\Delta^4}{\OmegaJC^4} }\\
    &\qquad \textcolor{orange}{+\frac{\Delta}{\OmegaJC}\left(3r\left(\frac{\Delta^2}{\OmegaJC^2}-\frac{6\omega_sN}{\OmegaJC}\right) - \frac{2\Delta g\sqrt{N}}{\OmegaJC^2}\sqrt{1-r^2}\cos(\Delta\Phi)\right)\Bigg]}.
\end{aligned}
\end{equation}
In the limit of large $N$, the maximum error in QNDness occurs for a state initialized in $r=-\sign(\Delta)$,
\begin{equation}\label{eq:errorNonDispersive_2}
\begin{aligned}
    \text{max}_{r,\Delta\Phi}(\text{error}(1-\text{QNDness}(t_r)))=\textcolor{brown}{\frac{1}{4}\frac{\omega_s}{\OmegaJC}\left(1+\frac{|\Delta|}{\OmegaJC}\right)^2} + \mathcal{O}(N^{-2}).
\end{aligned}
\end{equation}
Using \cref{eq:fidelityErrorNonDispersive}, the maximum error in the fidelity, $\max(1-\mathcal{F}(t_r))$, is four times smaller than $\max(1-\text{QNDness}(t_r))$, see \cref{eq:errorNonDispersive_2}, just as in the dispersive model, see text above \cref{eq:errorFidelityQNDnessDispersive}. Note that we have written all the expressions in this appendix for the JC model without the classical drive, \cref{eq:JCmodel}, to avoid overly long formulas. For the two-drive model \cref{eq:JCplusRabiModel}, simply send $N\to \Neff$ (defined in \cref{eq:effectiveTwoDrivePhotonNumber}), $\omega_s\to\bar{\omega}_s$ and $\OmegaJC\to\bar{\Omega}_{\text{JC}}$ (see definitions in Table I) in all the expressions.

\end{document}